\documentclass[12pt,leqno]{amsart}
\usepackage{amsmath,amsthm,amssymb}

\def\init{\setcounter{equation}{0}}
\newtheorem{theorem}{Theorem}[section]
\newtheorem{corollary}{Corollary}[section]

\newtheorem{lemma}[theorem]{Lemma}

\newcommand{\e}{{\varepsilon}}
\newcommand{\D}{\mathcal{D}}

\def\init{\setcounter{equation}{0}}

\newcommand{\R}{\mathbb{R}}

\newcommand{\C}{\mathbb{C}}

\newcommand{\Z}{\mathbb{Z}}

\newcommand{\T}{\mathbb{T}}

\newcommand{\rw}{\rightarrow}

\usepackage{tikz}
\usetikzlibrary{decorations.pathreplacing}

\title{
Aharonov-Bohm effect revisited
}

\author{
Gregory Eskin}

\begin{document}
\maketitle 
\begin{center}
{\it Department of Mathematics, UCLA,
Los Angeles,\\ CA 90095-1555, USA,
\\
eskin@math.ucla.edu}

\end{center}

\renewcommand{\abstractname}{}

\begin{abstract}
Aharonov-Bohm effect is a quantum mechanical phenomenon that attracted the
attention of many physicists and mathematicians since the publication of
the seminal paper of Aharonov and Bohm [1] in 1959.

We consider different types of Aharonov-Bohm effect such as magnetic AB effect,
electric AB effect,  combined electromagnetic AB effect,  AB effect for the Schr\"odinger
equations with Yang-Mills potentials,  and the gravitational analog  of AB effect.

We shall describe different approaches to prove the AB effect based on the inverse 
scattering problems, the inverse boundary value problems in the presence of obstacles,
spectral asymptotics, and the direct proofs of the AB effect.
\\
\\
  {\it Keywords:}  Aharonov-Bohm effect,  Schr\"odinger equation, gauge equivalence.
\\
\\
Mathematics Subject Classification 2010: 35R30, 35P25, 81P15
\end{abstract}

 \tableofcontents
 
\section{Introduction}
\init
The Aharonov-Bohm effect was 
discovered by Aharonov-Bohm in the famous paper [1].  Consider the Schr\"odinger
equation
\begin{equation}                                                  \label{eq:1.1}
-ih\frac{\partial u}{\partial t}+\frac{1}{2m}\sum_{j=1}^n\Big(-ih\frac{\partial}{\partial x_j}
-\frac{e}{c}A_j(x)\Big)^2 u+eV(x)u=0
\end{equation}
in the plane domain 
$(\R^2\setminus\Omega_1)\times(0,T)$,
where 
\begin{equation}                                         \label{eq:1.2}
u\Big|_{\partial\Omega_1\times(0,T)}=0,
\end{equation}
\begin{equation}                                         \label{eq:1.3}
u(x,0)=u_0(x), \ \ x\in \R^2\setminus\Omega_1.
\end{equation}
Here $\Omega_1$ is a bounded domain in $\R^2$  called an obstacle.
Equation (\ref{eq:1.1}) describes a nonrelativistic  quantum electron 
in a classical electromagnetic field with time-independent magnetic potential 
$A(x)=(A_1(x),A_2(x))$  and electric potential $V(x)$.

Assume 
that the magnetic field $B(x)=\mbox{curl}\, A(x)=0$  in 
$\R^2\setminus\Omega_1$,  i.e. the magnetic field is shielded in $\Omega_1$.
Aharonov and Bohm have shown that despite the absence of the magnetic field in
$\R^2\setminus \Omega_1$  the magnetic potential $A(x)$  has a physical impact.

They proposed the following physical experiment to demonstrate this fact:

Let a coherent beam of electrons splits into two parts  and each part passes on the opposite
sides of the obstacle $\Omega_1$.  Then both beams merge at the interferometer behind
the obstacle
$\Omega_1$.  The interference of these two beams allows to measure
\begin{equation}                        \label{eq:1.4}
\alpha=\frac{e}{hc}\int\limits_\gamma A(x)\cdot dx \ \ (\mbox{modulo\ \ }2\pi n)
\end{equation}
where $n\in \Z$.

Here $\gamma$  is a simple contour encircling $\Omega_1$.  The integral $\alpha$  is called
the magnetic flux.

When two potentials $A^{(1)}(x)$  and $A^{(2)}(x)$
 are such that $\mbox{curl}\, A^{(1)}=\mbox{curl}\, A^{(2)}=0$  but
$$  
\frac{e}{hc}\int\limits_\gamma (A^{(1)}(x)-A^{(2)}(x))\cdot dx\neq 2\pi n,\ \ n\in\Z,
$$
the magnetic potentials 
$A^{(1)}(x)$ and $A^{(2)}(x)$  make a different physical impact,  since  the measurements
of the interferometer are different.  This phenomenon is called the Aharonov-Bohm (AB) effect.

We shall present,  following Wu and Yang [52],
a more general formulation 
of the AB effect that can be applied to  more general situations:

Let $G(\R^2\setminus \Omega_1)$ be the group on $\R^2\setminus\Omega_1$  consisting 
of all smooth complex-valued functions $g(x)$  such  that
$|g(x)|=1$  in $\R^2\setminus\Omega_1$  and
$g(x)=e^{ip\theta(x)}\big(1+O\big(\frac{1}{|x|}\big)\big)$  
when $|x|\rw\infty$.  Here
$p\in \Z,\ O=(0,0)\in\Omega_1$   and $\theta(x)$  is the polar angle of $x$.  The group
$G(\R^2\setminus \Omega_1)$  is called the gauge group.  The map $u'=g^{-1}(x)u$ is called 
the gauge transformation.  If $u(x,t)$ satisfies (\ref{eq:1.1})
and $u'=g^{-1}u$,  then $u'(x,t)$  satisfies (\ref{eq:1.1}) with $A(x)$  replaced  by
\begin{equation}                                            \label{eq:1.5}
\frac{e}{c}A'(x)=\frac{e}{c}A(x)+ihg^{-1}\frac{\partial g}{\partial x}.
\end{equation}
Two magnetic potentials related  by (\ref{eq:1.5}) are called gauge equivalent.
Any two magnetic potentials belonging to the same gauge equivalence class represent
the same physical reality and cannot be distinguished in any physical experiment.

The  Aharonov-Bohm effect  is the statement that 
two magnetic potentials belonging to different gauge equivalent classes make 
a different physical impact.

The first mathematical proof of AB effect was given by Aharonov and Bohm in 
the original paper [1].  They found explicitly  the scattering amplitude
in the case when the obstacle $\Omega_1$ is a point $O$ and 
$A(x)=\frac{\alpha}{2\pi}\big(\frac{-x_2}{|x|},\frac{x_1}{|x|}\big)$.
They have shown that the scattering cross-section is proportional to $\sin^2\frac{\alpha}{2}$
where $\alpha$  is the magnetic flux (\ref{eq:1.4}).  Their result was extended by
Ruijsenaars [43]  to the case when $\Omega_1$ is the disk $|x|\leq R$.

The  
further progress was
done in the solution of the inverse scattering problem of defining the gauge equivalence 
class of magnetic
potential knowing the scattering matrix (amplitude). We shall mention only the works
when obstacles are present.
Nicoleau [39],  Weder [50], and Ballesteros and Weder [2] proved that the scattering matrix asymptotics for high
 energies in the dimensions 2 and 3  determines all integrals
 \begin{equation}                                   \label{eq:1.6}
\exp\Big(\frac{ie}{hc}\int\limits_{-\infty}^\infty A(x_0+t\omega)\cdot \omega dt\Big),
\end{equation}
where $x=x_0+t\omega$ is any straight line that does not intersect the obstacle $\Omega_1$.
When the obstacle is convex they used the  $X$-ray transform to determine 
the gauge equivalence class of the magnetic potential,   in particular,  to determine the magnetic field
$B=\mbox{curl}\,A$.  Further,  using  the second term of high  energy  asymptotic of the scattering 
matrix and knowing $B(x)$,  they were able to determine  all integrals
$\frac{e}{h}\int_{-\infty}^\infty V(x_0+t\omega)dt$.  Thus,  the $X$-rays transform  allows to recover 
electric  potential  $V(x)$ outside a convex obstacle.  See also  a related work of Enss and Weder  [9].

In [24] Eskin, Isozaki and O'Dell studied  the inverse scattering problem 
 for any number of obstacles,  not necessary convex.

In [44], [45], [53] Yafaev,  and Roux and Yafaev  described the singularities of the scattering amplitude .  

More on the inverse scattering  problem see  \S2.4. 

Another class of inverse problems are inverse 
boundary value problems.

Consider the stationary Schr\"odinger equation
\begin{equation}                                       \label{eq:1.7}
\frac{1}{2m}\sum_{j=1}^n\Big( -ih\frac{\partial}{\partial x_j}-\frac{e}{c}A_j(x)\Big)^2w(x)
+eV(x)w=k^2w(x)
\end{equation}
in the domain $\Omega\setminus\Omega_1$, 
where 
\begin{equation}                                    \label{eq:1.8}
\begin{aligned}
&w\Big|_{\partial\Omega_1}=0,
\\
&w\Big|_{\partial\Omega} =f.
\end{aligned}
\end{equation}

The Dirichlet to Neumann (DN)  operator $\Lambda(k)$  is the map of 
the Dirichlet data $f=w\big|_{\partial\Omega}$  to the Neumann data 
$\big(h\frac{\partial w}{\partial \nu}-i\, \frac{e}{c}A\cdot \nu w\big)\big|_{\partial\Omega}$
for all smooth solutions of (\ref{eq:1.6}),  (\ref{eq:1.7}),  (\ref{eq:1.8}),  i.e.
\begin{equation}                                          \label{eq:1.9}
\Lambda(k)f=\Big(h\frac{\partial w}{\partial \nu}-i\, \frac{e}{c} A(x)\cdot \nu w\Big)\Big|_{\partial\Omega},
\end{equation}
where $\nu$  is the outward unit normal to $\Omega$.    Note that
the group $G(\overline \Omega\setminus \Omega_1)$
consists of all smooth
complex-valued $g(x)$  such that $|g(x)|=1$.

The inverse boundary value problem consists of determining the
 gauge equivalence class of the magnetic potential and
of determining the electric potential knowing the DN operator 
 $\Lambda(k)$ on $\partial\Omega$.

One can consider also the case of several obstacles
$\Omega'=\bigcup_{j=1}^r\Omega_j$
when $\overline\Omega_j\cap\overline\Omega_k=\emptyset$
where $j\neq k,\ \overline\Omega_j\subset\Omega,1\leq j\leq r$.  Then  $u\big|_{\partial\Omega'}=0$
in (\ref{eq:1.8})  instead
of $u\big|_{\partial\Omega_1}=0$.

The inverse boundary value problems 
were studied intensively in many papers  (see,  for example,  the monograph of
Isakov [33] and references there). 
The case of domain with obstacles was considered in
[11], [12], [13].
The strongest results were obtained by the reduction to the inverse boundary value problem
 for the hyperbolic equation $\big(\frac{h^2}{2m}\frac{\partial^2}{\partial t^2}+H\big)u=0$,
  where
$H$ is the operator in the left hand side of (\ref{eq:1.6}), 
and using the Boundary Control (BC)  method (see [5], [35], [36], [16], [17], [19]).  
This approach allows to solve the inverse boundary  value problem in the case of any
number of obstacles,  not necessary convex.  Moreover  it is enough to know the DN operator
only on an arbitrary open part of the boundary $\partial\Omega$.  Also BC method allows to recover
not only the gauge equivalent classes of magnetic potentials and the electric potentials,
but also allows to recover the number and location of obstacles (see more details in \S 2.1).

Assuming that $\mbox{curl}\,A=0$  in  $\Omega\setminus\Omega'$  we prove  the AB effect  
in \S 2.1.  Moreover,  we prove that always  when $A$  and $A'$ belong to distinct  gauge
equivalent classes  they have a different physical impact.

In [10], [15], [18], [47]  a more general class of Schr\"odinger equations with Yang-Mills
potentials was considered,  i.e.  the equations of the form 
\begin{equation}                                    \label{eq:1.10}
\sum_{j=1}^n\Big(-i\frac{\partial}{\partial x_j}I_m-A_j(x)\Big)^2u+V(x)u=k^2u,
\ \ x\in\Omega\setminus\Omega',
\end{equation}
where $u(x)=(u_1(x),...,u_m(x)$  is $m$-vector, $A_j(x),1\leq j\leq n,\ V(x)$  are
self-adjoint $m\times m$  matrices  called the Yang-Mills potentials,  $I_m$  is
the identity operator in $\C^m,\Omega'=\bigcup_{j=1}^r\Omega_j$.
We assume that 
$u\big|_{\partial\Omega'}=0$.

The Dirichlet-to-Neumann operator  has the form 
\begin{equation}                                 \label{eq:1.11}
\Lambda(u\big|_{\partial \Omega})=
\Big(\frac{\partial}{\partial x}-iA(x)\Big)\cdot\nu(x)u(x)\big|_{\partial \Omega},
\end{equation}
where $\nu(x)$  is the unit outward  normal to $\partial\Omega,A=(A_1,...,A_n)$.

The gauge group $G(\overline \Omega\setminus\Omega')$
consists of smooth unitary $m\times m$ matrices $g(x)$ on $\overline\Omega\setminus \Omega'$.
Two Yang-Mills potentials  $(A(x),V(x))$  and  $(A'(x),V'(x))$
are gauge equivalent if there exists $g(x)\in G(\overline \Omega\setminus\Omega')$ 
such that
\begin{equation}                                      \label{eq:1.12}
g^{-1}A(x)g(x)+i\frac{\partial g}{\partial x}g^{-1}=A'(x),\ \ g^{-1}V(x)g=V'(x).
\end{equation}

The Schr\"odinger equation with electromagnetic potentials  is a particular case when
$m=1$,

In [15] the BC method was applied to the equations of the form (\ref{eq:1.10})
and all results  for the equation (\ref{eq:1.6}) were extended to the equations of the form
(\ref{eq:1.10})  (see more in \S 2.2).

Note that  the DN operator is not gauge invariant.  

 The gauge invariant boundary data 
on $\partial\Omega$  were found in [11], [18]
    using the
probability density
$|w(x)|^2$  and  the probability current  
$S(w)=\Im \big(\frac{\partial w}{\partial x}-iA(x)w(x)\big)\overline{w}(x)$.
It will be shown in \S 2.3 that
\begin{align}                                   \label{eq:1.13}
&|w(x)|^2\big|_{\Gamma}=f_1(x),\ \ \ \frac{\partial}{\partial \nu}|w(x)|^2\big|_\Gamma=f_2(x),
\\
\nonumber
& S(w)\big|_{\Gamma}=f_3(x)
\end{align}
are gauge invariant boundary data that
uniquely determine the gauge equivalence class of magnetic potential 
$A(x)$ and the electric potential $V(x)$.
Here $\Gamma$  is any open subset  of $\partial\Omega$.
  Therefore if $A(x)$  and $A'(x)$  belong to
distinct gauge equivalence classes  then
corresponding gauge invariant boundary data (\ref{eq:1.13})
will be different.  This gives another proof of magnetic AB effect.

There is a close relationship between the inverse boundary value problems (IBVP)
and the inverse scattering problem (ISP).  We
will assume that the magnetic field $B=\mbox{curl}\, A$ and electric  potential $V(x)$
have  compact supports in the ball $B_R=\{|x|<R\}$.  If also $\mbox{supp}\, A(x)\subset B_R$  there 
is a general theorem (see \S 2.4)  that  the scattering amplitude
$a(\theta,\omega,k)$  given for all $|\omega|=|\theta|=1$ uniquely determine 
the DN operator $\Lambda(k)$  on $\{|x|=R\}$ and vice versa, i.e.  the IBVP and ISP  are equivalent.

When the flux  $\alpha\neq 0$ 
the magnetic potential is not compactly supported and  the relation between 
IBVP  and ISP  is more complicated  (see \S 2.4 for details).

Another venue to test the AB effect is the spectrum  of the magnetic  Schr\"odinger
operator.  The first result in this direction belongs to Helffer [28] (see also [37]).
He considered the magnetic Schr\"odinger operator  of the form (\ref{eq:1.7})  in 
$\R^2\setminus\overline\Omega_1$  where $\Omega_1=\{|x|<1\},\ \mbox{curl}\, A=0$  in
$\R^2\setminus\overline\Omega_1$  and $V(x)\rw +\infty$  when $|x|\rw \infty$.  He has shown
that the lowest Dirichlet eigenvalue depends on the cosine of the magnetic flux 
(\ref{eq:1.4}).  This proves the AB effect.  In [25]  the Schr\"odinger equation (\ref{eq:1.7}) 
in $\Omega\setminus\overline \Omega_1$  was considered where $\Omega=\{x:|x|<R\},\ R$  is
large,
with the zero Dirichlet conditions on $\partial\Omega$ and $\partial\Omega_1$
and $\mbox{curl}\, A=0$  in  $\Omega\setminus\Omega_1$.  It was proven that Dirichlet
spectrum also depends on $\cos\alpha$,  where $\alpha$  is the magnetic flux,  thus
proving the AB  effect.

Note that the AB effect holds always when the domain is not simply-connected  even
if there are no obstacles.  For example,  in [25]  the AB effect is demonstrated 
for the Schr\"odinger  operator of the form (\ref{eq:1.6}) on the torus  (see [25] and
\S 2.5).

All methods to prove the AB effect  described above  are quite complicated.

A direct and simple proof of the AB  effect 
was proposed in [21].  It  essentially
mimics the AB experiment 
(see \S 2.6  and Remark 3.1 in [21],  see also papers of Ballesteros and Weder [3], [4]  on
the justification of AB experiment).

The AB effect
holds also for $n\geq 3$
dimensions,  for example, when the domain is $\R^3\setminus \Omega_1$,
where $\partial\Omega_1=T^2$  is the two dimensional torus and the magnetic field is
zero outside $\Omega_1$ (see  \S 2.6).  Note that the  most accurate AB type experience
was done by Tonomura et al [T]  for such domain.

It is important also to study the case
of several obstacles $\Omega_1,...,\Omega_m$  in $\R^2$  where
$\overline \Omega_j\cap\overline \Omega_k=\emptyset$  when
$j\neq k$.  Suppose we have the magnetic field shielded inside
$\Omega_j,1\leq j\leq m,$  and $B=\mbox{curl}\,A=0$  outside of all obstacles.
Let $\alpha_j=\frac{e}{hc}\int_{\gamma_j}A\cdot dx$  be the fluxes corresponding
to each obstacles $\Omega_j$.   Here $\gamma_j$  is a simple contour encircling 
$\Omega_j$  only.  Suppose that some $\alpha_j\neq 2\pi n,\forall n\in \Z,$
but the total flux $\sum_{j=1}^m\alpha_j=0$ (modulo $2\pi n)$.  Suppose  that the obstacles
are close to each other and therefore 
we can not perform  AB experiment separately  for each $\Omega_j$.
From other side the treatment of $\bigcup_{j=1}^m\Omega_j$ as  one obstacle  does
not reveal  the AB effect since the total flux is zero modulo  $2\pi n$ .
The AB effect in this case was proven in [13], [21] using broken rays solutions.  We
were able to recover all magnetic  
fluxes $\alpha_j, j=1,2,...,m$,  up to a sign.

The magnetic AB effect is studied in the hundreds 
of papers (see the survey [42]).   In the original paper [1] 
Aharonov and Bohm discuss also the electric AB 
effect. They consider the Schr\"odinger equation with time-dependent electric 
potential and zero magnetic potential
\begin{equation}                                         \label{eq:1.14}
ih\frac{\partial u(x,t)}{\partial t}+\frac{h^2}{2m}\Delta u(x,t)-eV(x,t)u(x,t)=0.
\end{equation}
In contrast with hundreds  of papers on the magnetic AB effect  there are only
few papers dealing with the electric AB effect.  In particular,   in [51]  Weder studied  the electric AB  effect  
assuming  that the electric potential depends on a large  parameter.

Let domain $D\subset \R^n\times[0,T]$.  Denote by $D_{t_0}$ the
intersection of $D$  with the plane $t=t_0$  We assume that 
$$
u\big|_{\partial D_t}=0\ \ \mbox{for}\ \ 0<t<T \ \ \mbox{and}\ \ u(x,0)=u_0(x) \ \ \mbox{on}\ D_0.
$$
We
assume that the electric field $E=\frac{\partial V}{\partial x}=0$ in $D$.
If
$D_{t_0}$ is connected for all $t_0\in (0,T)$
then $\frac{\partial V(x,t_0)}{\partial x}=0$  implies  that $V(x,t_0)=V(t_0)$
is independent of $x$  in $D$.

Consider a gauge  transformation 
$$
w(x,t)=\exp\Big(i\frac{e}{h}\int\limits_0^t V(t')dt'\Big)u(x,t),
$$
where  $u(x,t)$  is  the solution of (\ref{eq:1.14}).  Then  $w(x,t)$  satisfies the
Schr\"odinger equation
$$
ih\frac{\partial w(x,t)}{\partial t}+\frac{h^2}{2m}\Delta u(x,t)=0.
$$
Note that $w\big|_{\partial D_t}=0$  for $0<t<T$  and $w(x,0)=u_0(x)$  on $D_0$.

Therefore  the electric  potential $V(x,t)$  is gauge  equivalent to zero
electric potential if $E=\frac{\partial V}{\partial x}=0$  in $D$  and $D_t$  are connected 
for all $t\in (0,T)$.  This explains why there was no neither experimental nor
mathematical evidence of AB effect in the situation when the domain $D$  has the form
$D=\Omega\times (0,T)$  where $\Omega$  is a domain in $\R^n$.  For the electric
AB effect  to take place one need to consider domains with moving boundaries,  i.e.
$D_t$ changes with $t$  and is connected for some $t$ and is disconnected for
other $t, t\in(0,T)$   (cf.  [21]  and \S 3).

In \S 4  we study  the Schr\"odinger equation with time-dependent electric and magnetic 
potentials.

Let $\Omega_j(t),1\leq j\leq r,$  be obstacles  in $\R^n$.  Let
$\Omega_0\supset\overline\Omega_j(t),\ \forall t\in [0,T], 1\leq j\leq r,\ \Omega_0$  be
a simply-connected 
bounded domain in 
$\R^n, \Omega'(t)=\bigcup_{j=1}^r\Omega_j(t),\Omega'=\bigcup_{0\leq t\leq T}\Omega'(t)$.

Consider the Schr\"odinger equation
\begin{equation}                                 \label{eq:1.15}
\Big(ih\frac{\partial u}{\partial t}-Hu\Big)=0\ \  \mbox{in}\ \ 
(\Omega_0\times(0,T))\setminus\Omega',
\end{equation}
where
$$
H=\frac{1}{2m}\sum_{j=1}^n\Big(-ih\frac{\partial}{\partial x_j}-\frac{e}{c}A_j(x,t)\Big)^2
+eV(x,t),
$$
 $A(x,t)=(A_1,...,A_n)$  and $V(x,t)$  are  magnetic and electric potentials.

We assume that
\begin{equation}                            \label{eq:1.16}
u(x,0)=0\ \ \mbox{in}\ \ \Omega_0\setminus \Omega'(0),
\end{equation}
\begin{equation}                               \label{eq:1.17}
u\big|_{\partial\Omega'(t)}=0,\ \ 0\leq t\leq T,\ \ u\big|_{\partial\Omega_0\times(0,T)}=f.
\end{equation}
We first consider the inverse boundary
value problem for (\ref{eq:1.15}),  (\ref{eq:1.16}), (\ref{eq:1.17}).  The gauge group
$G((\overline\Omega_0\times[0,T])\setminus\Omega')$  consists of 
$g(x,t)\in C^\infty((\overline\Omega_0\times[0,T])\setminus\Omega')$
such that $|g(x,t)|=1$.  Since coefficients of (\ref{eq:1.15})  are time-dependent
we can not reduce (\ref{eq:1.15}) to the hyperbolic equation and apply BC method as
in \S 2.1.    We use a more traditional approach  (cf.  [18]) consisting of two steps:

a)  Construction of geometric optics type solution for the Schr\"odinger 
equation with time-depending 
coefficients  that are concentrated in a small neighborhood of a ray or a broken ray.
This part can be done under quite mild restrictions on the geometry of obstacles (cf. [15],
[18]).

b)  In the second step one needs to study the injectivity of the X-ray type transform in
the domain with obstacles.  The presence of obstacles makes 
the results quite restrictive.

If the geometric conditions on obstacles are satisfied one can prove (cf. [18] and \S 4.1)
that if there
are two Schr\"odinger equations  $\Big(ih\frac{\partial u_k}{\partial t}-H_ku\big)=0,
k=1,2,$ of the form (\ref{eq:1.15}) with initial and boundary conditions
(\ref{eq:1.16}),  (\ref{eq:1.17}) and if corresponding DN operators $\Lambda_k,k=1,2,$
are gauge equivalent  on $\partial \Omega_0\times (0,T)$ then
electromagnetic potentials $(A^{(1)},V^{(1)})$  and $(A^{(2)},V^{(2)})$  are also 
gauge equivalent.

Consider now the equation (\ref{eq:1.15})  in unbounded 
domain \linebreak $(\R^n\times (0,T))\setminus \Omega'$  with the initial condition
\begin{align}
\nonumber
& u(x,0)=u_0(x) \ \ \mbox{in}\ \ \R^n\setminus\Omega'(0),
\\
\nonumber
& u\big|_{\partial\Omega'(t)}=0,\ \ 0\leq t\leq T.
\end{align}
We will assume  that $u_0(x)=0$  in $\Omega_0\setminus\Omega'(0)$  as in (\ref{eq:1.16}).

In this case the gauge group $G((\R^n\times(0,T))\setminus\Omega')$ consists 
of $|g(x,t)|=1$  in $\R^n\times[0,T]\setminus \Omega'$  and we assume 
that $g(x,t)$  are independent of $t$  in $(\R^n\setminus\Omega_0)\times[0,T]$.
When $n\geq 3$  we also assume that $g(x)=\exp\big(\frac{i}{h}\varphi(x)\big)$
for $|x|>R$,  where $\varphi(x)$  is real-valued,  $\varphi(x)=O\big(\frac{1}{|x|}\big)$.

When $n=2$  we assume that $g(x)=e^{ip\theta(x)}\big(1+O\big(\frac{1}{|x|}\big)\big)$
for $|x|>R$.

Note that $(A^{(1)},V^{(1)})$  and $(A^{(2)},V^{(2)})$  are gauge equivalent if
\begin{align}                                            \label{eq:1.18}
&\frac{e}{c}A^{(2)}(x,t)=\frac{e}{c}A^{(1)}(x,t)+ihg^{-1}\frac{\partial g}{\partial x},
\\
\nonumber
& eV^{(2)}(x,t)=eV^{(1)}(x,t)-ihg^{-1}\frac{\partial g}{\partial t}.
\end{align}
Since coefficients of the equation (\ref{eq:1.15}) 
are time-dependent,  the scattering operator for $H$  is not defined.  We propose 
a new inverse problem in $(\R^n\times(0,T)\setminus\Omega'$  instead of 
the inverse scattering problem.

Let $u(x,t)$  be the solution of (\ref{eq:1.15})  in $(\R^n\times(0,T))\setminus \Omega'$,
and let $(A(x,t),V(x,t))$  be independent of $t$
for $|x|>R$.  We assume that
\begin{equation}                                  \label{eq:1.19}
u(x,0)\ \ \mbox{and}\ \ u(x,T)
\end{equation}
are known  on $\R^n\setminus B_R$.

Then (see Theorem \ref{theo:4.3} in \S 4.2)   these 
two times ($t=0$ and $t=T$)
 data determine
$u(x,t)$  in $((\R^n\setminus B_R)\times (0,T)$.

More precisely,  the following result holds:

Let  $ih\frac{\partial u_k}{\partial t}-H_ku_k=0, k=1,2,$  be two  equations  of the form 
(\ref{eq:1.15}) in $(\R^n\times (0,T))\setminus\Omega'$.
Suppose corresponding electromagnetic potentials $(A^{(k)},V^{(k)}),k=1,2,$  are
independent of $t$  for $|x|>R$  and gauge equivalent  with some gauge $g_0(x)$.

Suppose the two times ($t=0$  and $t=T$) data (\ref{eq:1.19}) of $u_1(x,t)$  and
$u_2(x,t)$  are  gauge equivalent,  i.e.
$$
u_2(x,0)=g_0(x)u_1(x,0),\ \ u_2(x,T)=g_0(x)u_1(x,T),\ \ x\in \R^n\setminus B_R.
$$
Then 
\begin{equation}                                \label{eq:1.20}
u_2(x,t)=g(x)u_1(x,t) \ \ \mbox{in}\ \ (\R^n\setminus B_R)\times (0,T).
\end{equation}
The relation (\ref{eq:1.20}) implies that the DN operators $\Lambda_1$  and $\Lambda_2$
are gauge equivalent  on $\partial B_R\times (0,T)$.  Then assuming that 
the geometric  conditions 
on obstacles  formulated in Theorem \ref{theo:4.1} are satisfied,  the electromagnetic potentials $(A^{(1)},V^{(1)})$
and $(A^{(2)},V^{(2)})$  are gauge equivalent.

Note that as in the case of time-independent  magnetic and electric potentials it is naturally
to consider  the gauge invariant boundary data as in \S 2.3.

We shall mention also  the inverse boundary value problems for the time-dependent 
Yang-Mills potentials. The powerful BC method    used 
 in the case of time-independent Yang-Mills 
potentials can not be applied here. However,  we can solve the inverse boundary 
problem using the method of Non-Abelian Radon transforms developed in [11], [14], [40].
This method does not work,  unfortunately, if the obstacles are present.

Now we shall consider the AB effect for time-dependent  electromagnetic potentials  assuming 
that $B=\mbox{curl}\, A(x,t)=0$  and $E=-\frac{1}{c}\frac{\partial A(x,t)}{\partial t}
-\frac{\partial V(x,t)}{\partial x}=0$  in $(\Omega_0\times (0,T))\setminus \Omega'$.

Let $\alpha_\gamma=\int_\gamma A(x,t)\cdot dx- V(x,t)dt$  be  electromagnetic flux
where $\gamma$  is a closed  contour  in $(\Omega_0\times(0,T))\setminus\Omega'$.  It follows
from $E=B=0$  and the Stoke's theorem  that $\alpha_\gamma$  depends only on the
homotopy class  of $\gamma$  in $(\Omega_0\times(0,T))\setminus \Omega'$.

Let $\gamma_1,...,\gamma_m$  be  the basis of the homology  group of 
$(\Omega_0\times (0,T))\setminus \Omega'$,   
  i.e.  any  closed contour in $(\Omega_0\times(0,T))\setminus \Omega'$  is homotopic to a linear 
combination 
of $\gamma_1,...,\gamma_m$  with integer coefficients.

Denote  $\alpha_{\gamma_k}=\frac{e}{hc}\int\limits_{\gamma_k}A dx -V dt,\ \  \ 1\leq k\leq m$.

Two electromagnetic potentials   $(A^{(1)},V^{(1)})$  and  $(A^{(2)},V^{(2)})$
are gauge equivalent  if and only if
\begin{equation}                                     \label{eq:1.21}
\alpha_{\gamma_p}^{(1)}=\alpha_{\gamma_p}^{(2)}\ (\mbox{mod}\ 2\pi n,n\in\Z),\ \ 1\leq p\leq m.
\end{equation}
Here $\alpha_{\gamma_p}^{(k)}=\frac{e}{hc}\int\limits_{\gamma_p}A^{(k)}dx-V^{(k)}dt,\ \ k=1,2.$
Therefore to demonstrate the electromagnetic AB effect  we will need to check only a finite 
number of relations (\ref{eq:1.21}).

Thus we do not need to prove the injectivity of the X-ray transform
to demonstrate the AB effect.  Therefore we can relax the restriction
on the geometry of obstacles imposed in Theorem \ref{theo:4.1}.
Moreover,  we can consider a more general class of obstacles.  

We shall consider a class of domains $D^{(1)}$  with  obstacles that  may move  and may merge 
or  split at some times $t_k,\ 1\leq k\leq l$  (see Fig. 4). The intersections  of $D^{(1)}$
with $t=t_0$  are connected  for each  $t_0\in (0,T)$.  We denote by $D^{(2)}$ a more
general class of domains obtained  from $D^{(1)}$  by making holes in some obstacles\
(cf. \S 4.5).  Now the intersection  of $D^{(2)}$  with $t=t_0$  may be not connected 
for some $t_0\in (0,T)$ and hence the combined AB effect  takes place.  

A simple examples of domains of the type $D^{(1)}$  
is the following domain  $D_0^{(1)}$:  let
 $\Omega_0=\{x_1^2+x_2^2<r^2\}, D_0^{(1)}\cap\{t=t_0\}=\Omega_0\setminus\Omega_1(t_0)$,  
where $\Omega_1(t_0)$  is the obstacle  moving with the speed $v_1$  along $x_1$-axis:
$\Omega_1(t)=\{(x_1-v_1t)^2+x_2^2<r_1^2\},r_1\leq r$  and small,  $0\leq t\leq T$.
We assume that $\Omega_1(T)\subset\Omega_0$.

Let $\omega=\{(x_1-\frac{r_1}{2})^2+(x_2-\frac{r_1}{2})^2<\frac{r_1^2}{16}\}$.
The hole $H$ in $D_0^{(1)}$  is  the intersection  of the cylinder 
$\omega\times(0,T)$  with $\bigcup_{0\leq t\leq 1}\Omega_1(t)$.   Therefore the 
domain of class $D^{(2)}$  is $D_0^{(1)}\cup H$.

If $B=\mbox{curl}\, A=0$
and $E=-\frac{1}{c}\frac{\partial}{\partial t}-\frac{\partial V}{\partial x}=0$  in $D_0^{(1)}$
then  $\alpha_\gamma=\int_\gamma A(x,t)\cdot dx - V(x,t)dt$  is  the same for any closed
contour $\gamma$  in $D^{(1)}$  encircling the obstacles.  Any such $\gamma$  is homotopic to
a contour $\gamma_0$  in the plane $t=\mbox{const}$  encircling the obstacle,  i.e.   
$\gamma_0$  is the basis of the homology group in $D_0^{(1)}$.
In the case of $D_0^{(2)}$  
there are two contours  that form  the basis for the homology group in
$D_0^{(2)}$.  One of them is $\gamma_0$  and the second  is
any  
  closed contour $\gamma_1$
that is passing through the hole $H$ and not shrinking to a point.

In $D_0^{(1)}$ the potentials $(A^{(1)},V^{(1)})$  and $(A^{(2)},V^{(2)})$
are  having a different physical impact if
\begin{equation}                                        \label{eq:1.22}
\alpha=\frac{e}{hc}\int\limits_{\gamma_0}(A^{(1)}-A^{(2)})dx \ \mbox{or}\ \ -\alpha
\ \ \mbox{are not equal to}\ \ 2\pi p,\ \forall  p\in\Z.
\end{equation}
In $D^{(2)}$
$(A^{(1)},V^{(1)})$  and $(A^{(2)},V^{(2)})$
are having a different physical impact if either (\ref{eq:1.22})  holds or
$A^{(1)}$  and $A^{(2)}$  are  gauge equivalent  and  
\begin{equation}                            \label{eq:1.23}
\frac{e}{h}\int\limits_{\gamma_1}\frac{A^{(1)}-A^{(2)}}{c}\cdot dx-(V^{(1)}-V^{(2)})dt\neq
2\pi p,\ \forall p\in \Z.
\end{equation}
These two  examples  are  a particular case of general results in \S4.4 and \S 4.5.
 
An important part of the proof 
of the AB efect
is the construction of geometric optics type solution in
$D^{(1)}=(\Omega_0\times(0,T))\setminus\Omega'$  
similar  to the solutions for the solving inverse boundary value problem
(see \S 4.1).

These geometric optics type solutions are  the solutions of  (\ref{eq:1.15})
in $D^{(1)}$ only  and have  nonzero  Dirichlet data on 
$\partial\Omega_0\times(0,T)$.  It is not 
clear what is their  physical meaning. From the other side,
the solutions of  (\ref{eq:1.15}),
(\ref{eq:1.16}), (\ref{eq:1.17})
 in $(\R^n\times(0,T))\setminus\Omega'$  describe  
the electron  in the magnetic field shielded by obstacles $\Omega'$
and therefore are physically meaningful.  It is proven in \S 4.5
(the density lemma \ref{lma:4.5}) that  any  solution  of (\ref{eq:1.15}) 
in $D^{(1)}=(\Omega_0\times(0,T))\setminus\Omega'$ can be approximated  by 
the restrictions  to $D^{(1)}$  of physically  meaningful  solutions of
(\ref{eq:1.15}),  (\ref{eq:1.16}),  (\ref{eq:1.17})
  in $(\R^n\times(0,T))\setminus\Omega'$.
This allows to complete the proof  of electromagnetic AB effect in \S 4.4.

The AB type effect  holds  not only  in quantum  mechanics but 
also in other branches of physics  (cf. [6], [7], [49]).  We shall consider the gravitational  analog
of AB effect extending  the results of Stashel [46].

First,  we reformulate the magnetic AB effect in $\R^2\setminus\Omega_1$ assuming, 
 for the simplicity  of notations,
that $h=e= c=1$.   Suppose $B=\mbox{curl}\, A=\frac{\partial A_2}{\partial x_1}-
\frac{\partial A_1}{\partial x_2}=0$  in $\R^2\setminus\Omega_1$.  
If  $\omega\subset\R^2\setminus\Omega_1$
is a simply connected subdomain of $\R^2\setminus\Omega_1$  then 
$\frac{\partial A_1}{\partial x_2}-
\frac{\partial A_2}{\partial x_1}=0$
in $\omega$  implies  that there exists $\Psi(x_1,x_2)$  in $\omega$
such that  $A_1=\frac{\partial\Psi}{\partial x_1}, A_2=\frac{\partial \Psi}{\partial x_2}$,
  i.e.
$A(x)$ is the gradient  of $\Psi(x_1,x_2)$.
Making the gauge transformation $u'=e^{i\Psi}u$  we get  the Schr\"odinger
equation with zero magnetic potential in $\omega$,  
i.e.  there is no AB effect in $\omega$.  The AB effect
takes place when $\mbox{curl}\, A=0$  but $A(x)$  is not a gradient in $\R^2\setminus\Omega_1$
and has a flux $\alpha\neq 2\pi n,\ \forall n\in \Z$.

Similar situation arise for the wave equation  corresponding to a pseudo-Riemannian 
metric $\sum_{j=0}^n g_{jk}(x)dx_jdx_k$ with  Lorentz signature,
where $x_0$  is the time variable,  $x=(x_1,...,x_n)\in\Omega=
\Omega_0\setminus\bigcup_{j=1}^m\Omega_j$.  We assume that $g_{jk}(x)$  are 
independent  of $x_0$,  i.e.  the metric is stationary.

Consider  the group  of transformations
\begin{align}                        \label{eq:1.24}
&x'=\varphi(x)
\\
\nonumber
&x_0'=x_0+a(x),
\end{align}
where $\varphi(x)$  is a diffeomorphism 
of $\overline\Omega$ onto $\overline\Omega'=\varphi(\overline\Omega)$  and 
$a(x)\in C^\infty(\overline\Omega)$.  
Two metrics 
$\sum_{j,k=0}^ng_{jk}(x)dx_jdx_k$  and $\sum_{j,k=0}^ng_{jk}'(x')dx_j'dx_k'$
are called isometric if
\begin{equation}                                     \label{eq:1.25}
\sum_{j,k=0}^ng_{jk}(x)dx_jdx_k
=\sum_{j,k=0}^ng_{jk}'(x')dx_j'dx_k',
\end{equation}
where $(x_0',x_0)$  and $(x',x)$ are related by (\ref{eq:1.24}).

The group of isometries plays the role of the gauge group for the electromagnetic
AB effect.

We shall prove (cf. Theorem \ref{theo:6.2}) that if two metrics are
locally isometric but globally not isometric, then they have  a different physical impact.

We also extend a result of [46]  that if a metric is locally static  but
not globally static,  then this fact also has a physical impact.  This is 
a gravitational analog of AB effect  (cf. \S 5  and [22]).

\section{Magnetic AB effect}
\init

In this section  we consider  the most  well-known magnetic AB effect and we will review the different approaches to 
study it.

\subsection{Inverse boundary value problems for the Schr\"odinger  equation  with
time-independent
electromagnetic potentials}
\
\\
Let  $\Omega_0$  be a smooth bounded 
domain  in $\R^n,$  and let $\Omega_j,1\leq j\leq r,$  be the smooth obstacles 
inside $\Omega_0,\ \overline\Omega_j\cap\overline\Omega_k=\emptyset$  when $j\neq k$.
Consider a stationary Schr\"odinger equation in $\Omega_0\setminus\Omega'$,  where
$\Omega'=\bigcup_{j=1}^r\Omega_j $:
\begin{equation}                                           \label{eq:2.1}
Hw  \overset{def}{=}\frac{1}{2m_j}\sum_{j=1}^n\Big(-ih\frac{\partial}{\partial x_j}
-\frac{e}{c}A_j(x)\Big)^2w(x) + eV(x)w(x)=k^2w(x),
\end{equation}
\begin{equation}                                    \label{eq:2.2}
w\big|_{\partial \Omega'}=0,
\end{equation}
\begin{equation}                                    \label{eq:2.3}
w\big|_{\partial \Omega_0}=f.
\end{equation}
If $k$  does not belong to a discrete set $N$ of the Dirichlet eigenvalues then  
the Dirichlet-to-Neumann (DN)  operator
\begin{equation}                                  \label{eq:2.4}
\Lambda(k)f=\Big( h \frac{\partial w}{\partial \nu}
-i\frac{e}{c}A(x)\cdot\nu(x)w\Big)\Big|_{\partial\Omega_0}
\end{equation}
is well-defined bounded operator from $H_{\frac{3}{2}}(\partial\Omega_0)$   to
$H_{\frac{1}{2}}(\partial\Omega_0)$,  where $H_s(\partial\Omega_0)$  
is a Sobolev space of order $s$  on $\partial\Omega_0$.
Note that $\Lambda(k)$  is analytic in $k$  on $\C\setminus N$.  Thus the knowledge
of $\Lambda(k)$  on any small interval $(k_0-\e,k_0+\e)$  determines $\Lambda(k)$  for all
$k\in \C\setminus N$.

Let $\Gamma\in \partial \Omega_0$  be an open subset of $\partial\Omega_0$.  We  say 
that $\Lambda(k)$  is given on $\Gamma$  if the restriction $\Lambda(k)f\big|_\Gamma$  is 
known for any $f$  with support in $\overline\Gamma$.
\begin{theorem}                                    \label{theo:2.1}
Suppose two Schr\"odinger equations $(H-k^2)w=0$  and $(H'-k^2)w'=0$  are given
in $\Omega_0\setminus  \bigcup_{j=1}^{r'}\Omega_j$  
and $\Omega_0\setminus \bigcup_{j=1}^{r'}\Omega_j'$
with electromagnetic potentials $(A(x),V(x))$  and $(A'(x),V'(x))$,  respectively.
Suppose the corresponding DN operators $\Lambda(k)$ and $\Lambda'(k)$
coincide on $\Gamma$ for $k\in(k_0-\e,k_0+\e)$.
Then $A'(x)$  and $A(x)$  are gauge equivalent with the gauge $g(x)$  equal to 1 on 
$\Gamma,\ V'(x)=V(x), r'=r$  and $\Omega_j'=\Omega_j,1\leq j\leq r.$
\end{theorem}  

The proof of Theorem \ref{theo:2.1}  is based on the reduction to the hyperbolic 
inverse boundary value problem and use of the powerful Boundary Control method for 
solving  such problems (cf. Belishev [5],  Kachalov-Kurylev-Lassas [35],
Eskin [16], [17]):

Consider the initial-boundary value problem  
for the hyperbolic equation
\begin{equation}                                       \label{eq:2.5}
\frac{h^2}{2m}\frac{\partial^2 v}{\partial t^2}+Hv=0,\ \ 
x\in\Omega_0\setminus\bigcup_{j=1}^r\Omega_j,\ 
0< t<+\infty,
\end{equation}
with zero initial conditions
\begin{equation}                                         \label{eq:2.6}
v(x,0)=\frac{\partial v}{\partial t}(x,0)=0,\ \ x\in \Omega_0\setminus\Omega',
\end{equation}
and boundary conditions
\begin{equation}                                          \label{eq:2.7}
v\big|_{\partial\Omega_j'\times(0,+\infty)}=0,\ 1\leq j\leq r,\ \ \
v\big|_{\partial\Omega_0\times(0,+\infty)}=\varphi(x',t),
\end{equation}
where $\varphi(x',t)$  has a compact support on $\Gamma\times(0,+\infty)$.

Define the hyperbolic DN  operator $\Lambda_H$   as 
\begin{equation}                                 \label{eq:2.8}
\Lambda_H \,\varphi=\Big(h\frac{\partial v}{\partial \nu}
-i\frac{e}{c} A(x)\cdot\nu v\Big)\Big|_{\Gamma\times(0,+\infty)}.
\end{equation} 
The following result  holds  (see,  for example,  Theorem 1.1  in [16]):
\begin{theorem}                                    \label{theo:2.2}
Consider  two hyperbolic equations $\big(
\frac{h^2}{2m}\frac{\partial^2 }{\partial t^2}+H\big)v=0,\linebreak  
\big(\frac{h^2}{2m}\frac{\partial^2 }{\partial t^2}+H'\big)v'=0$ 
in  $(\Omega_0\setminus\bigcup_{j=1}^r\Omega_j)
\times(0,+\infty),
\ (\Omega_0\setminus\bigcup_{j=1}^{r'}\Omega_j')\times(0,+\infty)$,  respectively,  with zero initial 
conditions (\ref{eq:2.6}) and with the boundary conditions
\begin{equation}                                  \label{eq:2.9}
\ \ \ \ v\big|_{\partial\Omega_0\times(0,+\infty)}=\varphi,\ \ v\big|_{\Omega_j\times(0,+\infty)}=0,
\ \ \ 1\leq j\leq r,
\end{equation}
and
\begin{equation}                                    \label{eq:2.10}
v'\big|_{\partial\Omega_0\times(0,+\infty)}=\varphi',\ \ 
v'\big|_{\Omega_j'\times(0,+\infty)}=0,\ \ 1\leq j\leq r',
\end{equation}
respectively,  where $\mbox{supp}\,\varphi\subset\overline\Gamma,
\mbox{supp}\,\varphi'\subset\overline\Gamma$.

If the hyperbolic DN operator  $\Lambda_h$  and $\Lambda_h'$ are equal  on 
$\Gamma\times(0,+\infty)$,  i.e.  $\Lambda_h\varphi=\Lambda_h'\varphi$  on 
$\Gamma\times(0,+\infty)$  for  any $\varphi$  with the support   in 
$\Gamma\times[0,+\infty)$,  then 
$A(x)$  and $A'(x)$  are gauge equivalent  with the gauge $g=1$  on $\Gamma,\ V(x)=V'(x), r=r'$
and 
$\Omega_j=\Omega_j',\ 1\leq j\leq r$.
\end{theorem}

{\bf Remark 2.1}  Theorem 1.1 in [16]  states
 that there exists  a diffeomorfism  $x'=\psi(x)$  of
$\overline\Omega_0\setminus\bigcup_{j=1}^r\Omega_r$  onto 
$\overline\Omega_0\setminus \bigcup_{j=1}^{r'}\Omega_j',\ \psi(x)=x$ 
 on  $\Gamma$  and  $\sum_{j=1}^n(dx_j)^2=\sum_{j=1}^n(dx_j')^2$,  where $x'=\psi(x)$.
This implies that $\psi=I$  and therefore $r=r'$  and  $\Omega_j=\Omega_j',\ j=1,...,r$.
\vspace{5pt} 

To prove Theorem \ref{theo:2.1}  we take the Fourier transform  
in $t$.  Then the equation (\ref{eq:2.5}) becomes the equation (\ref{eq:2.1})
and the hyperbolic DN operator  $\Lambda_h$  on $\Gamma\times(0,+\infty)$
becomes the DN  operator $\Lambda(k)$  on $\Gamma$.  

We shall use Theorem  \ref{theo:2.1}  to prove the magnetic 
AB effect.

Suppose $(H_k-\lambda^2)w_k=0,\ k=1,2,$  are two  Schr\"odinger equations of the form  (\ref{eq:2.1})
with electromagnetic potentials 
$A^{(k)}(x),V^{(k)}(x)),k=1,2,$  respectively.  Suppose $V^{(1)}=V^{(2)}=V$
and $\mbox{curl}\,A^{(k)}=0$  in  $\Omega_0\setminus\Omega',\ k=1,2.$
Fix a point  $x_0\in\Gamma$
and let $\omega$  be  a simply-connected  neighborhood  of  $x_0$.  
Let $\omega_+=\Omega_0\cap\omega$  and suppose 
$\Gamma=\partial\Omega\cap \omega$.  Since $\mbox{curl}\,A^{(k)}=0$  in  
$\overline \omega_+$  and $\omega_+$  is simply-connected,  there exists a smooth
$\Psi_k(x)$  in $\overline\omega_+$  
such  that  $A^{(k)}=\frac{\partial\Psi_k}{\partial x}$  in 
$\overline \omega_+,\ k=1,2.$  Let $\tilde\Psi_k$  be a  
smooth extension  of  $\Psi_k(x)$ to $\overline\Omega_0\setminus\Omega'$  and
let  $g_k(x)=e^{-\frac{ie}{hc}\tilde\Psi_k}$.  Then making  the gauge  transformation 
 with
the gauge $g_k(x)$  we transform  $H_k$  to $\hat H_k,\ k=1,2,$  where $\hat H_k$ has
electromagnetic 
potentials  $(\hat A^{(k)},\hat V^{(k)})$
such  that  $\hat V^{(k)}(x)=V(x),\  \hat A^{(k)}=0$  in $\overline\omega_+,\ k=1,2$.  Therefore
$\hat H_1=\hat H_2$  in $\omega_+$.  Now  we shall prove  the magnetic AB effect.

\begin{theorem}                                      \label{theo:2.3}
Magnetic   potentials
$\hat A^{(1)}$  and  $\hat A^{(2)}$ (and consequently  $A^{(1)}$  and $A^{(2)}$) are  not  gauge  equivalent if and only  if  there exists $f_0\in C_0^\infty(\Gamma)$  such that
\begin{equation}                                    \label{eq:2.11}
\hat\Lambda^{(1)}f_0\big|_\Gamma\neq \hat\Lambda^{(2)}f_0\big|_\Gamma,
\end{equation}
where $\hat\Lambda^{(k)}$ are  DN  operators corresponding  to  $\hat H_k$.
\end{theorem}  
It follows  from  (2.11) that when  $\hat A^{(1)}$ and  $\hat A^{(2)}$  are not  
gauge  equivalent  they have different  physical impact,  i.e.  AB  effect  holds.

{\bf Proof of Theorem 2.3:}
Suppose $\hat A^{(1)}$  and  $\hat A^{(2)}$ are not gauge equivalent.  If  (2.11)
does not hold,  i.e. $\hat\Lambda^{(1)}f=\Lambda^{(2)}f$  on $\Gamma$  for all
$f\in C_0^\infty(\Gamma)$ then  by  Theorem  \ref{theo:2.1} $\hat A^{(1)}$  and
$\hat A^{(2)}$  must be gauge equivalent,  and this is  a contradiction.  Vice versa,
suppose (2.11) holds  but $\hat A^{(1)}$  and  $\hat A^{(2)}$  are gauge equivalent
with some gauge $g(x)$.  Since  
$\hat A^{(1)}=\hat A^{(2)}=0$  in  $\overline\omega_+$  we get from (\ref{eq:1.5}) that  $g(x)=e^{i\alpha}$  in
$\overline \omega_+$,  where $\alpha$  is an arbitrary  real  constant. 
Let $\hat u_k$  be  the solutions of  $(\hat H_k-\lambda^2)\hat u_k=0,
\hat u_k\big|_{\partial\Omega_0}=f,\ f\in C_0^\infty(\Gamma),\ 
u_k\big|_{\partial\Omega'}=0$. 
 Since $g=e^{i\alpha}$  on  $\overline\omega_+$ and $\hat u_1\big|_{\partial\Omega_0} =\hat u_2\big|_{\partial\Omega_0}=f$,  
 we get that
$\hat u_1=\hat u_2$  in $\omega_+$.  
Therefore  
$\hat\Lambda^{(1)}f\big|_\Gamma=\hat\Lambda^{(2)}f\big|_\Gamma$ for all  
$f\in C_0^\infty(\Gamma)$,  and this contradicts (2.11).
\qed

\subsection{Inverse boundary value problems for the Schr\"odinger  equation  with
time-independent
Yang-Mills potentials}
\
\\
Consider  the Schr\"odinger equation with
Yang-Mills potentials  (cf.  (\ref{eq:1.10})):
\begin{equation}                              \label{eq:2.11}
\sum_{j=1}^n\Big(-i\frac{\partial}{\partial x_j}-A_j(x)\Big)^2w(x)+
V(x)w(x)=k^2w(x),\ \ x\in \Omega_0\setminus\bigcup_{j=1}^r\Omega_j,
\end{equation}
where Yang-Mills potentials $A_j(x), 1\leq j\leq n,\ V(x)$  are $m\times m$  self-adjoint 
matrices.
The gauge group $G(\overline\Omega_0\setminus\bigcup_{j=1}^r\Omega_j)$  consists of all
unitary  $m\times m$  matrices and two Yang-Mills potentials $(A(x),V(x))$  and $(A'(x),V'(x))$
are gauge equivalent  if there exists 
$g(x)\in G(\overline\Omega_0\setminus\bigcup_{j=1}^r\Omega_j)$
such that (\ref{eq:1.12})  holds.

We assume  that
\begin{equation}                                             \label{eq:2.12}
w\big|_{\partial\Omega_j}=0,\ \ 1\leq j\leq r,\ \ \ \ \ w\big|_{\partial\Omega_0}=f,\ \ \
\mbox{supp}\,f\subset\overline\Gamma.
\end{equation}
The following theorem  generalizes Theorem \ref{theo:2.1} for the case of
Yang-Mills potentials.

\begin{theorem}                                                \label{theo:2.4}
Let $(H-k^2I_n)w=0$  and $(H'-k^2I_m)w'=0$  be  two Schr\"odinger equations  corresponding  to
Yang-Mills potentials \linebreak $(A(x),V(x))$  and $(A'(x),V'(x))$,  respectively.
$(H-k^2)w=0$  is considered  in $\Omega_0\setminus\bigcup_{j=1}^r\Omega_j$  with 
boundary conditions
$w\big|_{\partial\Omega_j}=0,1\leq j\leq r,\ w\big|_{\partial \Omega_0}=f$  and  $(H'-k^2)w'=0$
is considered in 
$\Omega_0\setminus\bigcup_{j+1}^{r'}\Omega_j'$ 
with boundary conditions  $w'\big|_{\partial\Omega_0}=f',\ w'\big|_{\partial\Omega_j'}=0,\ 
1\leq j\leq r'$.
Let $\Gamma$  be an open subdomain of $\partial\Omega_0$.
Suppose  that DN  operators 
$\Lambda(k)f=\big(\frac{\partial w}{\partial \nu}-iA\cdot \nu w\big)\big|_\Gamma$  and
$\Lambda'(k)f=\big(\frac{\partial w'}{\partial v}-iA'\cdot \nu w'\big)\big|_\Gamma$  coincide
on $\Gamma$,  i.e.
$\Lambda'(k)f\big|_\Gamma=\Lambda(k)f\big|_\Gamma$  
for any  $f$  with  the support  in $\overline\Gamma$  and all
$k\in (k_0-\e,k_0+\e)$.  Then $(A'(x),V'(x))$ are gauge equivalent  to
$(A(x),V(x)),\  r'=r$  and $\Omega_j'=\Omega_j,1\leq j\leq r$.
\end{theorem}

It was shown in [15]  that   the proof 
of Boundary Control method,  given in [16], [17],
extends  to the hyperbolic equation  with Yang-Mills  potentials.   Therefore  analog of
Theorem \ref{theo:2.2} holds  and this implies that Theorem \ref{theo:2.4}  is also true.
\\
\

{\bf Remark 2.1}
In the Theorem \ref{theo:2.4}  we assumed that the DN operators $\Lambda(k)$  and $\Lambda'(k)$
are equal  on the interval $(k_0-\e,k_0+\e)$  and  therefore are  equal  for all $k$
because  they are analytic in $k$.

When $n\geq 3,\ \Gamma=\partial\Omega_0$  and there is no obstacles,  
a stronger results was proven in [10]  that the Yang-Mills potentials 
$(A(x),V(x))$  and $(A'(x),V'(x))$  are gauge equivalent if
$\Lambda'(k_0)=\Lambda(k_0)$  for a fixed $k_0$.   The proof  requires a different idea  
(see [10]
and some simplifications of the proof in [26]).

\subsection{Gauge  invariant boundary data}
\
\\
Let $u(x)$  be the solution of (\ref{eq:2.1}), (\ref{eq:2.2}).  There are two basic gauge
invariant quantities  in quantum mechanics:   the probability density $|u(x)|^2$  and
the probability current
\begin{equation}                                   \label{eq:2.14}
S(u)=\Im\Big( h\frac{\partial u}{\partial x}-i\frac{e}{c}Au\Big)\overline u.
\end{equation}
The probability density is obviously gauge invariant since $|u'|^2=|g^{-1}(x)u|^2=|u|^2$
for any $g\in G(\overline\Omega_0\setminus\bigcup_{j=1}^r\Omega_j)$.  For  the probability current  we have  
\begin{align}            
\nonumber
&S(u')=\Im\Big(h\frac{\partial}{\partial x}(g^{-1}u)-\frac{ie}{c}A'g^{-1}u\Big)g\overline u
\\
\nonumber
=&\Im\Big(h\frac{\partial u}{\partial x}g^{-1}-hg^{-2}\frac{\partial g}{\partial x}u
-i\Big(\frac{e}{c}A +ihg^{-1}\frac{\partial g}{\partial x}\Big)g^{-1}u\Big)g\overline u.
\end{align}
We used above that $\overline g=g^{-1}$  and  that $\frac{e}{c}A'=\frac{e}{c}A 
 +ihg^{-1}\frac{\partial g}{\partial x}$  (cf. (\ref{eq:1.5})).   Therefore  
$S'(u')=\Im\big(h\frac{\partial u}{\partial x}-i\frac{e}{c}Au\big)\overline u=S(u)$. 

Using  the probability density and the probability  current  we define gauge invariant
data on $\partial\Omega_0$  for any solution $u(x)$  of 
(\ref{eq:2.1}),  (\ref{eq:2.2}):
\begin{equation}                                    \label{eq:2.15}
|u(x)|^2\big|_{\partial\Omega_0}=f_1(x'),\ \ 
\frac{\partial}{\partial \nu}|u(x)|^2\big|_{\partial\Omega_0}=f_2(x'),\ \
S(u)\big|_{\partial\Omega_0}=f_3(x').
\end{equation}
\begin{lemma}                                       \label{lma:2.5}
Consider all $u(x)$  and  $u'(x)$  such  that $(H-k^2)u=0$  in 
$\Omega_0\setminus\Omega',\ u\big|_{\partial\Omega'}=0,$
and  $(H'-k^2)u'=0$ 
 in $\Omega_0\setminus\Omega', \ u'\big|_{\partial\Omega'}=0$,  respectively.  Let
$\Lambda,\Lambda'$  be the corresponding  DN operators.  Suppose that the set 
$(f_1,f_2,f_3)$  of all  gauge  invariant  boundary  data for $u(x)$  and  $u'(x)$  
is the same.

Then there exists $g_0(x)\in G(\overline\Omega_0\setminus\bigcup_{j=1}^r\Omega_j)$  such that 
\begin{equation}                                        \label{eq:2.16}
 g_0\big|_{\partial\Omega_0}\Lambda'\big((g_0^{-1}u)\big|_{\partial\Omega_0}\big)=
\Lambda\big(u\big|_{\partial\Omega_0}\big)
\end{equation}
for all  $u(x)$  such  that $(H-k^2)u=0,\ u\big|_{\partial\Omega'}=0$.
\end{lemma}
{\bf Proof:}
Consider  smooth  $u_0(x),u_0'(x)$  having the same boundary data (\ref{eq:2.15})
and such that $|u_0(x)|=|u_0'(x)|>0$ on $\partial\Omega_0$.
Let  $g_0(x)=\frac{u_0(x)}{u_0'(x)}$ near  $\partial\Omega_0$.  Extend $g_0(x)$
to the whole 
domain  $\Omega_0\setminus\Omega'$  keeping  $|g_0(x)|=1$.  We have on 
$\partial\Omega_0$
$$
S(u_0')=\Im\big(hg_0^{-1}\frac{\partial u_0}{\partial x}-hg_0^{-2}\frac{\partial g_0}{\partial x}u_0
-i\frac{e}{c}A'(x)g_0^{-1}u_0\big)g_0\overline u_0
$$
$$
=S(u_0)+\Im\big(-hg_0^{-1}\frac{\partial g_0}{\partial x}
+i\frac{e}{c}(A(x)-A'(x)\big)|u_0|^2.
$$
Since  $S(u_0')=S(u_0)$  and since  $g_0^{-1}\frac{\partial g_0}{\partial x}$  
is imaginary,  we get
\begin{equation}                                    \label{eq:2.17}
-hg_0^{-1}\frac{\partial g_0}{\partial x}=i\frac{e}{c}(A'(x)-A(x))\ \ \mbox{when}\ \
x\in\partial\Omega_0.
\end{equation}
Analogously,  let  $u(x),u'(x)$  be any  solutions  of 
$(H-k^2)u=0,\ (H'-k^2)u'=0$  having  the same boundary  data and such that
$|u(x)|=|u'(x)|>0$.

Denote  $g(x)=\frac{u(x)}{u'(x)}$.   Then  $u'=g^{-1}u$  on  
$\partial\Omega$  and analogously  to (\ref{eq:2.17})
we get
$hg^{-1}\frac{\partial g}{\partial x}=i\frac{e}{c}(A(x)-A'(x))$.  Therefore
$g^{-1}\frac{\partial g}{\partial x}=g_0^{-1}\frac{\partial g_0}{\partial x}$.  Thus
$\frac{\partial}{\partial x}\big(\frac{g}{g_0}\big)=0$  on $\partial \Omega_0$.
Hence $g=e^{i\alpha}g_0$  where  $\alpha$  is a constant.

We have
\begin{align}
\nonumber
&\big(\Lambda'u'\big|_{\partial\Omega_0}\big)\overline u'\big|_{\partial\Omega}
=\Big( h\frac{\partial u'}{\partial \nu}
-i\frac{e}{c}A'\cdot \nu u'\Big)\overline u'\Big|_{\partial\Omega_0}
\\
\nonumber
 &=h\frac{1}{2}\frac{\partial|u'|^2}{\partial\nu}
+i\Im\Big(h\frac{\partial u'}{\partial \nu}
-i\frac{e}{c}A'\cdot \nu u'\Big)\overline u'\Big|_{\partial\Omega_0}
\\
\nonumber
 &=\Big(h\frac{1}{2}\frac{\partial}{\partial \nu}|u'|^2
+iS'(u')\nu\Big)\Big|_{\partial\Omega_0}.
\end{align}
We used above that  $\Re\frac{\partial u'}{\partial x}\overline u'=
\frac{1}{2}\frac{\partial |u'|^2}{\partial x}$.

Analogously,
$$
\Lambda\big(u\big|_{\partial\Omega_0}\big)\overline  u\big|_{\partial\Omega_0}
=\Big(h\frac{1}{2}\frac{\partial|u|^2}{\partial \nu}
+i S(u)\cdot\nu\Big)\Big|_{\partial\Omega_0}.
$$
Since $S'(u')=S(u)$  we get
$$
\Lambda'\big(u'\big|_{\partial\Omega_0}\big)\overline u'\big|_{\partial\Omega_0}
=\Lambda\big(u\big|_{\partial\Omega_0}\big)\overline u\big|_{\partial\Omega_0}
$$
for all $u,u'$  having the same boundary data  and  $|u|=|u'|>0$  on
$\partial\Omega_0$.  Since  $u'=e^{-i\alpha}g_0^{-1}u$
we get,  cancelling $\overline u$  and $e^{-i\alpha}$  that 
$g_0\Lambda'\big(g_0^{-1}u\big|_{\partial\Omega_0}\big)=\Lambda\big(u\big|_{\partial\Omega_0}\big)$.
Since  $u,\ |u|>0$  on $\partial\Omega_0$  are dense in $L_2(\partial\Omega_0)$  we
have that (\ref{eq:2.16}) 
holds  for all $u(x)$,
i.e.
Lemma \ref{lma:2.5}  is proven.

We shall call DN operators  $\Lambda$ and $\Lambda'$ 
satisfying  (\ref{eq:2.16}) gauge equivalent with the gauge $g_0$.

If potentials $A(x)$ and $A'(x)$  are gauge equivalent with gauge $g$ then DN  operator
$\Lambda$  and $\Lambda'$  are also gauge equivalent 
with the same gauge.  Indeed,  on $\partial\Omega_0$ we have
\begin{align}
\nonumber
&\Lambda'\big(u'\big|_{\partial\Omega_0}\big)
=\Big(h\frac{\partial u'}{\partial x}-i\frac{e}{c}A'u'\Big)\cdot\nu\Big|_{\partial\Omega_0}
\\
\nonumber
&=\Big(hg^{-1}\frac{\partial u}{\partial x}
-hg^{-2}u\frac{\partial g}{\partial x}\Big)\nu\Big|_{\partial\Omega_0}
-i\Big(\frac{e}{c}A(x)
+ih\frac{\partial g}{\partial x}g^{-1}\Big)\nu g^{-1}u\Big|_{\partial\Omega_0}
\\
\nonumber
&=\Big(h\frac{\partial u}{\partial x}
-i\frac{e}{c} Au\Big)\cdot \nu g^{-1}\Big|_{\partial\Omega_0}
=g^{-1}\big|_{\partial\Omega_0}\Lambda\big(u\big|_{\partial\Omega_0}\big),
\end{align}
i.e.  $\Lambda'$  and $\Lambda$  are gauge equivalent.
The converse statement is also true.

\begin{lemma}                                  \label{lma:2.6}
Suppose DN  operators $\Lambda$ and $\Lambda'$ 
are gauge equivalent with gauge $g_0$,  i.e.  (\ref{eq:2.16}) holds.
Then magnetic potentials $A(x)$ and $A'(x)$  are also gauge equivalent  with some gauge
$g(x)$ and $V(x)=V'(x)$.
\end{lemma} 
{\bf Proof:}
Consider Schr\"odinger equations $(H-k^2)u=0,\ (H'-k^2)u'=0$  corresponding  to potentials 
$(A,V),\ (A',V')$,  respectively.   Let $\Lambda,\Lambda'$   be the corresponding  DN  operators.
In $(H-k^2)u=0$  make the gauge transformation $u_0=g_0^{-1}u$.   Then we obtain  
the Schr\"odinger operator  $(H_0-k^2)u_0=0$  where  $(A_0,V_0)$  are
gauge equivalent  to $(A,V)$.  Note that
the DN  operator corresponding to $H_0$
has the form  $\Lambda_0\big(u_0\big|_{\partial\Omega_0}\big)
=g_0^{-1}\big|_{\partial\Omega_0}\Lambda (g_0u_0)\big|_{\partial\Omega_0}$.  It follows from 
(\ref{eq:2.16})  that $\Lambda'=\Lambda_0$. 
Therefore the DN operator for $(H'-k^2)u'=0$  and $(H_0-k^2)u_0=0$  are the same.
By Theorem \ref{theo:2.1}  the potentials $(A',V')$ and $(A_0,V_0)$  are gauge
equivalent with some gauge  $g_1$.

Therefore the potentials $(A,V)$  and $(A',V')$  are  also gauge equivalent
with gauge $g_1g_0$.

Combining Lemmas \ref{lma:2.5} and \ref{lma:2.6}  we get that if gauge invariant data for
$(A,V)$  and $(A',V')$  
are equal as in Lemma \ref{lma:2.5} then  $(A,V)$  and $(A',V')$  are gauge equivalent.
\\
\

{\bf Remark 2.2.}  Lemmas \ref{lma:2.5} and \ref{lma:2.6}
hold when we replace $\partial\Omega_0$  by any open  subset $\Gamma\subset\partial\Omega_0$.
Thus we have the following theorem:
\begin{theorem}                                  \label{theo:2.7}
Let $u(x),u'(x)$  and  $\Lambda,\Lambda'$ be the same as  in Lemma \ref{lma:2.5}.
If  the set of 
the gauge  invariant boundary  data on $\Gamma$ for  $u(x)$  and $u'(x)$  is the same,
then the magnetic potentials $A(x)$  and  $A'(x)$ are gauge equivalent and
$$
V(x)=V'(x).
$$
\end{theorem}
The converse statement  is obvious:  if $(A(x),V(x)$  are gauge equivalent  to  $(A'(x),V'(x))$
then  the set  of  boundary data  (\ref{eq:2.15}) on $\Gamma$  is the same  for
$u(x)$  and $u'(x)$  because the boundary data (\ref{eq:2.15})  are gauge invariant.

Theorem \ref{theo:2.7}  has the  corollary  that gives another proof of the magnetic AB effect:
\begin{corollary}
Suppose  $\mbox{curl}\,A=0,\ \mbox{curl}\,A'=0$  and $V(x)=V'(x)$.
If  $A(x)$ and $A'(x)$  are not  gauge  equivalent then  the sets of boundary data (\ref{eq:2.15})
are  different  for $u(x)$  and $u'(x)$.  This implies  that $A(x)$ and $A'(x)$  have a
different physical impact,  i.e.  the AB effect holds.
\end{corollary}

\subsection{Inverse scattering problems}
\
\\
We consider the Schr\"odnger equation (\ref{eq:2.1})  in
 $\R^n\setminus\bigcup_{j=1}^r\Omega_j$
assuming that
$$
u\big|_{\partial\Omega_j}=0,\ \ 1\leq j\leq r.
$$
In problems related to AB effect  the magnetic field 
$B=\mbox{curl}\,A$ is shielded inside the obstacles $\Omega_j,\ 1\leq j\leq r$,  and  
therefore has a compact support in $\R^n$.  The electric potential  $V(x)$   plays
no role in magnetic AB effect and could be taken even equal to zero.
We assume that  $V(x)$  also has a compact  support.  The magnetic potential $A(x)$
may have or may have not a compact support  if $B(x)$ 
has a compact support.

\begin{lemma}                                        \label{lma:2.8}
Let $B(x)$ has a compact support, $\mbox{supp}\,B(x)\subset B_R$.  If  $n\geq 3$  or if
$n=2$  and 
\begin{equation}                                          \label{eq:2.18}
\iint\limits_{|x|<R}B(x)dx_1dx_2=0,
\end{equation}
then there exists a magnetic potential  $A(x)$  with 
compact support  such that $\mbox{curl}\, A=B$  in $\R^n$  and  
$\mbox{supp}\,A(x)\subset  B_R$. 
\end{lemma}

{\bf Proof:} Consider first the case $n=2$  and
$\iint_{|x|<R}B(x)dx_1dx_2=0$.\linebreak  Let $\tilde B(\xi)=\int_{\R^2}B(x)e^{-ix\cdot\xi}dx$
be the Fourier transform  of $B(x)$.  Since $\mbox{supp}\,B(x)\subset B_R,
\ \tilde B(\xi_1,\xi_2)$
is an entire function  of $(\xi_1,\xi_2)\in\C\times\C$  and
$|\tilde B(\xi)|\leq Ce^{R|\Im\xi|}$,  where $\Im\xi=(\Im\xi_1,\Im \xi_2)$. 
 Since
$\iint_{|x|<R}B(x)dx_1dx_2=0$  
we have $\tilde B(0,0)=0$.
Applying  the mean value theorem  we have
\begin{equation}                                      \label{eq:2.19}
\tilde B(\xi_1,\xi_2)=\xi_1\tilde B_1(\xi)+\xi_2\tilde B_2(\xi),
\end{equation}
where
 \begin{equation}                                    \label{eq:2.20}
\tilde B_j(\xi_1,\xi_2)=\int\limits_0^1\frac{\partial \tilde B}{\partial\xi_j}(t\xi_1,t\xi_2)dt,
\ \ j=1,2.
\end{equation}
Obviously,  $B_j(\xi_1,\xi_2)$  are also entire functions of $(\xi_1,\xi_2)$  and
\begin{equation}                                          \label{eq:2.21}
|\tilde B_j(\xi)|\leq C e^{R|\Im \xi|},\ j=1,2.
\end{equation} 
By the Paley-Wiener 
theorem the inverse Fourier transform $B_j(x)=F^{-1}\tilde B_j(\cdot)$ is 
also contained in  $B_R$. 
 We have $\frac{\partial A_2}{\partial x_1}- \frac{\partial A_1}{\partial x_2}=
B(x)$.  Making the Fourier transform and using (\ref{eq:2.19}) we can take
$\tilde A_1(\xi)=i\tilde B_2(\xi),\linebreak \tilde A_2(\xi)=-i\tilde B_1(\xi)$.  Therefore
$\mbox{supp}\,A_j\subset B_R$  and 
$\frac{\partial A_2}{\partial x_1}- \frac{\partial A_1}{\partial x_2}=B(x)$.

Now consider the case $n\geq 3$.  The equation $\mbox{curl}\,A=B$  has the following
form after  performing the Fourier transform
\begin{equation}                                            \label{eq:2.22}
\xi_2\tilde A_3(\xi)-\xi_3\tilde A_2(\xi)=-i\tilde B_1,\ \ 
-\xi_1\tilde A_3(\xi)+\xi_3\tilde A_1(\xi)=-i\tilde B_2,\ \ 
\end{equation}
\begin{equation}                                            \label{eq:2.23}
\xi_1\tilde A_2(\xi)-\xi_2\tilde A_1(\xi)=-i\tilde B_3.
\end{equation} 
Note that $\mbox{div}\, B=0$,  i.e.  
$\xi_1\tilde B_1(\xi)+\xi_2\tilde B_2(\xi)+\xi_3\tilde B_3(\xi)=0$.  In particular,
we have $\tilde B_3(0,0,\xi_3)=0$.  Therefore, as in (\ref{eq:2.19}),
we have
$\tilde B_3(\xi)=\xi_1\tilde B_{31}(\xi)+\xi_2\tilde B_{32}(\xi)=0$
and we choose $\tilde A_1(\xi)=i\tilde B_{32}(\xi),\ \tilde A_2(\xi)=-i\tilde B_{31}(\xi)$.
Substituting in (\ref{eq:2.22}) we get
\begin{equation}                                       \label{eq:2.24}
\xi_2\tilde A_3(\xi)=-i\tilde B_1-i\xi_3\tilde B_{31},
\end{equation}
\begin{equation}                                       \label{eq:2.25}
\xi_1\tilde A_3(\xi)=+i\tilde B_2+i\xi_3\tilde B_{32},
\end{equation}
Note that $\xi_1(-i\tilde B_1-i\xi_3\tilde B_{31}) =\xi_2(i\tilde B_2+i\xi_3B_{32})$  since
$\xi_1\tilde B_{31}+\xi_2\tilde B_2+\xi_3(\xi_1\tilde B_3+\xi_2\tilde B_{32})=0$.

Therefore
\begin{equation}                                      \label{eq:2.26}
\tilde A_3=\frac{-i\tilde B_1-i\xi_3\tilde B_{31}}{\xi_2}=
\frac{i\tilde B_2(\xi)+i\xi_3\tilde B_{32}}{\xi_1}.
\end{equation}
It follows from (\ref{eq:2.24}),  (\ref{eq:2.25})  that $\tilde A_3(\xi)$ is analytic when
$\xi_2\neq 0$  or $\xi_1\neq 0$.   Therefore by the theorem of removable singularity
for analytic functions of several  variables  $\tilde A_3(\xi)$  is 
an entire  analytic function.
Since estimates of the form (\ref{eq:2.19}) hold,  $A_3(x)=F^{-1}\tilde A_3(\xi)$ 
has  the support  in $B_R$.

Therefore 
we proved the existence of the magnetic potential with compact support such that
$\mbox{curl}\,A=B$.
\\
\

{\bf Remark 2.3.}  A more careful analysis allows to conclude that if 
$\mbox{supp}\,B \subset\Omega_0$,  where $\Omega_0$  is a convex domain,
 then $\mbox{supp}\,A\subset\Omega_0$.

\begin{lemma}                               \label{lma:2.9}
Let $n=2,\ \mbox{supp}\,B(x)\subset\Omega_0$,
where $\Omega_0$  is convex,
$(0,0)\in\Omega_0,$
  $\iint\limits_{\Omega_0}B(x)dx_1dx_2=\alpha_0\neq 0$  Then 
there exists a magnetic potential $A(x)$  in $\R^2$ such 
that $\mbox{curl}\,A=B$  and $A(x)=A_0(x)$  in  $\R^2\setminus\Omega_0$,  where
\begin{equation}                                      \label{eq:2.27}
A_0(x)=\frac{\alpha_0}{2\pi}\frac{(-x_2,x_1)}{|x|^2}.
\end{equation}

The potential (\ref{eq:2.27})  is called the AB potential  (cf. [1]).
\end{lemma}

{\bf Proof:} 
Note that  $\mbox{curl}\, A_0=\alpha_0\delta(x)$  in $\R^2$.  Let $A'(x)$  be a magnetic 
potential  such that $\mbox{curl}\,A'=B(x)-\alpha_0\delta(x)$  in $\R^2$.  Since
$\iint_{\Omega_0}(B(x)-\alpha_0\delta(x))dx=0$, by Lemma \ref{lma:2.8}
we can choose 
$A'(x)$ such that $\mbox{supp}\,A'\subset\Omega_0$.  Consider  $A(x)=A_0(x)+A'(x)$.
Then $\mbox{curl}\,A=B$ in $\R^2,\ A=A_0(x)$  for  $x\in \R^2\setminus\Omega_0$.
\qed

Consider now the inverse scattering problem for the case when 
$A(x)$  and $V(x)$  have  compact supports that are contained in $\{|x|< R-\e\}$.
We assume also that all $\overline\Omega_j\subset\{|x|<R-\e\},\ 1\leq j\leq r.$

A solution $w(x,k\omega)$  of the form
$$
w(x,k\omega)=e^{ik\omega\cdot x}+\frac{a(\theta,\omega,k)e^{ik|x|}}{|x|^{\frac{n-1}{2}}}+
O\Big(\frac{1}{|x|^{\frac{n+1}{2}}}\Big)
$$
is called
a distorted plane wave.  Here $|\omega|=1,\ \theta=\frac{x}{|x|},\ a(\theta,\omega,k)$ is called
the scattering amplitude.  The existence
of distorted plane wave is well-known (see, for example, [30] or [20]). For the case of
magnetic potentials in domains with obstacles see [12], [41].

We consider the inverse scattering problem of  determining of the gauge equivalence
class of $A(x)$  and of $V(x)$  knowing the scattering  
amplitude $a(\theta,\omega,k)$  for fixed $k$  and all $\theta\in S^{n-1},\ \omega\in S^{n-1}$.

Consider simultaneously the 
 inverse boundary value problem in the domain $B_R\setminus\bigcup_{j=1}^n\Omega_j$,
where $B_R=\{|x|<R\}$.  We assume that the Dirichlet problem in
 $B_R\setminus\bigcup_{j=1}^n\Omega_j$
has a unique solution.  Then DN operator  is well defined.

\begin{theorem}                               \label{theo:2.10}
Consider two equations $(H-k^2)u=0,\ (H'-k^2)u'=0$  in $\R^n\setminus \bigcup_{j=1}^r\Omega_j$.
Let $a(\theta,\omega,k)$  and $a'(\theta,\omega,k)$  be corresponding  scattering amplitudes
and let $\Lambda(k)$ and $\Lambda'(k)$  be the corresponding DN operators on 
$\partial B_R=\{|x|=R\}$.
If $a(\theta,\omega,k)=a'(\theta,\omega,k)$  
for fixed $k$  and for all $(\theta,\omega)\in S^{n-1}\times S^{n-1}$,
then  $\Lambda(k)=\Lambda'(k)$  for the same $k$.   Vice versa,  if 
$\Lambda(k)=\Lambda'(k)$,  then 
$a(\theta,\omega,k)=a'(\theta,\omega,k)$  for all $(\theta,\omega)\in S^{n-1}\times S^{-1}$.
\end{theorem}

{\bf Proof:}
Assume $a(\theta,\omega,k)=a'(\theta,\omega,k)$.  Let 
$w(x,k\omega)$  and $w'(x,k\omega)$ be corresponding distorted plane waves.  Since
$a=a'$  we have $w(x,k\omega)-w'(x,k\omega)=O\big(\frac{1}{|x|^{\frac{n+1}{2}}}\big),\ |x|>R$.  
  By the Rellich's lemma (see, for example, 
Lemma 35.2 in [20])  we get
$$
w(x,k\omega)-w'(x,k\omega)=0 \ \ \mbox{for}\ \ \ |x|\geq R.
$$
Differentiating  in $x$  we have 
$$
\frac{\partial}{\partial\nu}(w(x,k\omega)-\omega'(x,k\omega))\big|_{\partial B_R}=0,
$$
where $\frac{\partial}{\partial\nu}$  is the unit normal  to $\partial B_R$. 
Therefore we got  that $\Lambda(k) w=\Lambda'(k) w'$ 
on $\partial B_R$  
for any distorted plane wave.  It is known (see,  for example,  [20])
  that the restrictions  of the distorted
plane waves on $\partial B_R$  are dense in $L_2(\partial B_R)$.
Therefore   
taking the closure we get that $\Lambda(k)f=\Lambda'(k)f$  for  all $f\in L_2(\partial B_R)$
(cf. [20]).

The converse statement  is also true:

If $\Lambda(k)=\Lambda'(k)$  on $\partial B_R$  then
$a(\theta,\omega,k)=a'(\theta,\omega,k)$ for all $(\theta,\omega)\in S^{n-1}\times S^{n-1}$.
We shall omit the proof  (cf.  [20], [33]).
Therefore  combining the Theorem \ref{theo:2.10}  with  the Theorem \ref{theo:2.1}
for
$\Gamma=\partial B_R$  we get that if $a(\theta,\omega,k)=a'(\theta,\omega,k)$  for all 
$(\theta,\omega)\in S^{n-1}\times S^{n-1}$  then $A(x)$  and $A'(x)$  are gauge equivalent with the gauge
$g(x)=1$  for  $|x|\geq R,\ V(x)=V'(x),\ r=r',\ \Omega_j'=\Omega_j$  for $1\leq j\leq r$.
Note that 
in the case when $\mbox{supp}\,A(x)\subset B_R$
the gauge group in $\R^n\setminus\bigcup_{j=1}^r\Omega_j$  consists of
$|g(x)|=1,\ g(x)=1$  for  $|x|\geq R$.
\qed

Now consider the inverse scattering problem in the case $n=2$  and magnetic flux
$\alpha\neq 0$.  
We consider    magnetic potentials  of the form  (cf.  [24])
$A(x)=A_0(x)+A_1(x)$,  where  $A_0(x)$  has the form  (\ref{eq:2.27}) and
$A_1(x)=O\big(\frac{1}{|x|^{1+\e}}\big),\e>0$.  Note  that $\mbox{curl}\,A=B=0$  
for $|x|>R$.

We can choose inside the gauge equivalence class the magnetic
potential  equal  to 
$A_0(x)=\frac{\alpha(-x_2,x_1)}{2\pi|x|^2} $  for $|x|>R$
Since $A_0(x)=O\big(\frac{1}{|x|}\big)$  the scattering  
amplitude
is a distribution  and it  has the form 
(cf.[1], [43], [44], [45])
\begin{equation}                                           \label{eq:2.28}
a(\theta,\omega,k)=a_0(\theta-\omega)+a_1(\theta,\omega,k),
\end{equation}
where
\begin{multline}                                          \label{eq:2.29} 
a_0(\theta)=\cos\frac{\alpha}{2} \delta(\theta)+\frac{i\sin\frac{\alpha}{2}}{\pi}\
 p.v. \frac{e^{i\big[\frac{\alpha}{2\pi}\big]\theta}}{1-e^{i\theta}},
\ \ 
|a_1(\theta,\omega,k)|\leq C|\theta-\omega|^{-\e},
\\  0\leq \e <1.
\end{multline}
Here $[\alpha]$  is the smallest integer larger or equal to $\alpha$.

The following analog of Theorem \ref{theo:2.10}  holds  (cf [24]):
\begin{theorem}                            \label{theo:2.11}
Let $(H-k^2)u=0,\ (H'-k')u'=0$ be two Schr\"odinger operators in 
$\R^2\setminus \bigcup_{j=1}^r\Omega_j$.
Suppose $A_0=\frac{\alpha_0(-x_2,x_1)}{2\pi|x|^2},\ A_0'=\frac{\alpha_0'(-x_2,x_1)}{2\pi|x|}$
  for 
$|x|>R$.
If $a(\theta,\omega,k)=a'(\theta,\omega,k)$  and if $\alpha_0=\alpha_0'$  then
$A(x)$  and $A'(x)$  are gauge  equivalent.
\end{theorem}
Note that in Theorem \ref{theo:2.11}  we require not only that $a=a'$  but also  that
the magnetic fluxes $\alpha_0$  and $\alpha_0'$  are equal.

It was shown in [24] that  if there is only one convex obstacle, $\Omega_1=\Omega_1'$,  then
$a=a'$  and $\alpha_0\neq 2\pi n,\ \forall n\in\Z$,  implies that $\alpha_0'=\alpha_0$.

A similar  result  holds  for  the inverse boundary value  problem:   If $\Lambda(k)=\Lambda'(k)$
on $\partial B_R$  then  $A(x)$ and  $A'(x)$  are  gauge equivalent  in 
$B_R\setminus\bigcup_{j=1}^r\Omega_j$ and
$\alpha_0=\alpha_0'$.  Indeed, by Theorem \ref{theo:2.1}  $A(x)$  and  $A'(x)$  are  gauge equivalent    with the gauge $g(x)$  such that $g(x)=1$  on 
$\partial B_R$.  Thus 
$\int_{\partial B_R}g^{-1}\frac{\partial g}{\partial x}\cdot dx=0$  and 
therefore  $\alpha_0=\frac{e}{hc}\int_{\partial B_R}A(x)\cdot dx$  is equal  to 
$\alpha_0'=\frac{e}{hc}\int_{\partial B_R}A'(x)\cdot dx$.

Note that when $\alpha_0\neq 0$  the gauge group has the form 
$$
|g(x)|=1,\ x\in\R^2\setminus\bigcup_{j=1}^r\Omega_j,
\ \ g(x)=e^{i p\theta(x)}\Big(1+O\Big(\frac{1}{|x|}\Big)\Big),
$$
where $p\in \Z$.

When we make a gauge transformation, the scattering amplitude  changes
\begin{equation}                                        \label{eq:2.30}
a'(\theta,\omega,k)=e^{-ip\theta}a(\theta,\omega,k)e^{-ip(\theta+\pi)}.
\end{equation}
Consider  the gauge equivalence class of scattering amplitude for 
the operator $H-k^2$.   It was shown in [24]  that when $\Omega_1$  is a single 
convex obstacle and $\alpha\neq 2\pi n,\ \forall n\in\Z,$  then there is one-to-one 
correspondence between  gauge  equivalence  classes  of magnetic potentials and
gauge equivalence  classes of scattering  amplitudes.

\subsection{Aharonov-Bohm effect  and the spectrum  of the Schr\"odiger operator}
\
\\
In this subsection we shall show that $\cos\alpha$,  where $\alpha$  is a magnetic flux, 
 is determined by the spectrum.  Therefore  if $\cos\alpha_1\neq\cos\alpha_2$ for
two Schr\"odinger operators then their spectra are different.

Let $\Omega$  be a convex obstacle  in $\R^2$  containing the origin.  Let 
$B_R=\{|x|<R\}$,  where $R$  is large.  Consider 
the Schr\"odinger equation $(H-\lambda )u=0$  in the annulus domain 
$B_R\setminus\Omega$  with zero Dirichlet boundary conditions 
$u\big|_{\partial B_R}=0,\ u\big|_{\partial\Omega}=0$.  Let 
$\lambda_1\leq \lambda_2\leq \lambda_3 \leq ...$  be the Dirichlet spectrum and
let $E(x,y,t)$  be the hyperbolic Green function ,  i.e. 
\begin{align}
\nonumber
&\Big(\frac{\partial^2}{\partial t^2}  +H\Big)E(x,y,t)=0 \ \
\mbox{for}\ \ t>0,
\\ 
\nonumber
&E(x,y,t)\Big|_{\partial\Omega\times(0,+\infty)}=0, \ \
E(x,y,t)\Big|_{\partial B_R\times(0,+\infty)}=0,
\\
\nonumber
&E(x,y,0)=\delta(x-y),\ \ \ \frac{\partial E(x,y,0)}{\partial t}=0.
\end{align}
The following wave trace formula holds  (cf. [8])
\begin{equation}                                   \label{eq:2.31}
\mbox{Tr}(t)\stackrel{def}{=}\sum_{j=1}^\infty\cos\sqrt\lambda_j t=
\int\limits_{B_R\setminus\Omega}E(x,x,t)dx.
\end{equation}
It was proven in [8], [27]  that the singularities of the wave trace occur
at the time $t=T$,  where  $T$  is equal  to the length  of periodic null-geodesics.
In our geometry  the periodic null-geodesics are equilateral  $N$-gones 
inscribed in the circle $|x|=R$,
in particular,  equilateral triangles with the side $R\sqrt 3$. 
It was proven in [25]  that at $t=3R\sqrt 3$  the singularity
of $\mbox{Tr}(t)$  has the form
\begin{equation}                                     \label{eq:2.32}
-2^{-\frac{5}{2}}3^{\frac{1}{4}}R^{\frac{3}{2}}\cos\alpha(t-3R\sqrt 3)_+^{-\frac{3}{2}}+
O\big((t-3R\sqrt 3)^{-\frac{1}{2}}\big).  
\end{equation}
Here $\alpha=\int_\gamma A(x)\cdot dx$  is the magnetic flux, 
 $\gamma$  is any  simple  closed contour  between 
$\partial\Omega$  and $\partial B_R$ ($\alpha$  is
 independent of $\gamma$  since we assume that 
$\mbox{curl}\, A=0$
in $B_R\setminus\Omega),\ (t-3R\sqrt 3)_+^{-\frac{3}{2}}$  is a homogeneous 
of order 
$-\frac{3}{2}$  distribution  equal to zero  when  $t-3R\sqrt 3<0$.
Similar formula holds (cf [25])  when the triangle  is replaced by $N$-gone.
Therefore the spectrum  depends on the magnetic flux.

Aharonov-Bohm effect  holds when  the underlying manifold is not simply-connected  even
when there are no obstacles.

Consider the Schr\"odinger operator on the torus  (cf. [25]).
Let $L=\{m_1e_1+m_2e_2,\ m_1,m_2\in\Z\}$  be a lattice in $\R^2$  and let 
$L^*$  be the dual lattice  consisting of $\delta\in \R^2\ \ \mbox{such that}
\ \ \delta\cdot  d \in \Z\ \ \mbox{for all}\ \ d\in L$.
Consider the Schr\"odinger  operator 
\begin{equation}                                    \label{eq:2.33}
H=\Big(-i\frac{\partial}{\partial x}-A(x)\Big)^2 + V(x)\ \ \mbox{on the torus}\ \
\T^2=\R^2/L.
\end{equation}
The potentials $A(x)$ and $V(x)$  are periodic,  i.e.  
$A(x+d)=A(x),\ V(x+d)=V(x)$  for all $x\in R^2$  and $d\in L$  and therefore there are
defined on $\T^2=\R^2/L$. We assume that the magnetic field $B=\frac{\partial A_2}{\partial x_1}
-\frac{\partial A_1}{\partial x_2}=0$.   Let 
 $\gamma_1,\gamma_2$  be the basis of the homology group  of $\T^2$.
Denote
\begin{equation}                                      \label{eq:2.34}
\alpha_j=\int\limits_{\gamma_j}A(x)\cdot dx, \ \  j=1,2.
\end{equation}
The gauge group $G(\T^2)$  consists of $g(x)\in C^\infty(\T^2)$.
such that $|g(x)|=1$.  Any such $g(x)$ has the form $g(x)=e^{i\delta\cdot x+i\varphi(x)}$
where $\varphi(x)\in C^\infty(\T^2)$  and $\delta\in L^*$.

Two magnetic potentials $A(x)$  and $A'(x)$ 
are gauge  equivalent  if $A'=A+ig^{-1}(x)\frac{\partial g}{\partial x}$.
\begin{theorem}                                               \label{theo:2.12}
Let $H$  and $H'$  be two Schr\"odinger operators on $\T^2$
with electromagnetic potentials  $(A(x),V(x))$  and $(A'(x),V'(x))$. Suppose
$\mbox{curl}\,A=\mbox{curl}\,A'=0$.
Suppose that the spectrum  of $H$  and $H'$  are the same.  Then
\begin{align}                                    \label{eq:2.33}
&\cos\alpha_j=\cos\alpha_j',\ \ j=1,2,
\\
\nonumber
\mbox{where}\ \ \ 
\alpha_j=&\int\limits_{\gamma_j}A(x)\cdot dx,
\ \ \alpha_j'=\int\limits_{\gamma_j}A'(x)\cdot dx,\ \ 
j=1,2.
\end{align}
\end{theorem}
This demonstrates
the AB effect on torus since the magnetic fluxes make a physical impact.

\subsection{Direct proof of magnetic AB effect}
\
\\
Consider 
the nonstationary Schr\"odinger equation
\begin{equation}                                       \label{eq:2.36}
-ih\frac{\partial u}{\partial t}+
\frac{1}{2m}\sum_{j=1}^n\left(-ih\frac{\partial}{\partial x_j}
-\frac{e}{c}A_j(x)\right)^2u+e V(x)u=0,
\end{equation}
in $(\R^n\setminus\Omega')\times (0,T)$
where  $n\geq 2,\ \Omega'=\bigcup_{j=1}^r\Omega_j, $
\begin{equation}                                    \label{eq:2.37}
u(x,0)=u_0(x),
\end{equation}
\begin{equation}                                     \label{eq:2.38}
u\Big|_{\partial\Omega_j\times(0,T)}=0,\ \ 1\leq j\leq r.
\end{equation}
At first  we shall study  the case of one obstacles in $\R^3$.
Suppose $\Omega_1$ is a toroid and $\mbox{curl}\,A=0$  in $\R^n\setminus\Omega_1$.
It was shown in \S 2.4  that  we can choose  $A(x)$  having a compact  support.  
Let $x^{(0)}\not\in\overline\Omega_1$
and  $\theta\in S^2$  be a unit vector.  Suppose  $\theta_{\perp 1},\theta_{\perp 2}$  
are two unit vectors such that $\theta,\theta_{\perp 1},\theta_{\perp 2}$  
is an orthonormal basis in $\R^3$.  Let $\chi_0(s)\in C_0^\infty(\R^1),\chi_0(s)=1$
for $|s|<\frac{1}{2},\ \chi_0(s)=0$  for  $|s|>1,\ \chi_0(s)=\chi_0(-s)$.
It was proven in [21]  that there exists a solution  of  (\ref{eq:2.36})  of the form
\begin{align}                                       \label{eq:2.39}
&u(x,t,\theta)
=e^{-i\frac{mk^2t}{2h}+i\frac{mk}{h}x\cdot\theta}
\chi_0\Big(\frac{(x-x^{(0)})\cdot\theta_{\perp 1}}{\delta_1}\Big)
\chi_0\Big(\frac{(x-x^{(0)})\cdot\theta_{\perp 2}}{\delta_1}\Big)
\\
\nonumber
&\cdot\exp\Big(i\frac{e}{hc}\int_0^\infty \theta\cdot A(x-s'\theta)ds'\Big)
+O(\e),
\end{align}
where $t\in (0,T),\ T=O\big(\frac{1}{k^{\delta_1}}\big),\ k$  is large ,  $\delta_1$  is small,
$\e >0$ can be chosen arbitrary small  if $k$  is large  enough.

The support of $u(x,t,\theta)$  modulo  $O(\e)$  is  contained in a small
neighborhood of the line  $x=x^{(0)}+s\theta$.
\\
\\
\begin{tikzpicture}[scale=1]
\draw[->](0,5) -- (-2,0);
\draw(1.5,2.5) ellipse (1.8 and 1);
\draw(1.5,2.5) ellipse (0.9 and 0.5);

\draw[->](0,5) -- (1.8,2.0);
\draw[->](2.0,1.6) -- (3,0);
\draw(0.3,5.3)  node {$x^{(0)}$};
\draw(0.7,4.5)  node {$\theta$}; 
\draw(-0.5,4.5)  node  {$\omega$};

\end{tikzpicture}
\\ 
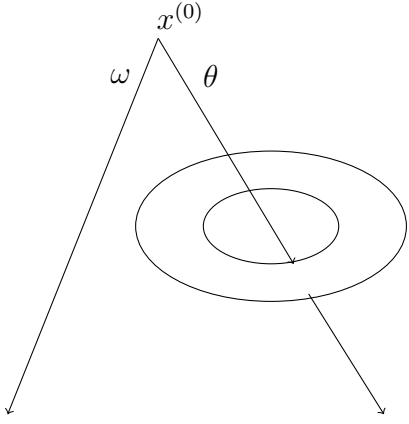
{\bf Fig. 1.} Two rays $x=x^{(0)}+s\theta,\ x=x^{(0)}+s'\omega,\ $
$0\leq s<+\infty,\ \linebreak 0\leq s'<+\infty$, intersect at point  $x^{(0)}$.
 Only  the ray $x=x^{(0)}+s\theta$  is passing  through the hole of
the toroid $\Omega_1$. 
\\
\\
We take  two  solutions $u(x,t,\theta)$  and  $v(x,t,\omega)$ 
 of the form  (\ref{eq:2.39})  corresponding to 
the directions $\theta$  and $\omega$,  respectively (cf.  Fig.1).  Let $U_0$
be a ball of radius $\e_0$ centered at  $x^{(0)}$.  We have  for $x\in U_0$
\begin{equation}                                     \label{eq:2.40}
|u(x,t,\theta)-v(x,t,\omega)|^2
=\big|1-e^{\frac{imk}{h}x(\theta-\omega)+i(I_1-I_2)}\big|^2+O(\e),
\end{equation}
where
\begin{equation}                                         \label{eq:2.41}
I_1=\frac{e}{hc}\int\limits_0^\infty\theta\cdot A(x-s\theta)ds,\ \ 
I_2=\frac{e}{hc}\int\limits_0^\infty\omega\cdot A(x-s\omega)ds.
\end{equation}
Since $A(x)$  has a compact support  and $\mbox{curl}\,A=0$
we have that  $I_1-I_2=\alpha$,  
where  $\alpha=\frac{e}{he}\int_\gamma A(x)\cdot dx$  is the magnetic flux,  
$\gamma$  is a closed curve passing  through  the hole  and not shrinking to a point. 
 Therefore
\begin{equation}                                          \label{eq:2.42}
|u(x,t,\theta)-v(x,t,\omega)|^2
=4\sin^2 \frac{1}{2}\big(\frac{mk}{h}x\cdot(\theta-\omega)+\alpha\big)+O(\e).
\end{equation}
Choose $k_n$  large and such that $\frac{mk_n}{h}x^{(0)}\cdot (\theta-\omega)=2\pi n,
n\in \Z$.
For $x\in U_0$  we have
\begin{equation}                                          \label{eq:2.43}
\big|\frac{mk_n}{h}(x-x^{(0)})\cdot(\theta-\omega)\big|\leq 2\frac{mk_n}{h}\e_0.
\end{equation}
Therefore,  choosing  $\e_0$  small enough  we get
\begin{equation}                                            \label{eq:2.44}
|u(x,t,\theta)-v(x,t,\omega)|^2=4\sin^2\frac{\alpha}{2}+O(\e).
\end{equation}
This proves  AB effect since  the probability density  
$|u(x,t,\theta)-v(x,t,\omega)|^2$                   
changes  with  the magnetic flux $\alpha$.   Note that here we cannot 
distinguish  between $+\alpha$  and $-\alpha$ modulo $2\pi n$.

Now consider the case  of several  obstacles  $\Omega_j, \ 1\leq j\leq r$ for
$n=2$.  Let
$\alpha_j=\frac{e}{hc}\int_{\gamma_j}A(x)\cdot dx,\ 1\leq j\leq r$,  where
$\gamma_j$  is a simple  contour  encircling  $\Omega_j$  only.
Let  $x^{(1)}\not\in \overline\Omega'=\bigcup_{j=1}^r\overline\Omega_j$.  
Denote by  $\gamma^{(0)}=
\gamma_0\cup\gamma_2\cup...\cup\gamma_d$  
the broken ray  starting at  $x^{(1)}$  and  reflecting  at 
$\Omega'$  at points $x^{(2)},...,x^{(d-1)}$. Let  $\omega_p,\ 1\leq p\leq d$,
be the directions of $\gamma_p$.  Note that
$\omega_{p+1}=\omega_p-2(\nu(x^{(p)})\cdot \omega_p)\nu(x^{(p)})$  where 
$\nu(x^{(p)})$  is  the outward  unit  normal  to $\Omega'$.  The last leg $\gamma_d$  of
this broken ray does not intersect  $\Omega'$ and can be extended  to infinity.
Let $x^{(0)}$  be some point on $\gamma_d$.  It was
proven in [21] that there exists  a solution  $u(x,t)$  of (\ref{eq:2.36})
satisfying  boundary 
conditions (\ref{eq:2.38})    and such that
  $\mbox{supp}\,u(x,t)$  is contained in a small neighborhood of the broken
ray  $\gamma^{(0)}$  and $u(x,t)$  has the following form  
in a small neighborhood of the point $x^{(0)}$:
\begin{equation}                                  \label{eq:2.45}                                      
u(x,t)=c_0(x,t')\exp i\Big(-\frac{mk^2t}{2h}+\frac{mk}{h}\psi_d(x)
+\frac{e}{hc}\int\limits_{\gamma(x(t'))} A(x)\cdot dx\Big)
+O\Big(\frac{1}{k}\Big),
\end{equation}
where
$$
\big|\frac{\partial\psi_d}{\partial x}\big|^2=1,\ \ \frac{\partial\psi_d(x^{(0)})}{\partial x}
=\omega_d,\ \  c_0(x,t')\neq0,\ \ t=\frac{t'}{k}.
$$
In (\ref{eq:2.45}) we denoted by $\gamma(x(t'))$  the broken
ray  starting near  $x^{(1)}$  at $t=0$  and ending at $x(t')$  near $x^{(0)}$.
In particular,
$\gamma(x(t^{(0)})=\gamma(x^{(0)})=\gamma^{(0)}$,  i.e. 
$x(t^{(0)})=x^{(0)}$  is the endpoint of  $\gamma^{(0)}$.
We assume that $\omega_1$  is the direction of the first leg 
of all broken rays  starting near  $x^{(1)}$
at $t=0$.  We choose 
the endpoints of $\gamma^{(0)},\ x^{(1)}$  and $x^{(0)}$,  far from the obstacles.  Thus,
the straight  ray $\beta$  starting at $x^{(1)}$  and ending at  $x^{(0)}$  does not
intersect the obstacles.  If  $x=\hat x(t)$  is the equation of $\beta$, where $t$  is the time parameter,
we assume that $\hat x(t_1)= x^{(1)}$  and  $\hat x(t^{(0)})=x^{(0)}$.  Thus  $t_1$  is equal to  
$t^{(0)}-|\beta|$
where $|\beta|$  is the length of $\beta$ (cf  Fig.2).  We can construct  a solution $v(x,t)$  such that
\begin{align}                                    \label{eq:2:46}
v(x,t)=
&\chi_0\Big(\frac{(x-x^{(2)})\cdot\theta_\perp}{\delta_1}\Big)
c_1(x,t')
\\
\nonumber
\cdot
&\exp\Big(-i\frac{mk^2t}{2h}+\frac{imk}{h}x\cdot\theta
+\frac{ie}{hc}\int\limits_{\beta(x(t'))} A(x)\cdot dx
+O\Big(\frac{1}{k}\Big),
\end{align}
where $t=\frac{t'}{k},(x,t')\in U_0$  where  $U_0$  is a neighborhood of $(x^{(0)},t^{(0)})$.

We choose the initial condition $c_1(x,t_1)$  such 
that $c_1(x,t')=c_0(x,t')$  at $(x^{(0)},t^{(0)})$.  

As in (\ref{eq:2.42})  we have (cf. [21])
\begin{align}                                          \label{eq:2.47}
&|u(x,t)-v(x,t)|^2
\\
\nonumber
=
&|c_0(x^{(0)},t^{(0)})|^2	
\Big(
4\sin^2 \frac{1}{2}\Big(\frac{mk}{h}(\psi_d(x)-\theta\cdot x)\Big)+
I_1-I_2\Big) +
O(\e)
\\
\nonumber
=&|c_0(x^{(0)},t^{(0)}|^24\sin^2\frac{1}{2}(I_1-I_2)+O(\e),
\end{align}
where $I_1$  and $I_2$  are integrals of $A(x)$  over  $\gamma(x^{(0)})$  and  $\beta(x^{(0)})$,
respectively.

Note that  $I_1-I_2=\alpha_\gamma$  where $\alpha_\gamma$ is the sum of magnetic fluxes of
all obstacles encircled by $\gamma^{(0)}$ and $\beta$.
\\
\\
\begin{tikzpicture}[scale=0.8]
\draw(-3,5) -- (7,4);
\draw (7,4) --(8,1);
\draw (-1,0)--(8,1);
\draw(-1,0) -- (-3,5);

\draw(7.7,5) circle (1.2);
\draw(9.1,0.5) circle (1.2);
\draw(5.5,2.5) circle (1.4);

\draw(-1.3,-0.3) node {$x^{(1)}$};
\draw(-3,5.3) node {$x^{(0)}$};
\draw(3.5,0.1) node {$\gamma_0$};
\draw(7.8,3) node {$\gamma_1$};
\draw(3,5) node {$\gamma_2$};
\draw(-2.5,3)  node {$\beta$};
\draw(5.5,2) node {$\Omega_1$};
\draw(7.7,4.8) node {$\Omega_3$};
\draw(9.1,0.2) node {$\Omega_2$};
\end{tikzpicture}
\\ 
\

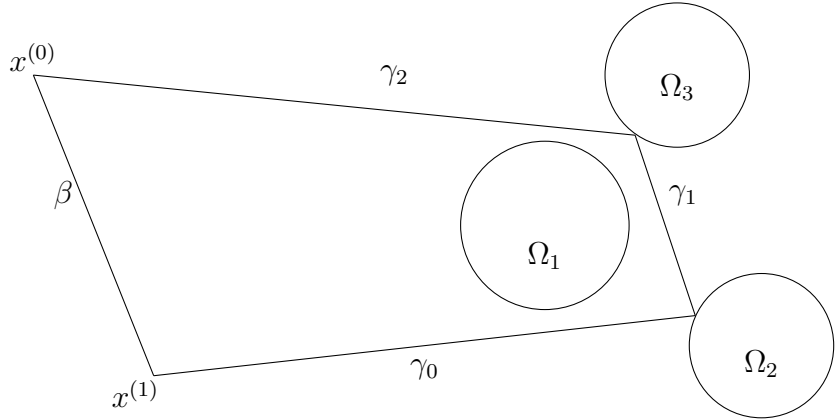
{\bf Fig. 2.} Broken ray $\gamma^{(0)}=\gamma_0\cup\gamma_1\cup\gamma_2$  starts at $x^{(1)}$
at  $t=0$,
reflects at $\Omega_2$  and  $\Omega_3$ and ends  at  $x^{(0)}$ at $t=t^{(0)}$.
The ray $\beta$ starts at $x^{(1)}$  at  $t=t_1$ and ends  at  $x^{(0)}$  at  $t=t^{(0)}$.
\\
\\
\\
Varying $\gamma^{(0)}$ and $\beta$ at least  $r$  times  we get  enough linear relations to recover
two  gauge  equivalence classes:  $\{\alpha_j\,(\mbox{mod}\ 2\pi n),\ 1\leq j\leq r\}$
and $\{-\alpha_j\,(\mbox{mod}\ 2\pi n),1\leq j\leq r\}$.

\section{Electric AB effect}
\init

In this section we shall study  the electric 
AB effect.  Consider
the Schr\"odinger  equation with electric potential  $V(x,t)$  and zero magnetic potential in 
$(\R^n\times[0,T])\setminus\Omega$  where $\Omega$  is a domain in 
$\R^n\times[0,T]$  that  we shall describe  below. We have
\begin{equation}                                \label{eq:3.1}
ih\frac{\partial u(x,t)}{\partial t}+\frac{h^2}{2m}\Delta u(x,t)-eV(x,t)u(x,t)=0
\end{equation}
with the initial and boundary conditions
\begin{equation}                                 \label{eq:3.2}
u(x,0)=u_0(x),\ \ x\in \R^n\setminus\Omega_0,
\end{equation}
\begin{equation}                                   \label{eq:3.3}
u\Big|_{\partial\Omega_{t_0}}=0,\ \ 0<t_0<T,
\end{equation}
where
$\Omega_{t_0}=\Omega\cap\{t=t_0\}$.  We assume that  the normals  
to $\Omega$  in $\R^n\times(0,T)$  are  not parallel
to the $t$-axis when $0<t<T$.

Consider the following domain:
Suppose $\Omega(\tau)$  in polar coordinates $(r,\theta)$  has the following form  (cf. Fig.3):
$$
\Omega(\tau)=\{(r,\theta)\ s.t.\ R_1<r<R_2,\ -\pi+\tau\leq \theta\leq \pi-\tau\},
$$
  where 
$0\leq \tau<\pi$.  Note that $\Omega(0)$ is the annulus domain $\{R_1<|x|<R_2\}$.
Let $\Omega_{t_0}=\Omega\cap\{t=t_0\}$  be equal to $\Omega(\e-t_0)$  when 
$0\leq t_0\leq \e,\ \Omega_{t_0}=\Omega(0)$  for $\e\leq t_0\leq T-\e,
\ \Omega_{t_0}=\Omega(t_0-T+\e)$  for  $T-\e\leq t_0\leq T$.  Thus $\Omega=
\bigcup_{0\leq t\leq T}\Omega_t$  is a time-dependent obstacle.

Let
$D=(\R^n\times(0,T))\setminus\Omega,\D_{t_0}=D\cap\{t=t_0\}$.
The domains $D_{t_0}$  are connected when $0\leq t_0<\e$  and  $T-\e<t_0\leq T$  and they 
are not connected when $\e\leq t_0<T-\e$:  there are two connected components:
$|x|>R_2$  and  $|x|<R_1$  for  $\e\leq t_0\leq T-\e$.

We consider two Schr\"odinger equations in $D$ of the form (\ref{eq:3.1}).  The first is when
$V_1(x,t)\equiv 0$  in $D$  and the second is when $V_2(x,t)=0$  outside 
$Q=\{(x,t):R_1\leq |x|\leq R_2,\e\leq t\leq T-\e\}$
and $V_2(x,t)=V_2(t)$  in $Q$.  Note that $E=\frac{\partial V_2(x,t)}{\partial x}=0$
in $D$.  We choose $V_2(t)$  such  that
\begin{equation}                                   \label{eq:3.4}
\frac{e}{h}\int\limits_\e^{T-\e}V_2(t)dt=\alpha\neq 2\pi p,\ \forall p\in \Z,\ \ 
V_2(t)=0 \ \mbox{near}\ \ t=\e \ \ \mbox{and}\ \ t=T-\e.
\end{equation}
We assume that $u_1(x,t)$  and  $u_2(x,t)$  have  the same  initial and boundary conditions where 
$u_1$  corresponds to $V_1=0$  and $u_2$  corresponds to $V_2(x,t)$.

We shall prove that $|u_1(x,t)|\not\equiv |u_2(x,t)|$   for  
$t>T-\e$  when (\ref{eq:3.4})  holds,  i.e.  electric AB  effect  takes place.

When $0<t<\e$  we have that $u_1(x,t)=u_2(x,t)$  since $V_1=V_2=0$  
for $(x,t)\in D,\ t<\e$.  For  $\e< t<T-\e$  we have that 
$$
u_2(x,t)=\Big(\exp\frac{ie}{h}\int_\e^t V_2(t')dt\Big)u_1(x,t)
$$ 
 for   $(x,t)\in Q, \ 
\e <t\leq T-\e$.

Let $x^{(2)}$  be such that $R_1<|x^{(2)}|<R_2$  and 
$u_1(x^{(2)},T-\e)\neq 0$  and let 
$|x^{(1)}|>R_2$ be such that  $u_1(x^{(1)},T-\e)\neq 0$.

We can choose the initial condition $u_0(x)$ such that this holds.  Indeed,  let 
$u_1(x)$  be any function such that $u_1(x^{(1)})\neq 0$  and $u_1(x^{(2)})\neq 0$.
Consider  the backward  initial  boundary value problem  for $0< t<T-\e,\ 
ih\frac{\partial u_1}{\partial t}+\frac{h^2}{2m}\Delta u_1=0$  for  $(x,t)\in D, 
\ u_1(x,T-\e)=u_1(x),\ u_1\big|_{\partial \Omega}=0$.

Then we take  $u_0(x)=u_1(x,0)$  as the initial condition  for the initial boundary value
problem (\ref{eq:3.1}),  (\ref{eq:3.2}),  (\ref{eq:3.3}).

We claim (cf. [21])  that $|u_1(x,t)|^2\neq |u_2(x,t)|^2$  in a neighborhood  $U_0$  of  
$(x_1^{(0)},T-\e)$  for $t>T-\e$.  Note that  $u_1(x,T-\e)=u_2(x,T-\e)$  for
$|x|>R_2$ 
since $u_1(x,\e) =u_2(x,\e)$  for  $|x|>R_2$  and $u_1,u_2$  have zero  boundary
conditions on $\{(|x|=R_2)\times(\e,T-\e)\}$.  In particular  $u_1(x,T-\e)=u_2(x,T-\e)$  in $U_0$.
Suppose  that  $|u_1(x,t)|=|u_2(x,t)|$ in $U_0$ for  $t>T-\e$.

Use the 
polar representation in $U_0\cap\{t>T-\e\}$:
$$
u_1(x,t)=R_1(x,t)e^{i\Phi_1(x,t)},\ u_2(x,t)=R_2(x,t)e^{i\Phi_2(x,t)}.
$$
Note that $R_1=R_2=R$.

Substituting in (\ref{eq:3.1})  we get
\begin{equation}                                    \label{eq:3.5}
-h\frac{\partial R}{\partial t}=\frac{h^2}{2m}(2\nabla R\cdot \nabla \Phi_j +  R\Delta\Phi_j),
\end{equation}
\begin{equation}                                    \label{eq:3.6}
h\frac{\partial \Phi_j}{\partial t} R=\frac{h^2}{2m}(\Delta R-R|\nabla \Phi_j|^2),\ \ j=1,2.
\end{equation}
Therefore $\Phi_1$  and $\Phi_2$  satisfy
the same first order partial differential equation  
(\ref{eq:3.6})
with the same initial 
condition in  $U_0\cap\{t=T-\e\}$.
$$
\Phi_1(x,T-\e)=\Phi_2(x.T-\e)
$$
since $u_1(x,T-\e)=u_2(x,T-\e)$.  
Therefore,  by the uniqueness of the Cauchy problem  we have 
$\Phi_1(x,t)=\Phi_2(x,t)$ in $U_0\cap\{t>T-\e\}$.  Thus
$u_1(x,t)=u_2(x,t)$  in $U_0\cap\{t>T-\e\}$.
Then 
by the unique continuation  property for the Schr\"odinger 
 equation  (cf. [32], Sect. 6) we get  that 
$u_1(x,t)=u_2(x,t)$ for $(x,t)\in D,\ T-\e<t<T$.  By the continuity in $t$  we conclude  that
$u_1(x,T-\e)=u_2(x,T-\e),\ R_1<|x|<R_2$.  Since $u_1(x^{(2)},T-\e)\neq 0$  and 
 $u_2(x^{(2)},T-\e)=u_1(x^{(2)},T-\e) \exp i\alpha$  we got
a contradiction since $\exp i\alpha\neq 1$  when $\alpha$  satisfies (\ref{eq:3.4}).
This concludes the proof  of electric AB effect.
\\

\begin{tikzpicture}[scale=0.8]
\draw  (-2.6,1.5) arc (-150:150:3 and 3);
\draw  (-1.7,2.0) arc (-150:150:2 and 2);

\draw (-2.6,1.5) -- (-1.7,2.0);
\draw (-2.6,4.5) -- (-1.7,4.0);
\draw (0,3) circle (0.02);

\draw (2.1, 4.0) node {$R_1$};
\draw (3.1, 4.2)  node {$R_2$};

\end{tikzpicture}
\\ 
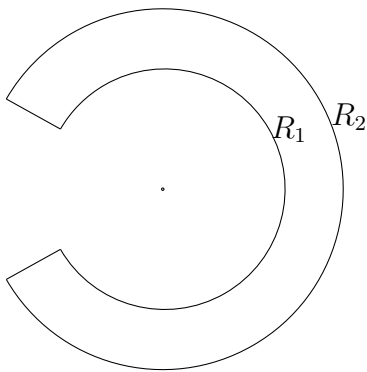
{\bf Fig. 3.} The intersection $D_{t_0}$  of  the domain $D$  with the plane $t=t_0$  
is the complement  in $\R^2$  of  $\Omega(\tau)=\{R_1\leq |x|\leq R_2,\ -\pi+\tau\leq \theta
\leq \pi-\tau\}$,  where $\tau$ depends on $t_0$.
When $\tau=0\ D_{t_0}$ has two connected components.

\section{The Schr\"odinger equation with time-dependent  magnetic and electric potentials}
\init

The case of time-dependent electromagnetic potentials is 
much harder than the case 
when  $A$  and $V$  are time-independent.  Many powerful tools such as the BC-method are not applicable.  
Therefore the results on the inverse boundary value problems are much weaker.  The study  of the 
Aharonov-Bohm effect also becomes more complicated.

\subsection{Inverse boundary value problem}
\
\\
Let $\Omega_j(t)\subset\R^n,0\leq t\leq T$,  be the obstacles,
$\overline \Omega_j(t)\cap \overline\Omega_k(t)=0,1\leq j,k\leq r$, 
and let $\Omega_0\supset\overline\Omega'=\bigcup_{k=1}^n\overline\Omega_j(t)$,
where  
$\Omega_0$ is a simply-connected  domain  in $\R^n$.   Let  
$\Omega'=\bigcup_{0\leq t\leq T}\bigcup_{g=1}^r\Omega_j(t)$. 
 Consider  in $(\Omega_0\times(0,T))\setminus\overline\Omega'$  the
Schr\"odinger  equation with time-dependent  magnetic and electric potentials
\begin{multline}                                               \label{eq:4.1}
ih\frac{\partial u(x,t)}{\partial t}-
\frac{1}{2m}\sum_{j=1}^n\Big(-ih\frac{\partial}{\partial x_j}-\frac{e}{c}A_j(x,t)\Big)^2u(x,t)
\\
-eV(x,t)u(x,t)=0,\ \ (x,t)\in (\Omega_0\times (0,T))\setminus\Omega'.
\end{multline}
We assume that 
\begin{equation}                                              \label{eq:4.2}
u\Big|_{\partial\Omega_j(t)}=0,\ \ t\in (0,T),\ \ j=1,...,r.
\end{equation}
We also assume that the normals to $\Omega'$  for
$0<t<T$ are not parallel to the $t$-axis.  This condition assures the existence 
of the solution of the initial-boundary value problem for (\ref{eq:4.1}).

The gauge group $G((\overline\Omega_0\times[0,T])\setminus\Omega')$  in this case consists 
of $g(x,t)\in C^\infty((\overline\Omega_0\times[0,T])\setminus\Omega')$
such that $|g(x,t)|=1$.  Thus the electromagnetic potentials  $(A(x,t),V(x,t))$  and
  $(A'(x,t),V'(x,t))$  are gauge  equivalent  if there exists $g(x,t)\in  
G((\overline\Omega_0\times[0,T])\setminus\Omega')$
such that
\begin{align}                                        \label{eq:4.3}
&\frac{e}{c}A'(x,t)=\frac{e}{c}A(x,t)+ihg^{-1}(x,t)\frac{\partial g(x,t)}{\partial x},
\\
\nonumber
&e V'(x,t)=e V(x,t)-ihg^{-1}(x,t)\frac{\partial g(x,t)}{\partial x}.
\end{align}
Now we shall describe  
the class of obstacles considered in this subsection.  Since the potential depends on 
the time variable  we cannot switch from the Schr\"odinger equation to the wave equation 
and use the Boundary Control 
method as in section 2.1.  We shall use instead the inversion  of the $X$-ray  transform
and this approach in the presence of obstacles imposes severe restrictions on the obstacles.

In the case 
of $n\geq 3$  variables we assume that  the following condition is satisfied
\begin{align}                                               \label{eq:4.4}
&\mbox{For each}\ \ t_0\in[0,T]\ \ \mbox{all obstacles}\ \ \Omega_j(t_0),\ 1\leq j\leq r,\ \
\\
\nonumber
&\mbox{are convex,  and for each point}\ \ x_0\in 
\Omega_0\setminus\bigcup\limits_{k=1}^r\overline\Omega_j(t_0)
\\
\nonumber
&\mbox{there exists a two dimensional plane}\ \
\Pi_{x_0}\subset\R^n,n\geq 3\ \
\\
\nonumber
&\mbox{that intersect at most one of the obstacles}\ \ \Omega_j(t_0).
\end{align}    
In the case of $n=2$  we assume that 
\begin{align}                                         \label{eq:4.5}
&\mbox{All obstacles}\ \ \Omega_j(t_0)\ \ \mbox{are convex in}\ \ \R^2
\ \
\mbox{for each}
\\
\nonumber
& t_0\in [0,T],\ 1\leq j\leq r. \ \ \mbox{If}\ \ r>1, \ \ \mbox{i.e. when}
\\
\nonumber
&\mbox{there are more then one obstacle, we assume}
\\
\nonumber 
&\mbox{that there is no trapped broken (reflected) rays in}
\\
\nonumber
& \Omega_0\setminus\bigcup_{j=1}^r\Omega_j(t_0),
\ \
\mbox{i.e.  any broken ray starting on}\ \ \partial\Omega_0 \ \ 
\\
\nonumber
&\mbox{returns to}\ \ \partial\Omega_0
\ \ \mbox{after a finite number of reflections}.
\end{align} 
When  the obstacles are smooth and $r\geq 2$  there are always 
trapped rays.  To have the situation when there are no trapped rays  we must require 
that  obstacles $\Omega_j(t_0),\ 1\leq  j\leq r,$  have a  finite  number  of corner points.

We consider  
only  the broken rays  avoiding corners points,   and we assume that the number  of reflections 
is uniformly bounded  for all broken rays.

As in \S 2.3
  we introduce
gauge  invariant  boundary  data  on \linebreak
$\partial\Omega_0\times(0,T)$:
\begin{equation}                                                \label{eq:4.6}
|u(x,t)|^2=f_1,\ \ \frac{\partial}{\partial \nu}|u(x,t)|^2=f_2,\ \
S(u)=\Im\Big(h\frac{\partial u}{\partial x}-i\frac{e}{c}A(x,t)u\Big)\overline u=f_3,
\end{equation}
where $(x,t)\in \partial\Omega_0\times (0,T)$.

The following theorem holds  (cf.  [12],  [13],  [18]).
\begin{theorem}                                  \label{theo:4.1}
Consider  two Schr\"odinger equations $\big(ih\frac{\partial}{\partial t}-H\big)u=0$
and $\big(ih\frac{\partial}{\partial t}-H'\big)u'=0$
of the form  (\ref{eq:4.1}) in $\Omega_0\times(0,T)\setminus\Omega'$  with  zero  
Dirichlet  boundary  conditions on $\partial\Omega'$  and zero  initial conditions on 
$\Omega_0\setminus\bigcup_{j=1}^r\Omega_j(0)$,  corresponding to electromagetic potentials  
$(A(x,t),V(x,t))$  and  $(A'(x,t),V'(x,t))$,  respectively.

Suppose obstacles $\Omega'$  satisfy  condition (\ref{eq:4.4})  when 
$n\geq 3$  and  the condition (\ref{eq:4.5}) when $n=2$.

If the sets of gauge invariant  boundary  data of $u$  and $u'$  are equal  on 
$\partial\Omega_0\times(0,T)$  then  the electromagnetic potentials $(A,V)$  and $(A',V')$
are gauge equivalent.
\end{theorem}
The proof of Theorem \ref{theo:4.1}  was given in [18].  Since  the case of time-dependent
potentials is relatively new we shall indicate the  main steps of the proof.

{\bf Proof:}
It was  shown  in \S2.3
that the equality of the gauge invariant  boundary data is equivalent  to the existence 
 of $g_0\in G((\overline\Omega_0\times[0,T])\setminus\Omega')$
such that  the corresponding  DN
operators $\Lambda$  and $\Lambda'$  are gauge equivalent  on 
$\partial\Omega_0\times(0,T)$,  i.e.
$\Lambda'v=g_{00}^{-1}\Lambda g_{00}v$  for  any  smooth $v$  
on $\partial\Omega_0\times(0,T)$.  Here 
$g_{00}$  is the restriction  of $g_0(x,t)$  to  $\partial\Omega_0\times(0,T)$.

Making the gauge transformation  $w=g_0^{-1}u''$  we get the Schr\"odinger  equation  
$\big(ih\frac{\partial}{\partial t} -H''\big)w=0$  with
electromagnetic potentials $(A''.V'')$  that are  gauge equivalent  to $(A',V')$.  Now we 
have that  $\Lambda=\Lambda''$  on $\partial\Omega_0\times(0,T)$  where  $\Lambda''$  is the
DN  operator corresponding  $\big(ih\frac{\partial}{\partial t}-H''\big)w=0$.

Consider first the more simple case of  $n\geq 3$ assuming that  (\ref{eq:4.4}) 
holds.
 
Let $\gamma(t_0)$  be  a ray in the domain $\Omega_0\setminus\bigcup_{j=1}^r\Omega_j(t_0)$
starting  and ending  on  $\partial\Omega_0$.    We shall construct  a solution  
$u(x,t,k)$  of  $\big(ih\frac{\partial}{\partial t}-H\big)u=0$  in 
$(\Omega_0\times(0,T)\setminus\Omega')$
depending on a large parameter $k$  and satisfying  the boundary  conditions
\begin{equation}                                        \label{eq:4.7}
u\Big|_{\partial\Omega'}=0,\ \ \ 1\leq j\leq r,
\end{equation}
initial conditions
\begin{equation}                                        \label{eq:4.8}
u\Big|_{\Omega_0\setminus\Omega'\cap\{t=0\}}=0,
\end{equation}
and concentrated  in a small  neighborhood $U_0\subset(\Omega_0\times(0,T))\setminus\Omega'$
of the ray $\gamma(t_0)$.

We are looking for $u(x,t,k)$  in the form
\begin{equation}                                   \label{eq:4.9}
u(x,t,k)=e^{-i\frac{mk^2}{2h}t+i\frac{mk}{h}x\cdot\omega}
\Big(\sum_{p=0}^N\frac{a_{p0}(x,t,\omega)}{(ik)^p}+O\Big(\frac{1}{k^{N+1}}\Big)\Big),
\end{equation}
where
\begin{align}                                    \label{eq:4.10}
&|\omega|=1,
\\
\nonumber
&\omega\cdot\Big(-ih\frac{\partial}{\partial x}-\frac{e}{c}A(x)\Big)a_{00}=0,
\\
\nonumber
&\omega\cdot\Big(-\frac{ih\partial}{\partial x}-\frac{e}{c}A(x)\Big)a_{p0}
=\Big(ih\frac{\partial}{\partial t}-H\Big)a_{p-1,0},\ \ p\geq 1.
\end{align}
We choose
\begin{multline}                                           \label{eq:4.11}
a_{00}
=\frac{1}{\e^{\frac{n}{2}}}\chi_0\Big(\frac{t-t_0}{ \e}\Big)
\prod_{j=1}^{n-1}\chi_0\Big(\frac{\tau_j-\tau_{0j}}{\e}\Big)
\\
\cdot\exp\Big(i\frac{e}{h c}
\int\limits_{s_0}^s A\Big(\sum_{j=1}^{n-1}\tau_j\omega_{\perp j}+s'\omega,t\Big)\cdot\omega ds',
\end{multline}
where $s=x\cdot \omega, \ \tau_j=x\cdot \omega_{\perp j},\ \chi_0$  is
 the same  as in  (\ref{eq:2.39}),    $\int_{-\infty}^\infty \chi_0^2(s)ds=1,\ s_0,r_{0j}$  are such  that  
$x^{(0)}=s_0\omega+\sum_{j=1}^{n-1}\tau_{0j}\omega_{\perp j}\not\in \Omega_0$,   where  $\omega\cdot\omega_{\perp j}=0,
1\leq j\leq n-1,\ \{\omega,\omega_{\perp 1}...,\omega_{\perp,n-1}\}$  is an orthogonal basis in $\R^n$. 
Also  the plane  $(x-x^{(0)})\cdot\omega=0$ does not intersect $\Omega_0$  (cf. [13], [18]).
We assume also  that
\begin{equation}                                        \label{eq:4.12}
a_{p0}(s,\tau,t)=0 \ \ \mbox{when}\ \ s=s_0,\ p\geq 1.
\end{equation}
Note  that the principal  term  (\ref{eq:4.10})  of (\ref{eq:4.9})  is the  same
as in the case of potentials  independent  of $t$.  However,  the lower other terms $a_{p0},\ p\geq 1$,
will pick up the derivatives of $A,V$  in $t$.

Solution  of the form  (\ref{eq:4.9})  is different from
the  geometric optics type solutions  in \S 2.6.  The latter solutions  describe
the propagation in the time  and (\ref{eq:4.9})  
propagates  in the plane  $t=t_0$  along  the  space direction $s=x\cdot\omega$.

We shall show below   that  solutions  (\ref{eq:4.9})  can be approximated
by physically  relevant  solutions.

Having  solutions  of 
the form  (\ref{eq:4.9})  we can conclude the proof of Theorem \ref{theo:4.1}
in two steps. First,  substituting the solutions  of  
$\big(ih\frac{\partial}{\partial t}-H\big)u=0$
and  $\big(ih\frac{\partial}{\partial t}-H''\big)w=0$  having  both the form (\ref{eq:4.9}),
in the Green's formula,  using that    
$\Lambda=\Lambda''$  on  $\partial\Omega_0\times(0,T)$ and  passing  to the limit when
$\e\rw 0$ we get
\begin{equation}                                              \label{eq:4.13}
\exp\Big(\frac{i e}{h c}\int\limits_{\gamma(t_0)}A(x,t_0)\cdot  dx\Big)=
\exp\Big(\frac{i e}{h c}\int\limits_{\gamma(t_0)}A''(x,t_0)\cdot  dx\Big)
\end{equation}
for all  rays $\gamma(t_0),\ t_0\in(0,T)$  is arbitrary,  but fixed.  Now,  using  
the Helgason's hole theorem
(cf.  [Hel]),
  we prove the uniqueness  of the $X$-ray transform to get 
that there exists $g(x,t)=e^{i\varphi(x,t)}
\in G((\overline\Omega_0\times[0,T])\setminus\Omega'),
\ g=1$  on $\partial\Omega_0\times(0,T)$,  such that
$$
\frac{e}{c}A''(x,t_0)=
\frac{e}{c}A(x,t_0)+ihg^{-1}\frac{\partial g(x,t_0)}{\partial x}.
$$
Here $t_0$  is a parameter.  Making  the gauge transformation  in 
$\big(ih\frac{\partial}{\partial t}-H''\big)w=0$  with  the gauge $g(x,t)$  we get
the Schr\"odinger equation 
$\big(ih\frac{\partial}{\partial t}-H'''\big)w_1=0$
with magnetic potential $A(x,t)$  and the electric  potential  
$eV'''\equiv eV''-ihg^{-1}\frac{\partial g}{\partial t}$.
Now apply again the Green's formula to  $\big(ih\frac{\partial}{\partial t}-H\big)u=0$
and $\big(ih\frac{\partial}{\partial t}-H'''\big)w=0$
using the  solution  of the form  (\ref{eq:4.9})  and  that  $\Lambda=\Lambda'''$.  
Since $H'$  and $H'''$    
 have the same magnetic potentials,
 their contribution  will cancel each other and we get that
\begin{equation}                                           \label{eq:4.14}
\int\limits_{\gamma(t_0)}\Big(eV(x,t_0)-eV''+ihg^{-1}\frac{\partial g}{\partial t}\Big)ds=0  
\end{equation}
for all  rays $\gamma(t_0)$.

Therefore,  the uniqueness theorem  of the $X$-ray transform  gives that 
$$
eV(x,t)-eV'''(x,t)+i h g^{-1}(x,t)\frac{\partial g}{\partial t}=0.
$$
Thus $(A,V)$  and $(A''',V''')$  are gauge equivalent.  Hence  
$(A,V)$  and $(A',V')$  are gauge equivalent  too.

Now consider a more difficult 
case $n=2$  and  (\ref{eq:4.5})  is satisfied.
As in \S 2.6 we will use  the broken  rays.

Let $\gamma(t_0)=\gamma_0(t_0)\cup\gamma_1(t_0)\cup...\cup\gamma_d(t_0)$
be a broken ray starting at some point $x^{(0)}\in \partial\Omega_0$
reflecting  at some obstacles $\Omega_j(t_0),1\leq j\leq r,$  and ending  on $\partial\Omega_0$.
We shall construct a solution   of $\big(i\frac{\partial}{\partial t}-H\big)u=0$
concentrated in a small neighborhood  of  $\gamma(t_0)$.
We are looking  for  $u(x,t,k)$  in the form 
\begin{equation}                                  \label{eq:4.15}
u(x,t,k)=\sum_{j=0}^d\sum_{p=0}^N\frac{a_{pj}(x,t.\omega)}{(ik)^p}
e^{-\frac{imk^2}{2h}t+i\frac{mk}{h}\psi_j(x,t,\omega) }   
\end{equation}
where $\psi_0(x,t_0,\omega)=x\cdot\omega,\ a_{p0}(x,t,\omega)$  are  the same  as in 
(\ref{eq:4.9}),  $\omega=\theta_1$ is the direction of  
$\gamma_0$,  
$\psi_j(x,t,\omega)$ satisfy the equations
\begin{align}                                 \label{eq:4.16}
&\Big|\frac{\partial\psi_j}{\partial x}\Big|=1,\ \ \
\psi_j(x,t,\omega)\Big|_{\partial\Omega'}=\psi_{j+1}(x,t,\omega)\Big|_{\partial\Omega'},
\\
\nonumber
&\frac{\psi_{j+1}(x_0^{j+1},t_0,\omega)}{\partial x}=\theta_{j+1},\ \ 0\leq j\leq d-1,
\end{align}
where 
$x_0^{(j+1)}$  is the point  of reflection of  $\gamma_j$  at  
$\partial\Omega'\cap\{t=t_0\}$  and  $\theta_{j+1}$  is the direction 
of $\gamma_{j+1}$.

Functions  $a_{pj}$  satisfy  the following  equations:
\begin{multline}                                \label{eq:4.17}
\frac{\partial a_{pj}}{\partial x}\cdot\frac{\partial \psi_j}{\partial x}
+\frac{1}{2}\Delta\psi_ja_{pj}-i\frac{e}{c}A(x,t)\cdot\frac{\partial\psi_j}{\partial x}a_{pj}
\\
=f_{pj}(x,t,\omega)+i\frac{m}{h}\frac{\partial\psi_j}{\partial t}a_{pj},\ \ p\geq 0,
\end{multline}
where $f_{0j}=0, f_{pj}$  depends on $a_{0j},...,a_{p-1,j}$.

When obstacles are independent  of $t$ then  $\psi_j(x,t)$ are also  independent
of $t$.  We impose  also  the following  
conditions on $a_{pj}$
\begin{equation}                               \label{eq:4.18}
a_{pj}\big|_{\partial\Omega'}=-a_{p,j+1}\big|_{\partial\Omega'}
\end{equation}
This last condition 
implies  that $u\big|_{\partial\Omega'}=0$.
Inserting  (\ref{eq:4.15}) into the Green's formula   instead of (\ref{eq:4.9})
we get,  analogously to (\ref{eq:4.13}),  (\ref{eq:4.14}),  that
\begin{equation}                                 \label{eq:4.19}
\exp\Big[\frac{ie}{hc}\sum_{j=0}^d\int\limits_{\gamma_j(t_0)}
(A(x_0^{(j)}+s\theta_j,t_0)-
A''(x_0^{(j)}+s\theta_j,t_0))\cdot\theta_jds\Big]=1
\end{equation}
and
\begin{equation}                                 \label{eq:4.20}
\sum_{j=0}^d\int\limits_{\gamma_j(t_0)}
\Big(eV(x_0^{(j)}+s\theta_j,t_0)-eV'''(x_0^{(j)}+s\theta_j,t_0)
+ihg^{-1}\frac{\partial g}{\partial t}\Big)ds=0.
\end{equation}
Proving the uniqueness of $X$-ray transform   problem  for broken  rays 
is much harder.  It was shown in [13],  [15]  that  (\ref{eq:4.19}),  (\ref{eq:4.20})
imply  that the electromagnetic potentials $(A,V)$ and $(A',V')$  are gauge 
equivalent.

This concludes the proof of Theorem \ref{theo:4.1}.

\subsection{Inverse  boundary  value problems for the Schr\"odinger operator with 
time-dependent  Yang-Mills potentials}
\
\\
Consider  the Schr\"odinger  equation of  the form
\begin{equation}                                           \label{eq:4.21}
-i\frac{\partial u(x,t)}{\partial t}
+\sum_{j=1}^n\Big(I_m\Big(-i\frac{\partial}{\partial x_j}\Big)-
A_j(x,t)\Big)u(x,t)
+
V(x,t)u(x,t)=0,
\end{equation}
where $A=(A_1,...,A_n),\ A_j(x,t),V(x,t), 1\leq j\leq n$,  are
$m\times m$ self-adjoint matrices.  
It is convenient to consider  $u(x,t)$  also as $m\times m$  matrix.
$I_m$  is $m\times m$  identity  matrix.

We consider  (\ref{eq:4.21}) in $\Omega_0\times(0,T)$  with initial conditions
$$
u(x,0)=0,\ \ x\in \Omega_0,
$$ 
and the boundary conditions 
$$
u\Big|_{\partial\Omega_0\times(0,T)}=f.
$$
Let  $\Lambda f= \big(I_m\frac{\partial}{\partial \nu}
-iA\cdot \nu\big)u\big|_{\partial\Omega_0\times(0,T)}$ be the DN  operator.
The gauge group consists  of $m\times m$ unitary matrices,  smooth in 
$\overline\Omega_0\times[0,T]$.  We assume  that there is no obstacles in this 
subsection.

Yang-Mills potentials  $(A,V)$ and $(A',V')$  are gauge equivalent  if
there is $g\in G(\overline\Omega_0\times[0,T])$  such
that
\begin{align}                                   \label{eq:4.22}
&A_j'=g^{-1}A_jg+ig^{-1}\frac{\partial g}{\partial x_j},\ \ 1\leq j\leq n
\\
\nonumber
&V_j'=g^{-1}V_jg-ig\frac{\partial g}{\partial t}
\end{align}
\begin{theorem}                               \label{theo:4.2}
Consider two equations $\big(-i\frac{\partial}{\partial t}+H\big)u=0,\ 
\big(-i\frac{\partial}{\partial t}+H'\big)u'=0$ of the form
(\ref{eq:4.21})
with  Yang-Mills potentials $(A,V),(A',V')$,  respectively.
Suppose that DN  operators $\Lambda$  and $\Lambda'$,
corresponding to   
$\big(-i\frac{\partial}{\partial t}+H\big)u=0$  and 
$\big(-i\frac{\partial}{\partial t}+H'\big)u'=0$ 
are gauge equivalent  on  $\partial\Omega_0\times(0,T)$  with some gauge  $g_0(x)$, 
 i.e. $g_{00}^{-1}\Lambda g_{00}=\Lambda'$,  
where $g_{00}=g_0\big|_{\partial\Omega_0\times(0,T)}$.  
Then  $(A,V)$  and  $(A',V')$ are gauge equivalent too.
\end{theorem}
The beginning of the proof  of Theorem \ref{theo:4.2}  is similar to the
proof of Theorem \ref{theo:4.1} in the case $n\geq 3$.

We construct  a solution of (\ref{eq:4.21}) similar to (\ref{eq:4.9})
\begin{equation}                                   \label{eq:4.23}
u_\e(x,t,k)=e^{-i k^2t+i k x\cdot\omega}
\Big(\sum_{p=0}^N\frac{a_{p}(x,t,\omega)}{(ik)^p}+O\Big(\frac{1}{k^{N+1}}\Big)\Big),
\end{equation}
where
$$
a_0(x,t,\omega)=\frac{1}{\e^{\frac{n}{2}}}\chi_0\Big(\frac{t-t_0}{\e}\Big)
\Pi_{j=1}^{n-1}\chi_0\Big(\frac{\tau_j-\tau_{j0}}{\e}\Big)c(x,t,\omega),
$$
$t_0,\tau_{j0},\tau_j$  are  the same  as in (\ref{eq:4.11}),
$c(x,t,\omega)$  satisfies  the equation
\begin{equation}                                       \label{eq:4.24}
\omega\cdot\frac{\partial c}{\partial x}-iA(x,t)\cdot\omega c=0
\ \ \mbox{for}\ \ s>s_0,\ \ c=I_m\ \ \mbox{when}\ \ s=s_0,
\end{equation}
$a_{p}$  satisfy equations similar to (\ref{eq:4.10}) and $a_{p}(s,\tau,t)=0$ 
when  $s=s_0,\ p\geq 1$.

Applying  gauge  $g_0$  to $\big(-i\frac{\partial}{\partial t}+H'\big)u'=0$
we get an equation \linebreak 
$\big(-i\frac{\partial}{\partial t}+H''\big)u''=0$, gauge equivalent to
$\big(-i\frac{\partial}{\partial t}+H'\big)u'=0$
and such that $\Lambda''=\Lambda$.

Using the Green's formula and passing  the limit 
as $\e\rw 0$ we get,  similarly to 
Theorem \ref{theo:4.1} that
\begin{equation}                                    \label{eq:4.25}
c_0(+\infty,y',t_0,\omega)=c_0''(+\infty,y',t_0,\omega),
\end{equation}
where $y_1=x\cdot\omega,\ y'=x-(x\cdot\omega)\omega,\ c_0(y_1,y',t_0,\omega)$
and $c_0''(y_1,y',t_0,\omega)$  are matrices $c(x,t_0,\omega)$  and  
$c''(x,t_0,\omega)$  (cf.  (\ref{eq:4.24})) in $(y_1,y')$  coordinates,  $c$  
corresponds  to $\big(-i\frac{\partial}{\partial t}+H\big)u=0, \ \ c''$  corresponds
to
$\big(-i\frac{\partial}{\partial t}+H''\big)u''=0$.

Note that  (\ref{eq:4.25})  is the analog  of  
(\ref{eq:4.13})  when $m>1$.

The matrix $c_0(+\infty,y',t_0,\omega)$  
is called  the non-Abelian  Radon transform  of $A(x)$. 
The problem of the recovery of $A(x)$  
from the non-Abelian Radon transform is much more difficult
then in the case 
of electromagnetic potentials, i.e.  when  $m=1$.   This was done in [11], [14],  [40].
The recovery  of   $V(x,t)$    was also  done in [11],  [14]].  Note 
that  the most  difficult
case is $n=2$.  The extension to $n\geq 3$  dimensions is relatively easy (cf.  [18]).

\subsection{An inverse  problem  for the time-dependent  Schr\"odinger  equation  in an 
unbounded domain}
\
\\
When  the Schr\"odinger operator  with time-independent
coefficients  is studied    in $\R^n$  
outside the obstacles,   
it is natural  to consider  the scattering problem.  When the coefficients are
 time-dependent 
we propose a new  problem.

Consider  the Schr\"odinger equation of the form
\begin{equation}                                    \label{eq:4.26}
-ih\frac{\partial u}{\partial t}+Hu=0\ \ \ \mbox{in}\ \ 
(\R^n\times(0,T))\setminus\Omega',
\end{equation}
where   
$$
Hu=\frac{1}{2m}\sum_{j=1}^n\Big(-ih\frac{\partial}{\partial x_j}
-\frac{e}{c}A(x,t)\Big)^2u(x,t)+eV(x,t)u(x,t),
$$
$\Omega'=\bigcup\limits_{0\leq t\leq T}\bigcup\limits_{j=1}^r\Omega_j(t)$
are obstacles.
We assume that  the electromagnetic potentials  are independent  of $t$  for
$|x|\geq R$  where $R$  is such that $\Omega'\subset B_R= \{|x|<R\}$.
We assume also
that $V(x)=O\big(\frac{1}{|x|^{1+\e}}\big),\ A(x)=O\big(\frac{1}{|x|}\big).$
Here $V(x)=V(x,t),\ A(x)=A(x,t)$  for $|x|>R,  t\in[0,T]$.

Assume  that  $u(x,t)$  satisfies the initial  condition
on $\R^n\setminus\bigcup_{j=1}^r\Omega_j(0)$:
\begin{equation}                                         \label{eq:4.27}
u(x,0)=u_0(x),\ \ u_0(x)=0\ \ \mbox{on}\ \  B_R\setminus(\Omega'\cap\{t=0\}).
\end{equation}
The gauge group $G((\R^n\times[0,T])\setminus\Omega')$  is different  in the cases
  $n=2$  and  $n\geq 3$.

We assume  that  $|g(x,t)|=1$ in $(\R^n\times[0,T])\setminus\Omega'$
and  $g(x,t)=e^{i\frac{\varphi(x)}{h}},\ \varphi(x)=O\big(\frac{1}{|x|}\big)$  for
$n\geq 3$.
When $n=2$ we assume
 $g(x,t)=e^{ip\theta(x)}\big(1+O\big(\frac{1}{|x|}\big)\big)$
where $p\in \Z$  and  $\theta(x)$
is the polar angle.  We also assume that the origin  belongs to 
$\Omega'\cap\{t=0\}$.

Suppose  we are given initial conditions  for the  equation (\ref{eq:4.26})  for  $t=0$
and the  condition
\begin{equation}                                      \label{eq:4.28}
u(x,T)=u_1(x),\ \ |x|> R
\end{equation}
for  $t=T$.

We shall call (\ref{eq:4.27}),  (\ref{eq:4.28}) the two times data, $t=0$ and $t=T$, 
  for  the equation (\ref{eq:4.26}).
We shall prove  that these data  determines electromagnetic potentials for 
$|x|<R$  up to a gauge equivalence.  More precisely,   the following theorem holds:
\begin{theorem}                                   \label{theo:4.3}
Consider  two  equations $\big(-ih\frac{\partial}{\partial t}+H\big)u=0$
and
\linebreak
 $\big(-ih\frac{\partial}{\partial t}+H'\big)u'=0$ of the form  (\ref{eq:4.26})
in $(\R^n\times(0,T))\setminus\Omega'$ with  electromagnetic potentials 
$(A(x,t),V(x,t))$ and  $(A'(x,t),V'(x,t))$,  respectively.  Assume  that 
$(A,V)$ and  $(A',V')$ are 
independent  of $t$  for  $|x|>R$.

Suppose  $(A(x),V(x))$  and  $(A'(x),V'(x))$  are gauge  equivalent  for  
$|x|>R$,  i.e.  there exists $g_0(x),\ |g_0(x)|=1, $  such that  for  
$|x|>R$  we have
\begin{align}                                           \label{eq:4.29}
 &\frac{e}{c}A'(x)=\frac{e}{c}A(x)-ih g_0^{-1}(x)\frac{\partial g_0}{\partial x},
\\
\nonumber
&V'(x)=V(x).
\end{align}
Suppose  that  $u(x,t)$  and  $u'(x,t)$  have  gauge  equivalent  two times data 
\begin{align}                                           \label{eq:4.30}
&u(x,0)=g_0(x)u'(x,0),\ \ |x|>R,
\\
\nonumber
&u(x,T)=g_0(x)u'(x,T)\ \ \mbox{for}\ \ |x|>R.
\end{align}
Then  the DN  operators $\Lambda$  and  $\Lambda'$  are  gauge equivalent
on $\partial B_R\times (0,T)$, i.e.
$$
\Lambda' f=g_{00}^{-1}\Lambda g_{00}f
$$
for  all smooth $f$  on $\partial B_R\times(0,T)$  
and  $g_{00}(x)$  is the restriction  of  $g_0(x)$ 
to  $\partial B_R\times (0,T)$. 
\end{theorem}
Note  that combining Theorem \ref{theo:4.3}  with Theorem \ref{theo:4.1}
we get that $(A,V)$ and $(A',V')$  are gauge equivalent.

To prove Theorem \ref{theo:4.3}  we need two lemmas.
\begin{lemma}                                            \label{lma:4.4}
Let
\begin{equation}                                     \label{eq:4.31}
-ih\frac{\partial w}{\partial t} +
\frac{1}{2m}\sum_{j=1}^n\Big(-i\frac{\partial}{\partial x_j}-\frac{e}{c}A_j(x)\Big)^2w(x,t)
+eV(x)w(x,t)=0
\end{equation}
in $(\R^n\setminus B_R)\times(0,T)$,   where 
$A_j(x),V(x), 1\leq j\leq n$,  are independent  of $t$,
\begin{multline}                                 \label{eq:4.32}
\Big|\frac{\partial ^kA_j(x)}{\partial x^k}\Big|\leq
C_k(1+|x|)^{-1-|k|},\\ 
\Big|\frac{\partial^k V(x)}{\partial x^k}\Big|\leq C_k(1+|x|)^{-1-\e-|k|},
\ \ \e>0,\ \forall k.
\end{multline}
Suppose  $w(x,t)\in  C([0,T],L_2(\R^n\setminus B_R))$,
i.e. $w(x,t)$ is continuous in  $t$  on $[0,T]$  with values 
 in $L_2(\R^2\setminus B_R)$.

Suppose $w(x,0)=0,  \ w(x,T)=0$  for  $x\in \R^n\setminus B_R$.
Then $w(x,t)=0$  in $(\R^n\setminus B_R)\times(0,T)$.
\end{lemma}

{\bf Proof:}
Extend  $w(x,t)$  by zero  for  $t<0$  and for  $t>T$.
Let $\tilde w(x,\xi_0)$  be the Fourier  transform  of  $w(x,t)$  in $t$.
  Then  $\tilde w(x,\xi_0)\in L_2(\R^n\setminus B_R)$  for all $\xi_0$
 and
$$
h\xi_0\tilde w(x,\xi_0)+H\tilde w(x,\xi_0)=0\ \ \ \mbox{in}\ \ 
\R^n\setminus B_R.
$$
It follows  from  the H\"ormander [31]  that $\tilde w(x,\xi_0)=0$ 
 in  $\R^n\setminus B_R$  
if conditions (\ref{eq:4.32})  hold.

Therefore,  $w(x,t)=0$ in $(\R^n\setminus B_R)\times (0,T)$.
\\

{\bf Remark 4.1.}
In this paper  we mostly 
consider  the case when
$B=\mbox{curl}\,A=0$  for  $|x|>R$  and $V(x)=0$  for  $|x|>R$.
In such case
there is  a simpler  way  to prove Lemma \ref{lma:4.4}  without using [31].

If  $n\geq 3$  and $\mbox{curl}\,A=0,\ V=0$  for  $|x|>R$,  we can  make  a gauge 
transformation $g(x)$  such that 
$w'=g^{-1}w(x,t)$  satisfies  the equation
$$
\xi_0\tilde w'(x,\xi_0)-\frac{h^2}{2m}\Delta\tilde w'(x,\xi_0)=0\ \ \ 
\mbox{for}\ \ |x|>R,
$$
where  $\tilde w'(x,\xi_0)$  is the Fourier  transform  in $t$.
Since  $\tilde w'(x,\xi_0)\in L_2(\R^n\setminus B_R)$    we have  that
  $\tilde w'(x,\xi_0)=0$ by the classical  Rellich's lemma (see, for example,  [20]).

When $n=2$  and the magnetic  flux $\frac{eh}{c}\int_{|x|=R}A(x)\cdot dx=
\alpha\neq 0$  we can  make the gauge  transformation  $w'=g^{-1}(x)  w(x,t)$
such that  $A'(x)=\frac{\alpha}{2\pi}\frac{(x_2,-x_1)}{x_1^2+x_2^2}$
is the Aharonov-Bohm  potential  (cf. [1]).
Then 
making the Fourier transform in $t$  we shall  have in polar  coordinates
\begin{equation}                                         \label{eq:4.33}
h\tilde w'(r,\theta,\xi_0)-\frac{h^2}{2m}\Big[
\frac{\partial^2 \tilde w'}{\partial r^2}
+\frac{1}{r}\frac{\partial\tilde w'}{\partial  r}+
\frac{1}{r^2}\Big(\frac{\partial}{\partial\theta}
+i\alpha\Big)^2\Big]\tilde w'(r,\theta,\xi_0)=0, 
\end{equation}
where 
$\theta\in[0,2\pi],r>R$  and
$$
\int\limits_{|x|\geq R}|\tilde w'(r,\theta,\xi_0)|^2rdrd\theta <\infty\ \ 
\mbox{for any}\ \ \xi_0\in \R.
$$
The general solution  of (\ref{eq:4.33})  has the form  (cf. [1])
$$
\tilde w'(r,\theta,\xi_0)=\sum_{n=-\infty}^\infty w_n(r,\xi_0)e^{in\theta},
$$
where
$$
w_n(r,\xi_0)=a_n(\xi_0)J_{n+\alpha}(kr)+b_n(\xi_0)J_{-n-\alpha}(kr),\ \ 
k=\sqrt{\frac{2m}{h}(-\xi_0)}.
$$ 
We have 
\begin{equation}                                      \label{eq:4.34}
\int\limits_{|x|>R} 
|\tilde w'(x,\xi_0)|^2dx=\sum_{n=-\infty}^\infty\int\limits_{r>k}|w_n(r,\xi_0)|^2rdr.
\end{equation}
Using the asymptotics of the Bessel's functions  we get from  (\ref{eq:4.34}) that
$\int_{r>R} |\tilde w(r,\xi_0^2)|^2 rdr<+\infty$  iff  $a_n(\xi_0)=b_n(\xi_0)=0,\ \forall n$.  Therefore
$\tilde w(x,\xi_0)=0$  for  $|x|>R$.
\\
\

{\bf  Remark 4.2.}  If  the equation (\ref{eq:4.31})  holds in  
$(\R^n\setminus\Omega_0)\times(0,T)$,  where  $\Omega_0\subset B_R$,  and  if 
$w(x,t)=0$  in $(\R^n\setminus B_R)\times(0,T)$,
then  $w(x,t)=0$  in $(\R^n\setminus\Omega_0)\times (0,T)$  by the unique
continuation  principle  (cf.  [32]).
\qed

Assume  $u_0(x)\in  H_2(\R^n\setminus\Omega'(0)),\ u_0(x)=0$ in
$\Omega_0\setminus\Omega'(0),$
where
$\Omega'(0)=\Omega'\cap\{t=0\}$.  There exists  a unique solution 
$u(x,t)$  of  (\ref{eq:4.26}) 
with the initial data $u(x,0)=u_0(x)$  belonging  to the 
space \linebreak  
$C((0,T),H_2(\R^n\setminus\Omega(t))\cap\overset{\circ}{H}_1(\R^n\setminus\Omega(t))$ 
(cf.,  for example,  [18]),  where  $\Omega(t_0)=\Omega'\cap\{t=t_0\}$ and  
$C((0,T),H_2(\R^n\setminus\Omega(t))\cap\overset{\circ}{H}_1(\R^n\setminus\Omega(t))$
is the space of continuous functions  on $[0,T]$  
with values  in 
$H_2(\R^n\setminus\Omega(t))\cap\overset{\circ}{H}_1(\R^n\setminus\Omega(t))$,
$\ \overset{\circ}{H}_1(\R^n\setminus\Omega(t))$
consists of functions 
in $H_1(\R^n\setminus\Omega(t))$  equal  to zero on $\partial\Omega(t)$.

Initial-boundary value problem  (\ref{eq:4.26}),  (\ref{eq:4.27}), 
$u\big|_{\partial\Omega'}=0,$  describes an electron confined  
to the region  $\R^n\setminus\Omega(t),\ 0\leq t\leq T$.

We shall denote,  for the brevity,
$C((0,T),H_2(\R^n\setminus\Omega(t))\cap\overset{\circ}{H}_1(\R^n\setminus\Omega(t))$,
by  $W((\R^n\times(0,T))\setminus\Omega')$
and we shall call solutions  in  $W((\R^n\times(0,T))\setminus\Omega')$  the 
physically meaningful solutions.

Let $w(x,t)$  be the solution  of  (\ref{eq:4.26})  in 
$(\Omega_0\times (0,T))\setminus\Omega'$   belonging  to 
$C((0,T),H_2(\Omega_0\setminus\Omega(t))\cap\overset{\circ}{H}_1(\Omega_0\setminus\Omega(t))$
where $w(x,0)=0$ in $\Omega_0\setminus\Omega(0)$.

For  the  brevity,  we denote such solutions by $W((\Omega_0\times(0,T))\setminus\Omega')$.
It is  not clear what is  the physical  meaning  of  the solution of (\ref{eq:4.26})
defined  in $(\Omega_0\times(0,T))\setminus\Omega')$  only and  having  nonzero boundary values on
$\partial\Omega_0\times(0,T)$
unless they are the restrictions  to $(\Omega_0\times(0,T))\setminus\Omega'$
of the physically meaningful solution  from $W((\R^n\times(0,T))\setminus\Omega')$.
We shall denote the space of restrictions of $u\in W((\R^n\times(0,T))\setminus\Omega')$
to 
$((\Omega_0\times(0,T))\setminus\Omega')$  by $W_0$.

Fortunately,  $W_0$  is dense in $W((\Omega_0\times(0,T))\setminus\Omega')$.

\begin{lemma}[Density lemma]                                         \label{lma:4.5}
Let  $w(x,t)\in W((\Omega_0\times(0,T))\setminus\Omega'),\ w(x,0)=0$ in 
$\Omega_0\setminus\Omega'(0)$.  For any $\e$ 
there exists $u(x,t)\in W((\R^n\times(0,T))\setminus\Omega'),\ u(x,0)=0$
in $\Omega_0\setminus\Omega'(0)$  such  that the restriction  of $u(x,t)$  to 
$(\Omega_0\times(0,T))$ satisfies
$$
\sup_{0\leq t\leq T}[w(x,t)-u(x,t)]_0<\e,
$$
where $[v(x,t)]_0^2=\int\limits_{\Omega_0\setminus\Omega(t)}|v(x,t)|^2dx$.
\end{lemma}

{\bf Proof:}
Denote by $V$  the Banach  space of functions $u(x,t)$  in 
$(\Omega_0\times(0,T))\setminus \Omega'$ with  the norm
$\|u\|_V=\int_o^T[u]_0dt$.  Let  $V^*$  be the dual  space  with the norm  
$\|v\|_{V^*}=\sup_{0\leq t\leq T}[v]_0$.  Denote  by $\overline W_0\subset V^*$  
the closure in $V^*$  norm  of solutions  from $W_0$,  i.e.  the restrictions to  
$(\Omega_0\times(0,T))\setminus\Omega'$  of functions from 
$W((\R^n\times(0,T))\setminus\Omega').$

Let $\overline W_0^\perp$  be  the set of 
$v\in V$  such that $(u,v)_0=0$  for all  $u\in \overline W_0$  where $(u,v)_0$  is 
the inner product  in $L_2((\Omega_0\times (0,T))\setminus\Omega')$.
Let
$f$  be any element of $\overline W_0^\perp$ and  $f_0$  be the extension 
of  $f$  by zero in $(\R^n\setminus\Omega_0)\times(0,T)$.
Denote by  $w(x,t)$   the solution  of
\begin{align}
\nonumber
&-ih\frac{\partial w}{\partial t}+Hw=f_0\ \ \mbox{in}\ \ (\R^n\times(0,T))\setminus\Omega',
\\
\nonumber
&w(x,T)=0\ \ \mbox{in}\ \ \R^n\setminus\Omega'(T),
\\
\nonumber
&w\big|_{\partial \Omega'}=0.
\end{align}
Note that  $w(x,t)\in C((0,T),\overset{\circ}{H}_1(\R^n\setminus\Omega(t)))$
since $f_0\in L_1((0,T), \ L_2(\R^n\setminus\Omega(t)))$.
Let $(u,w)$  be  $L_2$-inner product  in $(\R^n\times(0,T))\setminus \Omega'$.
By the Green's formula  in $(\R^n\times(0,T))\setminus\Omega'$ we have 
$$
0=(u,f_0)=\big(u,\big(-ih\frac{\partial}{\partial t}+H\big)w\big)
=ih\int\limits_{\R^n\setminus\Omega_0}u(x,0)\overline w(x,0)dx,
$$
for any $u\in W((\R^n\times(0,T))\setminus\Omega'),$  
since $-ih\frac{\partial u}{\partial t}+Hu=0,\ 
u\big|_{\partial\Omega'}=0,\ u(x,0)=0$ for $\Omega_0\setminus\Omega(0)$.
Since  $u(x,0)\in H_2(\R^n\setminus\Omega')$  is arbitrary on $\R^n\setminus\Omega_0$
we get  that
$$
w(x,0)=0 \ \ \mbox{on}\ \ \R^n\setminus\Omega_0.
$$
Since $w(x,t)$  satisfies $-ih\frac{\partial w}{\partial t}+Hw=0$
in $(\R^n\setminus\Omega_0)\times(0,T)$  and $w(x,0)=w(x,T)=0$  for  
$x\in \R^n\setminus\Omega_0$, we get,  by Lemma \ref{lma:4.4},
that $w(x,t)=0$ in  $(\R^n\setminus\Omega_0)\times(0,T)$.
Therefore the restrictions of $w(x,t))$  and of $\frac{\partial}{\partial \nu}w(x,t)$
to  $\partial \Omega_0\times(0,T)$  are equal to zero in the distribution
 sense (see [20], \S 24).  Let $v$ be any function  from 
$W((\Omega_0\times(0,T))\setminus\Omega')$.  
Note that  $\mbox{supp}\,w(x,t)\subset 
(\overline\Omega_0\times[0,T])\setminus\Omega'$.

Hence applying  the Green's formula over 
$(\Omega_0\times(0,T))\setminus\Omega'$  we get
$$
(v,f)_0=\big(v,\big(-ih\frac{\partial}{\partial t}+H\big)w\big)_0=
\big(\big(-ih\frac{\partial u}{\partial t}+Hv\big),w\big)_0=0
$$
for any  $f\in \overline W_0^\perp$.  Here $(\,\ )_0$ is the $L_2$-inner  product  in 
$(\Omega_0\times (0,T))\setminus \Omega'$  and  we used  that 
$-ih\frac{\partial u}{\partial t}+Hu=0$  and all boundary terms are equal to zero.

Thus $v\in \overline W_0$,  i.e. for any  $\e>0$  there exists $u(x,t)\in W_0$
such  that  $\sup_{0\leq t\leq T}[v-u]_0<\e$.
\qed

Now we can  finish the proof of Theorem \ref{theo:4.3}.

Let $u''(x,t)=g_0(x)u'(x,t)$  where $g_0(x)$  is the same as in (\ref{eq:4.29}).
Then  $u''(x,t)$  satisfies 
$-ih\frac{\partial u''}{\partial t}+H''u''=0$  in $(\R^n\times(0,T))\setminus\Omega'$
and  $A''(x)=A(x),\ V''(x)=V(x)$  for 
$x\in (\R^n\setminus\Omega_0)\times(0,T)$  (cf. (\ref{eq:4.29}))
and  $\Lambda''=g_{00}^{-1}\Lambda'g_{00}$ where $g_{00}$  is  the restriction  of
$g_0$  to $\partial\Omega_0\times(0,T)$.

Let $w=u(x,t)-u''(x,t)=u(x,t)-g_0(x)u'(x,t)$.  Then
$\big(-ih\frac{\partial w}{\partial t}+Hw\big)=0$
  in $(\R^n\setminus\Omega_0)\times(0,T)$  and
$w(x,0)=w(x,T)=0$  on $\R^n\setminus\Omega_0$.  Hence,  by Lemma \ref{lma:4.4},
$w(x,t)=0$  in $(\R^n\setminus\Omega_0)\times(0,T)$.  Therefore,
\begin{equation}                                                                \label{eq:4.35}
u\big|_{\partial\Omega_0\times(0,T)}=u''\big|_{\partial\Omega_o\times(0,T)}
\ \ \mbox{and}\ \ 
\frac{\partial u}{\partial\nu}\Big|_{\partial\Omega_0\times(0,T)}=
\frac{\partial u''}{\partial \nu}\Big|_{\partial\Omega_0\times(0,T)}
\end{equation}
for all $u(x,t)$  and $u''(x,t)$  belonging  to $W((\R^n\times(0,T))\setminus\Omega')$.
Using the density lemma \ref{lma:4.5} we can extend  (\ref{eq:4.35})  to all $u,u''$  belonging
to $W((\Omega_0\times(0,T))\setminus\Omega')$.
Therefore $\Lambda=\Lambda''$ on  $\partial\Omega_0\times (0,T)$.  

\subsection{Aharonov-Bohm effect for time-dependent  electromagnetic potentials}
\
\\
When considering  AB effect we assume that 
$B=\mbox{curl}\,A=0,\ E=-\frac{1}{2}\frac{\partial A}{\partial t}
-\frac{\partial V}{\partial x}=0$  in $(\Omega_0\times(0,T))\setminus\Omega'$,
where  $B,E$  are  the magnetic and  electric fields,  $\Omega'$ is the union 
of all  obstacles $\Omega'(t)\subset\Omega_0,\ 0\leq t\leq T$.
Since $B=E=0$  we do not  need to deal with  the complicated $X$-ray  problems and  we can
substantially relax  the restrictions on the obstacles made  in Theorem 
\ref{theo:4.1}.

We shall  consider  the following  class of domains  in 
$\R^n\times(0,T)$ that we shall denote by $D^{(1)}$:

Let  $0=T_0<...<T_r=T$.  Denote  by $D_{t_0}$  the intersection
of $D$  with the plane  $t=t_0$.  
Then  for $t_0\in(T_{p-1},T_p),\ p=1,...,r,$   we have 
$D_{t_0}=\Omega_0\setminus\overline \Omega_p(t_0)$,  where  $\Omega_0$  is a 
simply-connected  domain in $\R^n,\ \Omega_p(t_0)=\bigcup_{j=1}^{m_p}\Omega_{pj}(t_0),
\ \overline\Omega_{pj}(t_0)\cap\overline\Omega_{pk}(t_0)=\emptyset$  for $j\neq k,\ 
\overline\Omega_{pj}(t_0)\subset\Omega_0,\ \Omega_{pj}(t_0)$  are amooth  domains 
(obstacles).  Note that $m_p$  may be different for  $p=1,2,...,r$.
We 
assume that $\Omega_p(t_0)$  depends smoothly on $t_0\in (T_{p-1}T_p)$.
We also  assume  that $D_{t_0}$  depends continuously on $t_0\in [0,T]$.

Note that some obstacles may  merge or split  when  $t_0$  crosses 
$T_p,\ p=1,...,r-1$ (cf.  Fig. 4).
\\
\\
\begin{tikzpicture}[scale=0.8]
\draw  (-6,0)  arc(180:360:  6 and 2); 
\draw[dashed] (6,0) arc(0:180:6 and 2); 
\draw (0,6) ellipse (6 and 2);
\draw (-6,0) -- (-6,6);
\draw (6,0) -- (6,6); 
\draw (1.5,0) ellipse (0.5 and 0.2);
\draw (4.5,0) ellipse (0.5 and 0.2);
\draw (1.5,6) ellipse (0.5 and 0.2);
\draw (4.5,6) ellipse (0.5 and 0.2);
\draw(1,0) -- (2,3);
\draw(5,0) -- (4,3);
\draw(1,6) -- (2,3);
\draw (4,3) -- (5,6);
\draw(2,0) -- (3,2.5);
\draw(4,0) -- (3,2.5);
\draw(3,3.5) -- (2,6);
\draw (3,3.5) -- (4,6);

\draw(3,3) ellipse (1 and 0.2);
\draw[dashed](3,3) ellipse (0.3  and 2.5);
\draw(3.3,0.3) node {$\gamma_5$};

\draw(-1.5,0) ellipse (0.5 and 0.2);
\draw(-4,3) node {$\Omega^{(0)}$};

\draw(0.7,5.2) node {$\Omega^{(1)}$};
\draw(5.2,5.4) node {$\Omega^{(2)}$};
\draw(0.8,0.8) node {$\Omega^{(3)}$};
\draw(5.35,0.43) node {$\Omega^{(4)}$};
\draw (-4.5,6) ellipse (0.5 and 0.2);
\draw (-2,0) -- (-5,6);
\draw(-1,0) -- (-4,6);

\end{tikzpicture}  
\\
\\
\\
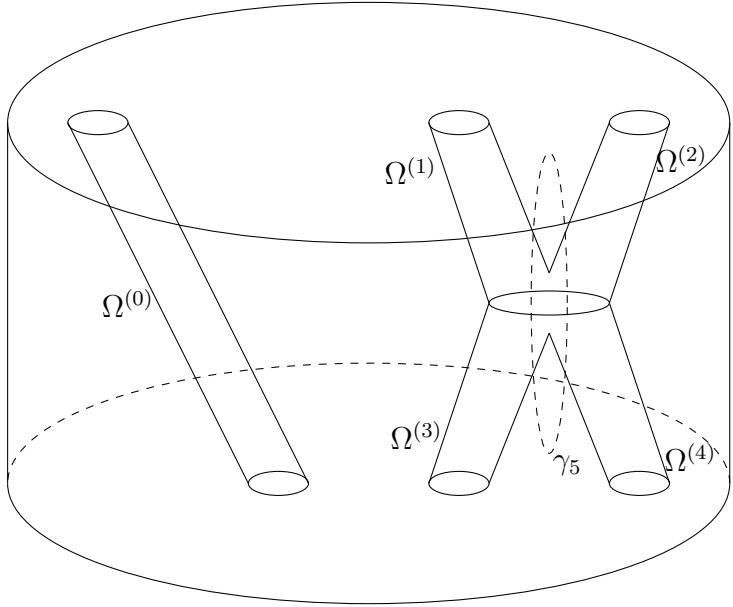
{\bf Fig. 4.} An example of a domain of class $D^{(1)}$. Obstacles $\Omega^{(3)}$
  and 
$\Omega^{(4)}$  merge,  
obstacles $\Omega^{(1)}$ and  $\Omega^{(2)}$ split.
\\
\\

Note that for each $t_0\in [0,T]$   the domains $D_{t_0}^{(1)}=D^{(1)}\cap\{t=t_0\}$
are connected.  Thus  the class of domains $D^{(1)}$  is too restrictive to
exhibit  the electric AB  effect.

We shall  introduce a more  general class of domains that we call $D^{(2)}$
such that $D_{t_0}^{(2)}=D^{(2)}\cap\{t=t_0\}$ may  be not connected  on some 
finite number of intervals in $(0,T)$.

An example  of a domain of type  $D^{(2)}$  is when  we make holes  in some
obstacles of  $D^{(1)}$.

We shall  prove  first the electromagnetic  AB  effect  in the case  of obstacles of
the class $D^{(1)}$.
Consider  the Schr\"odinger equation (\ref{eq:4.1})  in $D^{(1)}$,  where 
\begin{align}                                     
\nonumber
&u(x,0)=0,\ \ x\in D_0^{(1)}=D^{(1)}\cap\{t=0\},   
\\
\nonumber
&u\big|_{\partial\Omega_p(t_0)}=0,\ \ t_0\in[T_{p-1},T_p],\ p=1,...,r.
\end{align}

Let
\begin{equation}                                     \label{eq:4.36}
\alpha=\frac{e}{h}\int\limits_\gamma\frac{1}{c}A(x,t)\cdot dx-V(x,t)dt,
\end{equation}
where $\gamma$ is a closed curve in $D^{(1)}$.  Since we assume that $B=\mbox{curl}\,a=0,
\ E=-\frac{1}{c}\frac{\partial A}{\partial t}-\frac{\partial V}{\partial x}=0$,
the integral
(\ref{eq:4.36}), called the electromagnetic flux,  
depends only on the homotopy class of $\gamma$  in $D^{(1)}$.

Let $\gamma_1,..,\gamma_l$  
be the basis of the homology group  of $D^{(1)}$,  i.e.  any closed curve  in
$D^{(1)}$  is homotopic  to a linear  combination of 
$\gamma_1,...,\gamma_l$,  with integer coefficients.  Then the fluxes
$$
\alpha_j=\frac{e}{h}\int\limits_{\gamma_j}\frac{1}{c} A(x,t)\cdot dx-V(x,t)dt, \ \ \
j=1,...,l,
$$
determine the gauge equivalent  class of electromagnetic potentials \linebreak
$(A(x,t),V(x,t))$,  i.e. 
$(A(x,t),V(x,t))$ and  $(A'(x,t),V'(x,t))$  are gauge equivalent  iff
$\alpha_j-\alpha_j'=2\pi m_j,\ m_j\in\Z,$
where  $\alpha_j'=\frac{e}{h}\int_{\gamma_j}\frac{1}{c}A'(x,t)\cdot dx - V'(x,t)dt$.

As is \S 4.1.  we shall  introduce  localized  geometric optics type 
solutions $u(x,t)$  of the Schr\"odinger equation (\ref{eq:4.1})  in $D^{(1)}$  depending  
on a large parameter
$k$  and satisfying the zero initial  condition
\begin{equation}                                    \label{eq:4.37}
u(x,0)=0,\ \ \ x\in D_0^{(1)},
\end{equation} 
and zero boundary conditions on the boundaries of obstacles
\begin{equation}                              \label{eq:4.38}
u(x,t)\big|_{\partial\Omega'}=0,
\end{equation}
where $\Omega'\subset\R^n\times(0,T)$  is the union of all obstacles 
$\Omega_p(t),\ t\in [T_{p-1},T_p], p=1,...,r$,  and  
$D_0^{(1)}=\Omega_0\setminus\Omega_1(0),\ \Omega_1(0)=\Omega'\cap\{t=0\}$.
Such solutions were constructed in [18].  Suppose $t_0\in (T_{p-1},T_p),\ 1\leq p\leq r$.
Suppose $\gamma(x^{(1)},t_0)=
\gamma_0(t_0)\cup...\cup\gamma_{d-1}(t_0)\cup\gamma_d(x^{(1)},t_0)$  
is a broken ray  in $D_{t_0}$  with legs  $\gamma_0(t_0),...,\gamma_{d-1}(t_0),
\gamma_d(x^{(1)},t_0)$,
  starting  at 
point
$x^{(0)}\in \partial\Omega_0$,
 reflecting at $\partial\Omega_p(t_0)$
and
ending at  $x^{(1)}\in D_{t_0}^{(1)}$.

As in [18]  we can construct an asymptotic solution as $k\rightarrow\infty$ of the form
(\ref{eq:4.15}),
where
 $\mbox{supp\ }u_N(x,t,\omega)$ is contained in a
small neighborhood of $x=\gamma(x^{(1)},t_0),t=t_0$.
Note that (cf. [18])  one can find $u^{(N)}(x,t)$  such that $Lu^{(N)}=
-Lu_N=O(\frac{1}{k^{N+1}})$  in $D^{(1)}$,
 $u^{(N)}\big|_{t=0}=0,$  and $u^{(N)}\big|_{\partial\Omega'}=0,\ 
u^{(N)}\big|_{\partial\Omega_0\times(0,T)}=0$
and such that $u^{(N)}=O(\frac{1}{k^{N-2}})$. 
Here  $L$  is the left hand side  of (\ref{eq:4.1}). 
 Then 
\begin{equation}                               
\nonumber
u=u_N+u^{(N)}
\end{equation}
is the exact  solution of $Lu=0$  in $D^{(1)}$,
 $u\big|_{t=0}=0,\ x\in D_0^{(1)}$,  $u\big|_{\partial\Omega'}=0$.

Let  $t_0\in (T_p,T_{p+1})$  and  let $m_p$  be  the number of the obstacles  in 
$D_{t_0}^{(1)}$. 
It was proven  in [13], [18]  that $u(x,t)$  has  the following  form in
the neighborhood $U_0$  of $(x^{(1)},t_0)$:
\begin{multline}                                \label{eq:4.39}
u(x,t)
=c(x,t)\exp\Big(-i\frac{mk^2t}{2h}
+i\frac{mk}{h}\psi_d(x,t)
+\frac{ie}{hc}\int_{\gamma(x,t)}A(x,t)\cdot dx\Big)
\\
+O\Big(\frac{1}{k}\Big),
\end{multline}
Here $c(x^{(1)},t_0)\neq 0$
and $\gamma(x,t)$  is a broken ray in $D_t^{(1)}$ that starts at $(y,t)$,  $(y,t)$ is
close to $(x^{(0)},t_0)$,  and such that the first leg of $\gamma(x,t)$  has the
same direction  as $\gamma_0(t_0)$.

Note the difference between the asymptotic solution (\ref{eq:2.45}) in \S 2.6
and the asymptotic solution (\ref{eq:4.39}). 
The broken ray $\gamma=\bigcup_{k=0}^d\gamma_k$  in (\ref{eq:2.45})
is the  projection to $\R^2$ of
the broken  ray $\tilde\gamma=\bigcup_{k=1}^d\tilde\gamma_k$  in 
 $\R^2\times(0,+\infty)$  having the time variable $t$  as a parameter.
The solution  (\ref{eq:4.39}) 
corresponds  to a broken  ray $\bigcup_{k=0}^d \gamma_k(t_0)$  in the plane  $t=t_0$
with $s=x\cdot \omega_k$
as a parameter  on $\gamma_k(t_0)$. 
\qed

Let $\beta$  be the ray  $x=x^{(0)}+s\theta,\ s\geq 0,\ t=t_0,$  starting at
$(x^{(0)},t_0)$  and  ending at  $(x^{(1)},t_0)$.   Choose  $x^{(1)}\in \Omega_0$
 such that  $\beta$  does not intersect  $\Omega'\cap\{t=t_0\}$.
We assume  that $\Omega_0$  is large enough that  such $x^{(1)}$ exists (see Fig.5):  
\\
\begin{tikzpicture}[scale=0.7]

\draw(-3,5) -- (7,4);

\draw (7,4) --(8,1);
\draw (-1,0)--(8,1);
\draw(-1,0) -- (-3,5);


\draw(7.7,5) circle (1.2);



\draw(9.1,0.5) circle (1.2);

\draw(5.5,2.5) circle (1.4);

\draw(-1.3,-0.3) node {$x^{(0)}$};
\draw(-3,5.3) node {$x^{(1)}$};

\draw(3.5,0.1) node {$\gamma_0(t_0)$};

\draw(8.1,3) node {$\gamma_1(t_0)$};

\draw(3,5) node {$\gamma_2(t_0)$};

\draw(-3,3)  node {$\beta(t_0)$};

\draw(5.5,2) node {$\Omega_1(t_0)$};

\draw(7.7,4.8) node {$\Omega_3(t_0)$};

\draw(9.1,0.2) node {$\Omega_2(t_0)$};

\end{tikzpicture}
\\ 
\
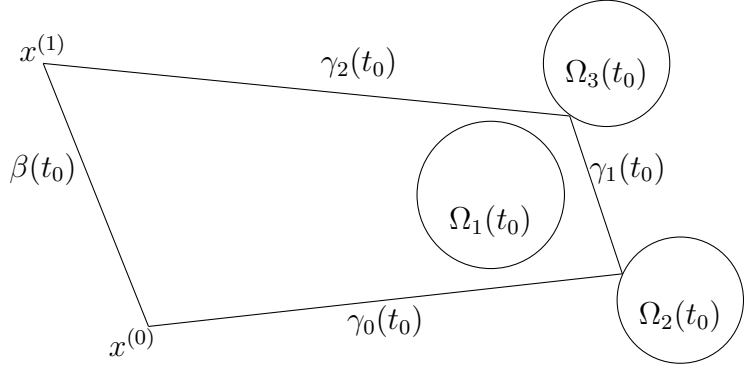
{\bf Fig. 5.} The broken ray $\gamma=\gamma_0(t_0)\cup\gamma_1(t_0)\cup\gamma_2(t_0)$ 
and the  ray $\beta(t_0)$  belong  to  $D_{t_0}^{(1)}=D^{(1)}\cap\{t=t_0\}$.
\\
\\
Let  $v(x,t)$  be a geometric optics type solution  
similar to
 (\ref{eq:4.9})
and 
corresponding to the ray $\beta$. 
  We have,  as in (\ref{eq:4.39}):
\begin{equation}                                \label{eq:4.40}
v(x,t)=c_1(x,t)\exp\Big(-i\frac{mk^2t}{2h}+i\frac{mk}{h}x\cdot\theta
+\frac{ie}{hc}\int_{\beta(x,t)}A(x,t)\cdot dx\Big)+O\Big(\frac{1}{k}\Big).
\end{equation}
We choose the initial value  for  $a_{0}(x,t,\theta)$  (cf.  (\ref{eq:4.39}))
near $(x^{(0)},t_0)$  such that
$$
c_1(x^{(1)},t_0)=c(x^{(1)},t_0).
$$
Consider  $|u(x,t)-v(x,t)|^2$  in a neighborhood 
$\{(x,t): |x-x^{(1)}|\leq\e_0,\ |t-t_0|<\e_0\}$.

As in \S 2.6 we get for a small neighborhood  of $(x^{(1)},t_0)$
\begin{equation}                                    \label{eq:4.41}
|u(x,t,\omega)-v(x,t,\theta)|^2=
|c(x^{(1)},t_0)|^2\ 4\sin^2\frac{\alpha(t_0)}{2}+O(\e),
\end{equation}
where
$$
\alpha(t_0)=\frac{e}{hc}\Big(\int_{\gamma(x^{(1)},t_0)}
A(x,t_0)\cdot dx-\int_{\beta(x^{(1)},t_0)}A(x,t_0)\cdot dx\Big).
$$
Note that $\alpha(t_0)$  is the sum of the fluxes  of those obstacles
$\Omega_{pj}(t_0),\ 1\leq j\leq m_p$,  that  are encircled by $\gamma\cup\beta$.
As in \S 2.6, 
varying $\gamma$  and $\beta$  at least  $m_p$  times we can recover 
 $\alpha_{pj}(t_0)\ (\mbox{modulo}\,2\pi n),\ 1\leq j\leq m_p$,  up to a sign,  where 
\begin{equation}                                   \label{eq:4.42}
\alpha_{pj}(t_0)
=\frac{e}{hc}\int_{\gamma_{pj}(t_0)}A\cdot dx, \ \ \ 1\leq j\leq m_p,
\end{equation}
and $\gamma_{pj}(t_0)$  is  a simple contour in $D_{t_0}^{(1)}$
encircling $\Omega_{pj}(t_0)$,
$1\leq j\leq m_p$.
Note that $\alpha_{pj}$  are  the same for any  $t_0\in (T_p,T_{p+1})$.
We  can repeat the same arguments for any $t_0\neq T_1,...,T_{p-1}$.

Our class of time-dependent obstacles is such that $D_{t_0}^{(1)}$  is connected for 
any $t_0\in [0,T]$.
It follows from this assumption that a basis of the homology group  of $D^{(1)}$  is
contained  in the set $\gamma_{pj}(t_p),\ 1\leq j\leq m_p,\ t_p\in (T_{p-1},T_{p}),\ 
 1\leq p\leq r$,  of ``flat"  closed  curves  that  are contained  in  the planes $t$=const.

Denote such basis by $\gamma^{(1)}(t^{(1)}),...,\gamma^{(l)}(t^{(l)})$.
Then any closed contour $\gamma$  in $D$  is homotopic to a linear combination 
$\sum_{j=1}^ln_j\gamma^{(j)}(t^{(j)})$  where $n_j\in \Z$.  Therefore the flux
\begin{equation}                                      \label{eq:4.43}
\frac{e}{h}\int_\gamma \frac{1}{c}A\cdot dx-Vdt=\sum_{j=1}^l n_j\alpha^{(j)}(t^{(j)}),
\end{equation}
where
$\alpha^{(j)}(t^{(j)})=\frac{e}{hc}\int_{\gamma^{(j)}(t^{(j)})}A\cdot dx$.

Thus
the fluxes $\alpha^{(j)}(t^{(j)}),\ 1\leq j\leq l,\ \mbox{mod\ }2\pi n,\ 
n\in \Z$,  determine the gauge equivalence class of $(A(x,t), V(x,t))$.  
Therefore computing the probability densities of appropriate solutions 
we are able  to determine  the gauge equivalence classes of electromagnetic 
potentials up to a sign.

The solution  $u(x,t,\omega)-v(x,t,\theta)$  in  (\ref{eq:4.41})
is a solution of the  Schr\"odinger  equation  in $D^{(1)}$  with nonzero  boundary  conditions 
on  $\partial\Omega_0\times(0,T)$.
The probability  density  $|u(x,t,\omega)-v(x,t,\theta)|^2$ depends on the flux $\alpha(t_0)$  
but  this does not prove  yet  that the magnetic flux  makes the physical
impact  since
$u(x,t,\omega)-v(x,t,\theta)$  may 
be not a physically  meaningful  solution.  
However,  by the density lemma \ref{lma:4.5}
there exists a physically meaningful  solution $v_{\e_1}\in W_0$  such that
$$
\max_{0\leq t\leq T}\int\limits_{D_{t_0}^{(1)}}|u-v-v_{\e_1}|^2dx <\e_1,
$$  
where $\e_1$  is much smaller  than $\e>0$  in (\ref{eq:4.41}).
Then
$$
\int\limits_{U_0}|v_\e(x,t_0)|^2=
|c^{(1)}(x^{(1)},t_0)|^2 4\sin^2\frac{\alpha(t_0)}{2}\mu(U_0)+O(\e),
$$
where $\mu(U_0)$ is the volume  of $U_0$,  i.e.  
$\int_{U_0}|v_\e(x,t_0)|^2dx$  depends  on  $\alpha(t_0)$.  Thus we proved AB effect
since $v_\e(x,t)$ is a physically  meaningful  solution.

{\bf Example 4.1}.
Consider the domain shown in Fig.4.  Let $\gamma_p, \ 0\leq p\leq 4,$
be simple closed curves encircling $\Omega^{(p)}$.
There is also a simple closed curve $\gamma_5$  that is not homotopic 
to any closed curve contained in the plane $t=\mbox{const}$.
Note that $\gamma_1+\gamma_2\approx \gamma_3+\gamma_4$  
where $\approx$  means homotopic.
Also $\gamma_5\approx \gamma_1- \gamma_3$.  Therefore
$\gamma_0,\gamma_1,\gamma_2,\gamma_3$  is a basis of the homology group of $D^{(1)}$.

Let $\alpha_j$ be the fluxes corresponding  to $\gamma_j$.  
Note  that if $\gamma_j$  is flat  then  $\alpha_j=\int_{\gamma_j} A\cdot dx$ 
is a magnetic flux. However   
$\alpha_5=\frac{e}{h}\int_{\gamma_5}\frac{1}{c}A\cdot dx-Vdt$
is an electromagnetic flux.
Since $\gamma_5\approx\gamma_1-\gamma_3$  we have that
$\alpha_5=(\alpha_1-\alpha_3)\ (\mbox{mod\ } 2\pi n),\ n\in \Z$.
\qed

\subsection{Combined electric and magnetic AB effect}  
\
\\
In this subsection  we consider  domains of
the class $D^{(2)}$  that will allow to study  the combined
electric and magnetic AB  effect (cf. Markovitch et al [38]  and [22]).

{\bf Example 4.2.}
We shall start with 
the example of the domain 
$D=(\R^n\times(0,T))\setminus\Omega_0$
of class $D^{(2)}$ that was considered  in \S 3 (see  Fig.3).
In \S3  we assumed  that $A(x)=0$.  Now  we shall consider the Schr\"odinger equation  of the form
(\ref{eq:4.26})  with $A(x,t), V(x,t)$  such 
that  $B=\mbox{curl}\,A=0,\ E=-\frac{1}{c}\frac{\partial A}{\partial t}
-\frac{\partial V}{\partial x}=0$  in $D$.

Denote by  $Q$  the cylinder $Q=\{R_1\leq |x|\leq R_2,\ \e\leq t\leq T-\e\}$.
Note that  $D_{t_0}=D\cap\{t=t_0\}$  is connected for  
$0\leq t_0<\e$  and $T-\e<t_0\leq T$  and has  two connected components when  
$\e\leq t_0\leq T-\e$,  one of them  being $Q=\{R_1\leq |x|\leq R_2\}$.

Since $Q$ is simply-connected  and $\mbox{curl}\,A=0$  in $Q$,
we can find $\varphi(x,t)\in C^\infty(Q)$  such  that 
$A(x,t)=\frac{\partial\varphi(x,t)}{\partial x}$ in  $Q$.

Making the gauge transformation  with the gauge
$e^{\frac{ie}{hc}\varphi(x,t)}$,  we can replace (\ref{eq:4.26})   with
a gauge equivalent  equation such  that  $\hat A(x,t)=0$  in $Q$  and $\hat V(x,t)=
V(x,t)+\frac{1}{c}\frac{\partial \varphi}{\partial t}$.  
Thus  $\frac{\partial \hat V(x,t)}{\partial x} =0$
in $Q$   since  $E=0$,   and we get that  $\hat V(x,t)=\hat V_0(t)$  in $Q$.

Therefore,  without  loss  of
generality  we can, from the beginning, assume  
$V(x,t)=V_0(t)$  in $Q$,  $A=0$ in $Q$.

The  basis  of the  homology  group for  $D$  consists of 
$\gamma_1=\{|x|=R_2+1\}$  and  of a closed  curve  $\delta_1$  in the $(x_1,t)$  plane that
encircles  the rectangle $\{R_1\leq x_1\leq R_2,\e\leq t\leq T-\e,x_2=0\}$.

Potentials  $(A,V),(A',V')$  are gauge equivalent  if 
$$
\alpha_1-\alpha_1'=2\pi n_1,\ n_1\in\Z,\ \ \alpha_2-\alpha_2'=2\pi n_2,\ n_2\in \Z,
$$
where 
$
\alpha_1=\int_{\gamma_1}A(x)\cdot dx,  \  \alpha_1'=\int_{\gamma_1}A'(x)\cdot dx,
\  \alpha_2=\int_{\delta_1}A(x)\cdot dx - Vdt,
\ 
\alpha_2'=\int_{\delta_1}A'(x)\cdot dx-V'dt.
$

We shall prove  that $(A,V),(A',V')$ made a different  physical impact
if either $\alpha_1-\alpha_1'\neq 2\pi n,\ \forall  n\in\Z, $
and $\alpha_1+\alpha_1'\neq 2\pi n, \ \forall n$,
  or if
$\alpha_1-\alpha_1'=2\pi n_1,$  and 
$\alpha_2-\alpha_2'\neq 2\pi n, \ \forall n\in\Z$.
This will prove the combined AB effect.

It follows from the results  of  \S4.4  
 that $(A,V),(A',V')$  have a different 
physical impact 
if $\alpha_1-\alpha_1'\neq 2\pi n,\ \forall n\in \Z$,
and $\alpha_1+\alpha_1'\neq 2\pi n,\ \forall n\in \Z$. 

Suppose $\alpha_1-\alpha_1'=2\pi n_1,\ n_1\in \Z$

Then  for each  $t_0\in [0,T]$  there exists $g(x,t_0)$  such that
$$
\frac{e}{c}A(x,t)=\frac{e}{c}A'(x,t)-ih\frac{\partial g}{\partial x}g^{-1}.
$$
Using the gauge $g(x,t)$ we transform  $\big(ih\frac{\partial}{\partial  t}-H'\big)u'=0$ 
to  $\big(ih\frac{\partial}{\partial t}-H''\big)u''=0$,
where  $A''(x,t)= A(x,t)$,  i.e.  $H$  and  $H''$  have the same magnetic potentials
in $D\setminus Q$.  
 Since  
$E=-\frac{1}{c}\frac{\partial A}{\partial t}-\frac{\partial V}{\partial x}=0$
and $E''=-\frac{1}{c}\frac{\partial A''}{\partial t}-\frac{\partial V''}{\partial x}=0$
we get that
$\frac{\partial V}{\partial x}-\frac{\partial V''}{\partial x}=0$.  
Hence  $V-V''=0$  in $D\setminus Q$
since  $V=V''=0$ for large  $|x|$.  Thus $A=A'',V=V''$  outside  of $Q,\ 
A=A''=0$  in  $Q,\ V=V_0(t),\ V''\equiv V_0''(t)$  in $Q$.

Note  that  $\alpha_2-\alpha_2'=\frac{e}{h}\int_\e^{T-\e}(V_0(t)-V_0''(t))dt$
since  $H=H''$  outside  of $Q$.   If 
 $\frac{e}{h}\int_\e^{T-\e}V_0(t)dt-
\frac{e}{h}\int_\e^{T-\e}V_0''(t)dt
\neq 2\pi n,  \ \forall n\in R,$
 we shall prove that $V_0(t),\ V_0''(t)$  have different  physical  impacts.

We shall use the same arguments  as in \S3.  The difference is  that  
$A\neq 0,\ V\neq 0$  outside of $Q$  for $u$ and $u''$.
We have  $u(x,T-\e)=u''(x,T-\e)$  for 
$|x|>R_2$  since 
$u(x,t)$  and  $u''(x,t)$  satisfy  the same equation, and
the   initial and boundary conditions  for $u$  and $u''$ are
the same when
$t<T-\e,|x|>R_2$.  Suppose $|u(x,t)|=|u'(x,t)|=R$  in  $U_0$,
where $U_0$  is the same as in \S 3.
   Using  the polar
representation $u=Re^{i\Phi},u''=Re^{i\Phi''(x,t)}$,  separating  the real part,
we get
$$
h\Phi_t=\frac{h^2}{2 m}(\Delta R-R|\nabla\Phi|^2)+\frac{eh}{mc}A\cdot \nabla\Phi R
+\Big(\frac{e^2}{2mc^2}A^2+eV\Big)R
$$  
in  $U_0$.
The equation  for $\Phi''$  is the same  since $R''=R$.  
The initial condition  in $U_0\cap\{t=T-\e\}$  for $\Phi$  and $\Phi''$  are also the 
same since 
$u(x,T-\e)=u''(x,T-\e)$.  Therefore  $\Phi=\Phi''$  in $U_0$.
The continuation of the proof is the same as in 
\S3.
\qed

Consider now  the equation  (\ref{eq:4.26}) in a general domain
of the form $D^{(2)}, B=E=0$  in $D^{(2)}$.
Let  $Q_j, j=1,2,...,d$,  be such that $Q_{jt_0}=Q_j\cap\{t=t_0\}$
is a bounded connected component  of $D_{t_0}^{(2)}=D^{(2)}\cap\{t=t_0\}$
for  $\e_j\leq  t_0\leq T_j',\ 0<T_1'<T_2'<...<T_d'<T,\ j=1,...,d$.

As in  the previous  example,  we may assume that  $A=0$  in $Q_j,\ V=V_j(t)$  in 
$Q_j,\ 1\leq j\leq d$.

The basis  of the homology  group   
$D^{(2)}$ consists of
the basis $\gamma_1,...,\gamma_l$  of 
the connected  domain $D^{(2)}\setminus\bigcup_{j=1}^dQ_j$ and curves 
$\delta_1,...,\delta_d$  similar  to $\delta_1$  in  Example 4.2,  passing through the holes 
$Q_1,...,Q_d$.

Let  $\alpha_j=\frac{e}{h}\int_{\gamma_j}\frac{1}{c}A\cdot dx-Vdt,\ 
\beta_k=\frac{e}{h}\int_{\delta_k}\frac{1}{c}A\cdot dx-Vdt$
be  the electromagnetic  fluxes.

We shall  show  that $(A,V)$ and  $(A',V')$  have a different  
physical  impact  if 

a) either  $\alpha_j-\alpha_j'\neq 2\pi n,\ \forall n\in\Z$ 
and  $\alpha_j+\alpha_j'\neq 2\pi n,\ \forall n\in Z$, for
 some  $j,\ 1\leq j\leq l$,
\\
or

b)  $\alpha_j-\alpha_j'=2\pi n_j,\ j=1,...,l$  and
$\beta_k-\beta_k'\neq 2\pi n_k,\ \forall n_k\in \Z$,  for  some 
 $k,\ 1\leq k\leq d$.  
\\
Here $\alpha_j',\beta_k'$  
are fluxes  for $(A',V')$.

Assertion a)  follows from the results of \S4.4.

If $\alpha_j-\alpha_j'=2\pi n_j,\ 1\leq j\leq l$,  we can,  as in Example 4.2,
  replace  $(A',V')$  by a gauge  equivalent  $(A'',V'')$  such that $A=A'',V=V''$ 
in  $D^{(2)}\setminus\bigcup_{j=1}^dQ_j$.

If  $\beta_1-\beta_1'\neq 2\pi n, \forall n\in \Z$, then we obtain, as  in the proof 
of Example 4.2,
 that  $|u(x,t)|^2\neq |u''(x,t)|^2$ for  $t>T_1'$ and thus we prove the AB effect.

If $\beta_1-\beta_1'=2\pi n,  n\in \Z$,  but  $\beta_2-\beta_2'\neq 2\pi n,
\ \forall n\in \Z$,
 we get that  $u(x,t)=u''(x,t)$  for  
$T_1'<t<T_2'$,
but  $|u(x,t)|\neq |u''(x,t)|$  for  $t>T_2'$,  etc.  Thus  AB effect holds if
$\beta_j-\beta_j'\neq 2\pi n,\ \forall n$,  for  one  of $1\leq j\leq d$.

\section{Gravitational  AB effect}
\init

In this section we shall study the gravitational analog of the quantum mechanical AB effect.
\subsection{Global isometry}
\
\\
Consider  a pseudo-Riemannian metric
$\sum_{j,k=0}^n g_{jk}(x)dx_{jk}dx_k$  with Lorentz signature in $\Omega$,
where $x_0\in \R$  is the time variable,  $x=(x_1,...,x_n)\in\Omega,\ \Omega=
\Omega_0\setminus \cup_{j=1}^m\overline\Omega_j,\ 
\Omega_0$ is simply connected,  $\overline\Omega_j\subset\Omega_0, 
\ \Omega_j, \ 1\leq j\leq m$,
are obstacles  (cf.  subsection 3.3).  We assume that $g_{jk}(x)$  are independent
of $x_0$,  i.e.  the metric is stationary.

Consider a group  of transformations  (changes of variables)
\begin{eqnarray}                                      \label{eq:6.1}
x'=\varphi(x),
\\
x_0'=x_0+a(x),
\nonumber
\end{eqnarray}
where $x'=\varphi(x)$  is a diffeomorphism  of $\overline \Omega$  onto
$\overline{\Omega'}=\varphi(\overline \Omega)$  and
$a(x)\in C^\infty(\overline\Omega)$.  Two metrics
$\sum_{j,k=0}^n g_{jk}(x)dx_jdx_k$  and
$\sum_{j,k=0}^n g_{jk}'(x')dx_j'dx_k'$
are called isometric if
\begin{equation}                                           \label{eq:6.2}
\sum_{j,k=0}^n g_{jk}(x)dx_jdx_k=\sum_{j,k=0}^n g_{jk}'(x')dx_j'dx_k',
\end{equation}
where
$(x'_0,x')$  and  $(x_0,x)$  are related by  (\ref{eq:6.1}).

The group
of isomorphisms (isometries) will play  the same role as the gauge group for
 the magnetic AB effect.

Let 
$$
\qed_gu(x_0,x)=0\ \ \mbox{in}\ \ \R\times\Omega
$$
be the wave equation corresponding to the metric $g$,   i.e.
\begin{equation}                                           \label{eq:6.3}
\qed_gu\overset{def}{=}\sum_{j,k=0}^n\frac{1}{\sqrt{(-1)^ng_0}}
\frac{\partial}{\partial x_j}
\Big(\sqrt{(-1)^n}g_0g^{jk}(x)\frac{\partial u}{\partial x_k}\Big)=0,
\end{equation}
where $g_0=\det[g_{jk}]_{j,k=0}^n,\ [g^{jk}(x)]=[g_{jk}]^{-1}$.

Solutions of (\ref{eq:6.3})  are called gravitational waves on the background of
the space-time  with the metric $g$.

Consider the initial boundary value problem for (\ref{eq:6.3}) in
$\R\times\Omega$  with  zero initial conditions
\begin{equation}                              \label{eq:6.4}
u(x_0,x)=0\ \ \mbox{for}\ x_0 \ll 0,\ x\in\Omega,
\end{equation}
and the boundary condition
\begin{equation}                               \label{eq:6.5}
u\big|_{\R\times\partial\Omega_0}=f,\ \ u\big|_{\R\times\partial\Omega_j}=0,
\ \ 1\leq j\leq m,
\end{equation}
where  $f\in C_0^\infty(\R\times\partial\Omega_0)$.  Let  $\Lambda_g$  
be the Dirichlet-to-Neumann  (DN)  operator,
i.e.  $\Lambda_g f=\frac{\partial u}{\partial \nu_g}\big|_{\R\times\partial\Omega_0}$,
where
\begin{equation}                                    \label{eq:6.6}
\frac{\partial u}{\partial \nu_g}=
\sum_{j,k=0}^ng^{jk}(x)\nu_j(x)\frac{\partial u}{\partial x_k}
\Big(\sum_{p,r=0}^ng^{pr}(x)\nu_p\nu_r\Big)^{-\frac{1}{2}}.
\end{equation}
Here 
$u(x_0,x)$  is the solution of (\ref{eq:6.3}),  (\ref{eq:6.4}),  (\ref{eq:6.5}),
$\nu(x)=(\nu_1,...,\nu_n)$  is the outward  unit normal to 
$\partial\Omega_0,\ \nu_0=0$.

Let  $\Gamma$  be an open subset  of  $\partial\Omega_0$.  We shall say that  
boundary measurements are taken  on  $(0,T)\times \Gamma$  
if we know the restriction $\Lambda_g f\big|_{(0,T)\times\Gamma}$  
for any  $f\in C_0^\infty((0,T)\times\Gamma)$.

Consider metric $g'$  in $\Omega'$ 
and the corresponding
 initial-boundary value problem
\begin{eqnarray}                                             \label{eq:6.7}
\Box_{g'}u'(x_0',x')=0\ \ \mbox{in}\ \ \R\times\Omega',
\\                                                          \label{eq:6.8}
u'(x_0',x')=0\ \ \mbox{for}\ \ x_0' \ll 0,\ x'\in \Omega',
\\                                                          \label{eq:6.9}
u\big|_{\R\times\partial\Omega_0'}=f,\ \ \ u'\big|_{\R\times\partial\Omega_j'}=0,
\ 1\leq j\leq m',
\end{eqnarray}
where 
$\Omega'=\Omega_0'\setminus\bigcup_{j=1}^{m'}\overline{\Omega_j'}.$

We assume that $\partial\Omega_0\cap\partial\Omega_0'\neq \emptyset$.  Let  $\Gamma$  be
an open subset  of $\partial\Omega_0\cap\partial\Omega_0'$.

The following theorem  was proven  in [19] (see [19], Theorem 2.3).
\begin{theorem}                                   \label{theo:6.1}
Suppose  $g^{00}(x)>0,\ g_{00}(x)>0$  in $\overline\Omega$  and 
$(g')^{00}>0,\ g_{00}'>0$  in $\overline{\Omega'}$.  
Suppose  $\Lambda_gf\big|_{(0,T)\times\Gamma}=
\Lambda_{g'}f\big|_{(0,T)\times\Gamma}$  for  all $f\in C_0^\infty((0,T)\times\Gamma)$.
Suppose  $T>T_0$,  where $T_0$  is sufficiently large.   Then  metrics $g$  and $g'$  are isometric,  i.e.
there exists a change of variables (\ref{eq:6.1})  such that (\ref{eq:6.2})  holds.  
Moreover,  $\varphi\big|_\Gamma=I,\ a\big|_\Gamma=0$.
\end{theorem}

If two metrics $g$  and $g'$  in $\Omega$  and $\Omega'$,  respectively,  are isometric,  then the solutions $u(x_0,x)$  and $u(x_0',x')$ of 
 the corresponding wave equations are the same after 
 the change of variables (\ref{eq:6.1}).
Therefore isometric metrics have the same physical impact.

Suppose two metric  $g$  and  $g'$  are isometric in some  neighborhood  $V\subset\Omega,\overline V\cap\partial\Omega\neq \emptyset$.  Let
$\Gamma\subset\overline V\cap\partial\Omega$.  There exists a local isomorphism
\begin{equation}                                  \label{eq:6.10}
x'=\varphi_V(x),\ \ x_0'=x_0+a_V(x)
\end{equation}
that transforms  $g'$  to the metric $\hat g$  in  $\overline V$  such that 
$\hat g=g$  in $V$.  Extend  the isometry (\ref{eq:6.10}) from 
$\overline V$  to 
$\overline\Omega$  and denote by $\hat g$  the image of $g'$  under this map.
Thus $\hat g$  isometric to $g'$  in 
$\Omega$  and $\hat g=g$  in $\overline V$.

\begin{theorem}                                     \label{theo:6.2}
The metrics $g$ and $\hat g$  are not isometric if and only if  the boundary measurements 
\begin{equation}                                     \label{eq:6.11}
\Lambda_g f\big|_{(0,T)\times\Gamma}\neq\Lambda_{\hat g}f\big|_{(0,T)\times\Gamma}\ \ \
\mbox{for some}\ \ \ f\in C_0^\infty((0,T)\times\Gamma),
\end{equation}
\end{theorem}
{\bf Proof} (cf. \S2.1):
Suppose  $g$ and $\hat g$ are not isometric.  
If  
$\Lambda_gf\Big|_{(0,T)\times\Gamma}=
\Lambda_{\hat g}f\Big|_{(0,T)\times\Gamma}$  for all  $f\in C_0^\infty((0,T)\times\Gamma)$  
then  by Theorem  \ref{theo:6.1}  there  exists  a map  of  the form (\ref{eq:6.1})
that  transforms $\hat g$  to $g$  and 
such that 
\begin{equation}                                   \label{eq:6.12}
\varphi\Big|_\Gamma=I,\ \ \ a\Big|_\Gamma=0.
\end{equation}
Since $g=\hat g$  in $\overline V$  any such map satisfies  (\ref{eq:6.12}).
Thus $g$  and  $\hat g$  are isometric,  i.e.  we got a contradiction.
Therefore if $g$  and $\hat g$ are  not isometric then (\ref{eq:6.11})  holds.

Vice versa,  suppose $g$  and $\hat g$ are  isometric,  i.e.  (\ref{eq:6.1})  holds.
Then for all solutions $u(x_0,x)$  and  $\hat u(\hat x_0,\hat x)$ of 
equations (\ref{eq:6.3}), (\ref{eq:6.4}),  (\ref{eq:6.5})  
and (\ref{eq:6.7}), (\ref{eq:6.8}),  (\ref{eq:6.9}),  respectively,
 we have $u(x_0,x)=\hat u(\hat x_0,\hat x)$,
where  $(x_0,x)$  and  $(\hat x_0',\hat x)$  are  related  by  (\ref{eq:6.1}).
Note that (\ref{eq:6.12}) also holds since  $g=\hat g$  in $\overline V$.
Thus we have  $\Lambda_gf\big|_{(0,T)\times\Gamma}=
\Lambda_{\hat g}f\big|_{(0,T)\times\Gamma}$ 
for all $f\in C_0^\infty((0,T)\times\Gamma)$.  Therefore  
if
(\ref{eq:6.11})  holds then
$g$ and  $\hat g$  are not isometric.

It follows from (\ref{eq:6.11}) that  non-isometric metrics $g$  and  $\hat g$  (and therefore 
$g$ and $g'$) have  different physical impacts.

Note that the open set $\Gamma$ can be arbitrary small.  However the time interval 
$(0,T)$  must be large enough: $T>T_0$.

\subsection{Locally static  stationary metrics}

Let $g$ and $g'$ be isometric.  Substituting 
$dx_0'=dx_0+\sum_{j=1}^na_{x_j}(x)dx_j$  and  taking into account  that 
$dx_0$  is arbitrary,  
we get from (\ref{eq:6.1}) and (\ref{eq:6.2})  that
\begin{equation}                                \label{eq:6.13}
g_{00}'(x')=g_{00}(x),
\end{equation}
\begin{equation}                                            \label{eq:6.14}
2g_{00}'(x')\sum_{j=1}^na_{x_j}(x)dx_j+2\sum_{j=1}^ng_{j0}'(x')dx'
=2\sum_{j=1}^ng_{j0}(x)dx_j.
\end{equation}
Using (\ref{eq:6.13}) we can rewrite (\ref{eq:6.14})  in the form
\begin{equation}                                  \label{eq:6.15}
\sum_{j=1}^n\frac{1}{g_{00}'(x')}g_{j0}'(x')dx'
=\sum_{j=1}^n\frac{1}{g_{00}(x)}g_{j0}(x)dx_j-\sum_{j=1}^na_{x_j}(x)dx_j.
\end{equation}
Let $\gamma$  be an arbitrary  closed curve  in $\Omega$,  and let
$\gamma'$  be the image  of $\gamma$  in $\Omega'$   under the map
(\ref{eq:6.1}).
Integrating (\ref{eq:6.15}) we get
\begin{equation}                                  \label{eq:6.16}
\int_{\gamma'}\sum_{j=1}^n\frac{1}{g_{00}'(x')}g_{j0}'(x')dx'
=\int_\gamma\sum_{j=1}^n\frac{1}{g_{00}(x)}g_{j0}(x)dx_j,
\end{equation}
since
$\int_\gamma\sum_{j=1}^na_{x_j}(x)dx_j=0$.
Therefore the integral
\begin{equation}                                  \label{eq:6.17}
\alpha=\int_\gamma\sum_{j=1}^n\frac{1}{g_{00}(x)}g_{j0}(x)dx_j
\end{equation}
is the same for all isometric metrics.
\qed

A stationary metric $g$  is called static in $\Omega$  if  it has the form
\begin{equation}                                        \label{eq:6.18}
g_{00}(x)(dx_0)^2+\sum_{j,k=1}^n g_{jk}(x)dx_jdx_k,
\end{equation}
i.e.   when $g_{0j}(x)=g_{j0}(x)=0,\ 1\leq j\leq n,\ x\in\Omega$.

Suppose  the 
stationary  metric  $g(x)$  
in $\Omega$ is locally static,  i.e.  for any  point in $\Omega$  there   
is a neighborhood $V$  such  that the isometry $x_0'=x_0+a_V(x),x'=x $
transforms  the metric $g$  restricted  to $V$ to some  static metric
$g_{00}'(x)(dx_0')^2+\sum_{j,k=1}^ng_{jk}'(x)dx_jdx_k$, i.e.
$g_{j0}'(x)=g_{jk}(x)-a_{Vx_j}(x)=0,1\leq j\leq n,x\in V$.

Suppose  that metric $g$ is not globally static in $\Omega$,  i.e.
there is no $a(x)\in C^\infty(\overline\Omega)$  such
that  $x_0'=x_0+a(x), x'=x,$  transforms  $g$  to a static metric  $g'$  globally in
$\Omega$,  i.e.  $g$  and $g'$  are not isometric.
Then Theorem \ref{theo:6.2} implies that
$\Lambda_gf\big|_{\Gamma\times(0,T)}\neq 
\Lambda_{g'}f\big|_{\Gamma\times(0,T)}$  for some $f\in C_0^\infty(\Gamma)$,  i.e.
metric $g$  and $g'$  have a different physical impact.
This proves the gravitational AB  effect.

Note  that  $\int_\gamma\sum_{j=1}^n\frac{1}{g_{00}(x)}g_{j0}(x)dx_j=0$ 
for any $\gamma\subset V$  if  $g$  is locally isometric to  a static  metric in $V$.
If  $g$  is not globally  isometric to a static metric then  integral (\ref{eq:6.17})
may be not zero.  It plays a role  of the magnetic flux for
the magnetic AB  effect and $\alpha$  in (\ref{eq:6.17})  depends only on
the homotopic class of $\gamma$ when $g$ is locally static.

This  formulation  of the gravitational AB effect  was given  by Stachel in [46]  who proved 
it  for some explicit class  of locally  static  but  globally  not  static  metrics.

\subsection{A new inverse problem for the wave equation}

 Let  $g$ and $g'$  be two stationary metrics in 
$\R^n\setminus\cup_{j=1}^m\Omega_j$  such that
\begin{equation}                                      \label{eq:6.19}
g_{jk}(x)=g_{jk}'(x) \ \ \mbox{for} \ \ |x|>R,
\end{equation}
where $R$  is large.  Assume also   that
\begin{equation}                                         \label{eq:6.20}
g_{jk}(x)=\eta_{jk}+h_{jk}(x)\ \ \mbox{for} \ \ |x|>R,
\end{equation}
where 
\begin{align}
\nonumber
&
h_{jk}(x)=O\Big(\frac{1}{|x|^{1+\e}}\Big)\ \ \mbox{for}\ \ 
|x|>R,\e>0,
\\ 
&\sum_{j,k=1}^n\eta_{jk}dx_jdx_k=dx_0^2-\sum_{j=1}^ndx_j^2
\nonumber
\end{align}
is the Minkowski  metric  and $h_{jk}(x)=O(\frac{1}{|x|^{1+\e}}),\ \e>0,$
for $|x|>R$.

The following theorem is analogous to Theorem \ref{theo:4.3}.
 
\begin{theorem}                                  \label{theo:6.3}
Let $\Box_gu=0$  and $\Box_{g'}u'=0$  in $(0,T)\times(\R^n\setminus\bigcup_{j=1}^m\Omega_j)$,
where  $T>T_0$  (cf.  Theorem  \ref{theo:6.1}). 
Suppose (\ref{eq:6.19})  and (\ref{eq:6.20})  hold.
 Consider  two  initial-boundary
value  problems  with the same initial conditions
\begin{align}
\nonumber
&u(0,x)=u_0(x),\ \ \ \ u'(0,x)=u_0(x),
\\
\nonumber
&u_t(0,x)=u_1(x),\ \  u_t'(0,x)=u_1(x),\ \ x\in\R^n\setminus\bigcup_{j=1}^n\overline{\Omega_j},
\\
\nonumber
&u\big|_{(0,T)\times\partial\Omega_j}=0,\ \ u'\big|_{(0,T)\times\partial\Omega_j}=0,
\ \ 1\leq j\leq m,
\\
\nonumber
&u_0(x)=u_1(x)=0\ \ \mbox{in} \ B_R\setminus\bigcup_{j=1}^m\Omega_j,
\end{align}
where $B_R=\{x:|x|<R\}$.
Suppose  $g_{00}(x)>0,\ g_{00}'(x)>0,\ g^{00}(x)>0,\ (g')^{00}>0$  in
$\R^n\setminus\bigcup_{j=1}^n\Omega_j$.
If  $u_0(x)\in \overset{\circ}{H}_1(\R^n\setminus \bigcup_{j=1}^m\Omega_j)),\ u_1(x)
\in L_2(\R^n\setminus \bigcup_{j=1}^m\Omega_j)$  and if
$$
u(T,x)=u'(T,x),\ \ u_{x_0}(T,x)=u_{x_0}'(T,x),\ \ \ x\in\R^n\setminus B_R,
$$
for all $u_0(x)$  and $u_1(x)$,
then  metrics  $g$  and $g'$  are isometric  in $\R^n\setminus\cup_{j=1}^n\Omega_j$.
\end{theorem}

{\bf  Proof:}
It follows from  the existence and uniqueness theorem  that the solutions 
$u(x_0,x)$  and  $u'(x_0,x)$  belong  to 
$H_1((0,T)\times(\R^n\setminus\bigcup_{j=1}^n\Omega_j))$.
Let  $v=u(x_0,x)-u'(x_0,x)$.  Then 
$\Box_g v=0$  in $(0,T)\times(\R^n\setminus B_R)$
and  $v(0,x)=v_{x_0}(0,x)=0,\ v(T,x)=v_{x_0}(T,x)=0$  for  $x\in \R^n\setminus B_R$.
Extend  $v(x_0,x)$  by  zero  for $x_0>T$  and  $x_0<0$  and make the Fourier  
transform in 
$x_0: \tilde v(\xi_0,x)
=\int_{-\infty}^\infty v(x_0,x)e^{-ix_0\xi_0}dx_0$.  Then 
$\tilde v(\xi_0,x)$  belongs to  $L_2(\R^n\setminus B_R)$  for  all  $\xi_0\in \R$
and  satisfies the equation
$$
L\big(i\xi_0,\frac{\partial}{\partial x}\big)\tilde v(\xi_0,x)=0,\ \ 
x\in \R^n\setminus B_R,
$$
where $L\big(i\xi_0,i\xi\big)$  is the symbol  of $\Box_g$.

It follows  from H\"ormander ([31])  that  $\tilde v(\xi_0,x)=0$  in $\R^n\setminus B_R$  for 
all $\xi_0$.  Therefore  $u(x_0,x)=u'(x_0,x)$  for 
$x_0\in (0,T),\  x\in \R^n\setminus B_R$.
Then  $u\big|_{(0,T)\times\partial B_R}=u'\big|_{(0,T)\times\partial B_R}
\in H_{\frac{1}{2}}((0,T)\times \partial B_R)$
and $\frac{\partial u}{\partial\nu_g}\big|_{(0,T)\times\partial B_R}=
\frac{\partial u'}{\partial\nu_g}\big|_{(0,T)\times\partial B_R}
\in H_{-\frac{1}{2}}((0,T)\times \partial B_R)$  (cf. [20], \S 23),
i.e.  the boundary measurements of $u$  and $u'$  on $(0,T)\times\partial B_R$
are the same. 

Analogously to the proof of Lemma \ref{lma:4.5}  one can show that
$u\big|_{(0,T)\times\partial B_R}$
and $u'\big|_{(0,T)\times\partial B_R}$  are dense 
in $H_{-\frac{1}{2}}((0,T)\times\partial B_R)$.
Hence  the DN operators $\Lambda$  and $\Lambda'$  are equal on
$(0,T)\times\partial B_R$.
 Thus Theorem \ref{theo:6.1}  implies  that  $g$  and $g'$  are isometric.


\begin{thebibliography}
{20pt}

\bibitem[1]{} Y. Aharonov and D. Bohm,  Significance  of electromagnetic potentials
in quantum theory, Phys. Rev., 115, 485-491 (1959)

\bibitem[2]{} M. Ballesteros  and R. Weder,  High-velocity estimates  for the scattering  
operator 
and Aharonov-Bohm effect in three dimensions, Comm. Math.  Phys., 283, 345-398 (2009)

\bibitem[3]{} M. Ballesteros  and R. Weder, The Aharonov-Bohm effect and Tonomura et al.
experiments. Rigorous results, J. Math.  Phys., 50 (2009), no. 12, 122108,  54 pp         

\bibitem[4]{} M. Ballesteros  and R. Weder, Aharonov-Bohm effect and high-velocity estimates
of solutions  to the Schr\"odinger equations,  
Comm. Math. Phys. 303 (2011), 175-211



\bibitem[5]{} Belishev, M., 1997, Boundary control in reconstruction
of manifolds and metrics (the BC method),
Inverse Problems 13, R1-R45

\bibitem[6]{} Berry, M., Chambers, R., Large, M., Upstill, C., 
Walmsley, J., 1980, Eur. J. Phys. 1, 154-162

\bibitem[7]{} Cook, R., Fearn, H., Millouni, P., 1995, 
Am. J. Phys. 63, 705-710

\bibitem[8]{}
 J.J. Duistermaat, V. Guillemin,
 The spectrum of positive elliptic operator and periodic bicharacteristics,
Invent. Math. {29} (1975), 39-79

\bibitem[9]{}  Enss, V.,  Weder, R.,  The geometrical approach  to multidimensional  inverse scattering,  J. Math. Phys.,
36,  3902-3921 (1995)

\bibitem[10]{} Eskin, G., 2001, Global uniqueness in the inverse scattering
problem for the Schr\"{o}dinger operator with external Yang-Mills potentials,
Commun. Math. Phys. 222, 503-531

\bibitem[11]{} Eskin, G., 2003, Inverse boundary value problems 
and the Aharonov-Bohm effect, Inverse Problems 19, 49-62

\bibitem[12]{} Eskin, G., 2003, Inverse problems for the Schr\"{o}dinger
operators with electromagnetic potentials in domains with obstacles,
Inverse Problems 19, 985-998

\bibitem[13]{} Eskin, G., 2004,  Inverse boundary value problems in
domains with several obstacles, Inverse Problems 20, 1497-1516

\bibitem[14]{} Eskin, G., 2004,  On non-abelian Radon transform,
Russian Journ.  of Math. Phys. 11,  391-408 

\bibitem[15]{} Eskin, G., 2005, Inverse problems for Schr\"{o}dinger
equations with Yang-Mills potentials in domains with obstacles
and the Aharonov-Bohm effect, Journal of Physics : Conference 
series 12,  23-31

\bibitem[16]{} Eskin, G., 2006,  A  new approach to the hyperbolic 
inverse problems,  Inverse Problems 22,  815-831

\bibitem[17]{} Eskin, G., 2007, A new approach to the hyperbolic
inverse problems II: global step,  
 Inverse Problems 23, 2343-2356

\bibitem[18]{} G. Eskin, Inverse problems for the Schr\"odinger equations with time-dependent 
electromagnetic potentials and the Aharonov-Bohm effect,    Journ. of Math. Phys.,  49, 022105, 18 pp (2008)

\bibitem[19]{} G. Eskin, 
Optical Aharonov-Bohm effect:  inverse hyperbolic problem approach,
Comm. Math. Phys. 284,  no 2, 317-343 (2008)

\bibitem[20]{} G. Eskin, Lectures on linear partial differential equations,  AMS,  
Providence,  RI  (2011)

\bibitem[21]{}  G. Eskin,
 A simple proof of magnetic and electric Aharonov-Bohm effect,
 Comm. Math. Phys. {321} (2013), 747-767

\bibitem[22]{}  G. Eskin, Remarks on magnetic and electric Aharonov-Bohm  effects,  
ArXiv: 1007.3979

\bibitem[23]{} G. Eskin and H. Isozaki, Gauge  equivalence and Inverse Scattering for
Louge-Range Magnetic Potentials, Russian Journal of Math. Phys., vol. 18, No 1(2010), 54-63

\bibitem[24]{} G. Eskin, H. Isozaki and S. O'Dell,  Gauge equivalence  and inverse 
scattering  for Aharonov-Bohm effect,
 Comm. in PDE, 35, 2164-2194 (2010)

\bibitem[25]{} G. Eskin, J. Ralston, The Aharonov-Bohm effect in spectral asymptotics,  Analysis 
\&
 PDE 7-1 (2014),  245-266 

\bibitem[26]{} G. Eskin, J. Ralston,  Inverse boundary  value problems  for   system of 
partial differential equations,  Recent development  in theory and numerics, 105-113,  
World Sci. Publ., River Edge,  NJ,  2003


\bibitem[27]{GM}
 V. Guillemin, R. Melrose,
 The Poisson summation formula for manifolds with boundary,
 Advances in Math., 32(1979), 204-232


\bibitem[28]{}  B. Helffer,  Effet d'Aharonov-Bohm sur un \'etat born\'e de
l'equation de Schr\"odinger,
Comm. Math. Phys. 119(1988), 315-329


\bibitem[29]{}  S. Helgason,  The Radon  transform,  2nd. ed.  (1999),  Boston: Birkhauser


\bibitem[30]{}  L. H\"ormander,  The
Analysis of Linear Partial Differential Operators, I-IV,
Springer-Verlag, Berlin (1985)


\bibitem[31]{}   L. H\"ormander, Uniqueness theorem  for second  order elliptic differential
equations,  Comm. PDE, 8(1983),  21-64  

\bibitem[32]{} V. Isakov,   Carleman type estimates in an anisotropic case  and applications,
J. Diff. Equations, 105 (1993), 217-238

\bibitem[33]{} V. Isakov, 1998, Inverse problems for partial differential 
equations,  Appl. Math. Studies, vol. 127, Springer, 284 pp.

\bibitem[34]{} Y. Kannai, Off diagonal short time asymptotics for fundamental
solutions of diffusions equations, Commun.  in PDE 2(1977), no. 8, 781-830

\bibitem[35]{} Katchalov, A., Kurylev, Y., Lassas, M., 2001,
Inverse boundary spectral problems (Boca Baton : Chapman\&Hall)

\bibitem[36]{} Kurylev, Y. and Lassas, M., 2000,
Hyperbolic inverse problems with data on a part of the boundary, 
AMS/1P Stud. Adv. Math, 16, 259-272

\bibitem[37]{}
 R. Lavine, M. O'Carrol,
Ground state property and lower bounds on energy levels of particle in
a uniform magnetic field and external potential,
J. Math. Phys. { 18} (1977), 1908-1912.

\bibitem[38]{}
S. Markovitch,  Y. Aharonov,  T. Kaufferr,  B. Reznik,
Combines  electric and magnetic Aharonov-Bohm effects,  Am.J. Phys.  75,  pp 1141-1145  (2007)

\bibitem[39]{} Nicoleau, F., An inverse scattering problem with
the Aharonov-Bohm effect, Journ. Math. Phys., 41, 5223-5237 (2000)

\bibitem[40]{} Novikov, R., (2002), On determination of a gauge field on $\R^d$
from its non-Abelian Radon transform along oriented straight lines,
J.  Inst.  Math.  Jussieu 1,  559-629  


\bibitem[41]{} O'Dell, S.,  Inverse scattering for the Laplace-Beltrami 
operators  with complex-valued electromagnetic potentials and
embedded obstacles,
Inverse problems 22 No 5 (2006),  1579-1603

\bibitem[42]{} Olariu, S.  and  I. Iovitzu Popescu, 1985,
The quantum effects of electromagnetic fluxes,  Review of Modern Physics,
vol.  57,  N2,  339-436 

\bibitem[43]{}
S.N.M. Ruijsenaars, The Aharonov-Bohm effect and scattering theory, 
Annals of Phys., 146 (1983), 1-34.

\bibitem[44]{}
Ph. Roux and D. Yafaev, The scattering matrix for the Schr{\"o}dinger operator with a 
long-range electro-magnetic potential, J. Math. Phys., (44) (2003), 2762-2786.

\bibitem[45]{} Ph. Roux and D. Yafaev, On the mathematical theory of the Aharonov-Bohm effect, J. Phys.
A: Math. Gen., 35, (2002), 7481-7492

\bibitem[46]{} J. Stachel,  Globally stationary, but locally static space-times:
a gravitational analog  of the Aharonov-Bohm effect,  Phys.   Rev. D,  vol. 26, No 4 (1981),
1281-1290

\bibitem[47]{} Sundrum, R. and Tassie, L., 1986, Non-Abelian
Aharonov-Bohm effect, Feynman paths, and topology, J. Math. Phys. 27,
1566-70

\bibitem[48]{}
A. Tonomura, N. Osakabe, T. Matsuda, T. Kawasaki, J. Endo, S. Yano, and H. Yamada, 
Evidence for Aharonov-Bohm effect with magnetic field completely shielded from electron wave, 
Phys. Rev. Lett., 56, (1986), 792-795

\bibitem[49]{} Varadarajan, V.S., 2003,Vector bundles and connections 
in physics and mathematics : some historical remarks, Trends Math, 502-541,
Birkhauser, Basel 

\bibitem[50]{}
R. Weder,  The Aharonov-Bohm effect and time-dependent inverse scattering theory, 
Inverse Problems, (18) (2002), 1041-1056.

\bibitem[51]{}  R. Weder,  The electric Aharonov-Bohm effect, 
J. Math. Phys. 52, no. 5, 052109,  17 pp. (2011)

\bibitem[52]{} Wu, T.T. and Yang, C.N., 
Concept of nonintegrable phase factors and  global formulation of gauge fields,
1975, Phys. Rev. D 12, 3845-3857


\bibitem[53]{}
D. Yafaev, Scattering by magnetic fields, St. Petersburg Math. J.,  (17) (2006), 675-695
 



\end{thebibliography}
\end{document}